\documentclass[
twocolumn,
tightenlines,
10pt,
longbibliography,
showpacs,
nofootinbib,
notitlepage,
superscriptaddress]{revtex4-2}

\usepackage{amsmath, amssymb, amsthm, braket, enumitem, dsfont}
\usepackage{makecell, graphicx, subfiles, multirow}
\usepackage[font=scriptsize]{subcaption}
\usepackage[font=scriptsize, justification=raggedright]{caption}
\usepackage[page]{appendix}
\usepackage[colorlinks, allcolors=black]{hyperref}
\usepackage[capitalize, nameinlink]{cleveref}
\usepackage{tikz}
\usetikzlibrary{shapes,arrows}
\usetikzlibrary{backgrounds,fit,decorations.pathreplacing}
\usetikzlibrary{arrows.meta}

\theoremstyle{definition}
\newtheorem{assumption}{Assumption}
\crefname{assumption}{Assumption}{Assumptions}

\setlist[itemize]{leftmargin=10pt}

\begin{document}

\title{A Benchmarking Study of Quantum Algorithms for Combinatorial Optimization}


\author{Krishanu~Sankar}
\affiliation{1QB Information Technologies (1QBit), Vancouver, BC, Canada}

\author{Artur~Scherer}
\affiliation{1QB Information Technologies (1QBit), Vancouver, BC, Canada}

\author{Satoshi~Kako}
\affiliation{Physics \& Informatics Laboratories, NTT Research Inc., Sunnyvale, CA, USA}

\author{Sam~Reifenstein}
\affiliation{Physics \& Informatics Laboratories, NTT Research Inc., Sunnyvale, CA, USA}

\author{Navid~Ghadermarzy}
\affiliation{1QB Information Technologies (1QBit), Vancouver, BC, Canada}

\author{Willem~B.~Krayenhoff}
\affiliation{1QB Information Technologies (1QBit), Vancouver, BC, Canada}

\author{Yoshitaka~Inui}
\affiliation{Physics \& Informatics Laboratories, NTT Research Inc., Sunnyvale, CA, USA}

\author{Edwin~Ng}
\affiliation{Physics \& Informatics Laboratories, NTT Research Inc., Sunnyvale, CA, USA}
\affiliation{E. L. Ginzton Laboratory, Stanford University, Stanford, CA, USA}

\author{Tatsuhiro~Onodera}
\affiliation{Physics \& Informatics Laboratories, NTT Research Inc., Sunnyvale, CA, USA}
\affiliation{School of Applied and Engineering Physics, Cornell University, Ithaca, NY, USA}

\author{Pooya~Ronagh}
\email[Corresponding authors:\\]{
pooya.ronagh@1qbit.com, \\
yoshihisa.yamamoto@ntt-research.com}
\affiliation{1QB Information Technologies (1QBit), Vancouver, BC, Canada}
\affiliation{Institute for Quantum Computing, University of Waterloo, Waterloo, ON, Canada}
\affiliation{Department of Physics \& Astronomy, University of Waterloo, Waterloo, ON, Canada}

\author{Yoshihisa~Yamamoto}
\email[Corresponding authors:\\]{
pooya.ronagh@1qbit.com, \\
yoshihisa.yamamoto@ntt-research.com}
\affiliation{Physics \& Informatics Laboratories, NTT Research Inc., Sunnyvale, CA, USA}

\date{\today}

\begin{abstract}

We study the performance scaling of three quantum algorithms for combinatorial optimization:
 measurement-feedback coherent Ising machines~(MFB-CIM), discrete adiabatic quantum computation~(DAQC), and the \mbox{D\"urr--H{\o}yer} algorithm for quantum minimum finding (DH-QMF) that is based on Grover's search. We use {\sc MaxCut} problems as a reference for comparison, and \mbox{time-to-solution~(TTS)} as a practical measure of performance for these optimization algorithms.
 For each algorithm, we analyze its performance in solving two types of \mbox{\sc MaxCut} problems: weighted graph instances with
 randomly generated edge weights attaining 21 equidistant values from $-1$ to $1$; and randomly generated \mbox{Sherrington--Kirkpatrick} (SK) spin glass instances.
 We empirically find a significant performance advantage for the studied \mbox{MFB-CIM} in comparison to the other two algorithms. 
 We empirically observe a sub-exponential scaling for the median TTS for the MFB-CIM, in comparison to the almost exponential scaling  for DAQC and the proven $\widetilde{\mathcal O}\left(\sqrt{2^n}\right)$ scaling for \mbox{DH-QMF}.
 We conclude that the \mbox{MFB-CIM} outperforms DAQC and \mbox{DH-QMF} in solving {\sc MaxCut} problems.
\end{abstract}

\maketitle

\section{Introduction}
\label{sec:intro}

Combinatorial optimization problems are ubiquitous in modern science, engineering, and medicine. These problems are often NP-hard, so the runtime of classical algorithms for solving them is expected to scale exponentially. One approach for tackling such hard optimization problems is to map them to the Ising spin glass model~\cite{barahona1982computational},
\begin{equation*}
\mathcal{H} = - \sum_{i<j} J_{ij} S_{i} S_{j} - \sum_{i} h_i S_{i}\,.
\end{equation*}
Here, each $S_{i}$ represents a classical Ising spin attaining a value of $\pm 1$,
$[J_{ij}]$ is an Ising coupling matrix, and $[h_i]$ is a vector of local field
biases on the spin sites. When all $h_i$ are zero,
the Ising model is equivalent to a (weighted) {\sc MaxCut} problem on a graph
with vertices corresponding to the spin sites and edge weights corresponding
to the Ising couplings between the spin sites. Various mathematical
programming problems, such as partitioning problems, binary integer linear
programming, covering and packing problems, satisfiability problems, colouring
problems, Hamiltonian cycles, tree problems, and graph isomorphisms can be
formulated in the Ising model, with the required number of spins scaling at most
cubically with respect to the problem size~\cite{lucas2014ising}. This has been
a primary motivation for the recent extensive study of various Ising solvers.
Several potential areas of industrial application of Ising solvers include drug
discovery and bio-catalyst development (e.g., in lead optimization or virtual
screening), compressed sensing, deep learning (e.g., in the synaptic pruning of
deep neural network), scheduling (e.g., resource allocation and
traffic control), computational finance, and social networks (e.g., community
detection).

Approximate algorithms and heuristics, such as semi-definite programming
(SDP)~\cite{goemans1995improved}, simulated annealing
(SA)~\cite{kirkpatric1983optimization,johnson1989optimization} and its
variants~\cite{aramon2019physics,takemoto20192}, and breakout local search
(BLS)~\cite{benlic2017breakout} have been widely used as practical tools for
solving {\sc MaxCut} problems.
However, even problem instances of moderate size require substantial
computation time and, in the worst cases, solutions cannot be found with such
approximate algorithms and heuristics. To overcome these shortcomings, a search
for alternative solutions using various forms of quantum computing has been
actively pursued. Adiabatic quantum computation~\cite{farhi2001quantum}, quantum
annealing~\cite{kadowaki1998quantum,brooke1999quantum}, and the quantum approximate
optimization algorithm (QAOA)~\cite{farhi2014quantum} using circuit model quantum computers have
been proposed. A coherent Ising machine (CIM) using networks of quantum
optical oscillators has also been studied and
implemented~\cite{wang2013coherent,yamamoto2017coherent}.

Given that the present circuit model quantum computers suffer from short
coherence times, gate errors, and limited connectivity among qubits, a fair comparison between them and modern heuristics is not yet possible~\cite{choi2008minor,
boixo2014evidence, hamerly2019experimental}. This situation raises the important
question of whether quantum devices can, even in principle, provide
sensible solutions to combinatorial optimization problems, assuming
all sources of noise and imperfections can be overcome and ideal quantum
processors are built in the
future. In order
to address this pressing question, we perform a comparative numerical study on
three distinct quantum approaches, ignoring the effects of noise, gate errors,
and decoherence, that is, we compare the ultimate theoretical limits of three
quantum approaches.

The first approach is based on the effects of constructive and destructive
quantum interference of amplitudes in a circuit model quantum computer that
utilizes only unitary evolution of pure states and projective (exact)
measurement of qubits. The approach uses Grover's search
algorithm~\cite{grover1996fast,grover1997quantum}
as a key computational primitive. We call this approach ``DH-QMF''
in reference to {D\"urr} and H{\o}yer's ``quantum minimum finding'' algorithm~\cite{durr1996quantum}.
Our scaling analysis of  \mbox{DH-QMF} is presented
in \cref{sec:res_qmf}; additional details are provided in \cref{app:app_qmf}.
A review of related literature and a discussion of how our analysis differs from previous work are given in \cref{app_related_work}.

The second approach is based on adiabatic quantum state preparation
implemented on a circuit model quantum computer. The underlying concept,
the quantum adiabatic theorem, goes as far back as the seminal
work of Born and Fock~\cite{born1928beweis}. Its application
to quantum computing and solving optimization problems was introduced
by Farhi et al.~\cite{farhi2001quantum}.
A Trotterized approximation to adiabatic evolution gives rise to a discrete
implementation suitable for the circuit model. We refer to this approach as
\lq\lq{}discrete adiabatic quantum computation\rq\rq{}~(DAQC).
This algorithm uses an iterative unitary evolution of pure states in a quantum circuit
according to a mixing Hamiltonian and a problem Hamiltonian, which in the framework
of adiabatic quantum computation correspond to the initial and final Hamiltonians of evolution, respectively. The coefficients in the exponents form the gate parameters, which can be treated as hyperparameters that follow a tuned schedule, and the overall number of Trotter steps directly pertains to the circuit depth of the algorithm.
To attain the ultimate theoretical performance limit,
we use pre-tuned DAQC schedules
and allow for quantum circuits of arbitrary depth.
Our scaling analysis of  \mbox{DAQC} is presented in \cref{sec:res_qaoa}; additional details are provided in \cref{app_hptuning}. In the presence of noise, the closely related NISQ-type “quantum approximate optimization algorithm” (QAOA) [12, 22] (see \cref{app_related_work}) deviates from DAQC in its use of (a) shallow (i.e., short-depth) quantum circuits (hence, attempting to perform ground-state preparation diabatically as opposed to adiabatically) and (b) an outer classical optimization routine to variationally optimize the diabatic evolution. We do not include QAOA in this study in view of its poor and unstable scaling, which we empirically observed in comparison to that of DAQC. This poor performance is  exacerbated especially if the overhead of the classical optimizer is taken into account. Our observations are consistent with the challenges of variational quantum algorithms in overcoming the barren plateau problem~\cite{holmes2021connecting,mcclean2018barren}. Further details are provided in~\cref{app_qaoaopt}.

The third approach is based on a measurement-feedback coherent Ising machine
(MFB-CIM)~\cite{leleu2019destabilization, kako2020coherent}. This algorithm
utilizes a quantum-to-classical  transition in an open-dissipative, non-equilibrium
network of quantum oscillators.
A critical phenomenon known as pitchfork bifurcation realizes the
transition of squeezed vacuum states to coherent states in the
optical parametric oscillator. The measurement-feedback circuit plays several important roles.
It continually reduces entropy and sustains a quasi-pure state in the quantum oscillator network in a controlled manner
using repeated approximate
measurements.
It, additionally, implements the Ising coupling matrix $[J_{ij}]$
and local field vector $[h_i]$ in an iterative fashion. Finally, it removes
the amplitude heterogeneity among the oscillators and destabilizes the machine
state out of local minima. \cref{table:1} summarizes the differences among the
three approaches studied in this paper.

\begin{table*}[t]
\centering
\begin{tabular}{l@{\hskip 0.2in}|l@{\hskip 0.2in}l@{\hskip 0.2in}l}
\hline\\
& DH-QMF  & DAQC & MFB-CIM \\ [0.5ex]
\hline\hline
&&&\\
Quantum dynamics
& \thead[l]{Closed-unitary}
& \thead[l]{Closed-unitary}
& \thead[l]{Open-dissipative}
\\
&&&\\
Operational principle
& \thead[l]{Amplitude amplification \\ by quantum interference}
& \thead[l]{Adiabatic quantum  evolution}
& \thead[l]{Quantum-to-classical transition}
\\
&&&\\
Information carrier
& \thead[l]{Digital (spin-1/2 particle)}
& \thead[l]{Digital (spin-1/2 particle)}
& \thead[l]{Analog (harmonic oscillator)}
\\
&&&\\
Decoherence time & $T_2\to\infty$ & $T_2\to\infty$ & $T_2\to\infty$ \\
&&&\\
Dissipation time & $T_1\to\infty$ & $T_1\to\infty$ & $T_1:$ finite
\\
&&&\\
Gate error
& \thead[l]{None}
& \thead[l]{None}
& \thead[l]{Vacuum noise limited}
\\
&&&\\
Spin-spin coupling
& \thead[l]{all-to-all}
& \thead[l]{all-to-all}
& \thead[l]{all-to-all}
\\[1ex]
\hline
\end{tabular}
\caption{Three approaches studied for {\sc MaxCut} problems: the D\"urr--H{\o}yer algorithm for
quantum minimum finding (DH-QMF) based on Grover's search,
the discretized adiabatic quantum computation algorithm (DAQC), and
the measurement-feedback coherent Ising machine (MFB-CIM).}
\label{table:1}
\end{table*}

When studying quantum algorithms, it is important to consider the effects of
noise and control errors, and the overhead needed to overcome them. Several previous
studies have investigated these effects on the performance of QAOA 
(here viewed as a NISQ-type, diabatic counterpart to DAQC). In references
\cite{xue2019effects,marshall2020characterizing}, various Pauli noise channels,
namely the dephasing, bit-flip, and the depolarizing noise channels, are
considered. These two papers report on the fidelity of the state prepared
by a noisy QAOA circuit to the state prepared by an ideal QAOA circuit, for
varying amounts of physical noise affecting the circuit. In contrast,
\cite{guerreschi2019qaoa} models noise via single-qubit rotations by an angle
chosen from a Gaussian distribution with variance
values of $T_\text{G}/T_2$, where $T_\text{G}$ is the
gate time and $T_2$ is the decoherence time of the qubits. All three papers
provide insight into how noise affects the expected energy of the prepared state.
Note that arbitrary-depth circuits are permitted in our study of
DAQC, and optimal circuit depths resulting in the best algorithmic performance can be 
substantially larger than the size of circuits suitable for NISQ devices.

DH-QMF circuits are much deeper than typical DAQC circuits; thus, their performance is significantly hampered by various sources of
noise unless the algorithm is run on a fault-tolerant quantum computer with
quantum error
correction~\cite{pablo1999noise, long2000dominant, azuma2002decoherence,
shenvi2003effects, shapira2003effect,salas2008noise, gawron2012noise,
reitzner2019grover}.
Different noise models have been used to study the
sensitivity of Grover's search by simulating small quantum circuits that
apply it to simple functions.
\cite{pablo1999noise} introduces random Gaussian noise on each step of
Grover's search. \cite{long2000dominant} studies the effect of gate
imperfections on the probability of success of the algorithm.
\cite{shapira2003effect} examines the effect of unbiased and isotropic unitary
noise resulting from small perturbations of Hadamard gates.
\cite{azuma2002decoherence}~models the effect of decoherence by introducing
phase errors in each qubit and time step and using a perturbative method.
\cite{salas2008noise} conducts a numerical analysis on the effects
of single-qubit and two-qubit gate errors and memory errors,
modelling decoherence using a depolarizing channel.
The impact of using a noisy oracle is examined in
\cite{shenvi2003effects}, wherein noise is modelled by introducing random phase
errors. The effects of localized dephasing are studied in
\cite{reitzner2019grover}. Finally, \cite{gawron2012noise}
investigates the effects of various noise channels using trace-preserving,
completely positive maps applied to density matrices.

In our benchmark study, by “solving” an optimization problem we mean finding an actual optimal solution with high probability (as opposed to an approximate, suboptimal solution). For a fair comparison, this notion pertains to all three algorithms considered in this work. As a practical measure of the algorithms’ performance, we use the time-to-solution (TTS) metric, which refers to the time required to find an optimal solution with high confidence. For the \mbox{MFB-CIM} and DAQC, the TTS is computed as the number of \lq\lq{}shots\rq\rq{} (i.e., trials) that must be performed to ensure a high probability (specified by a target probability of success, often taken to be 0.99) of observing an optimal solution at least once, multiplied by the time required for the execution of a single shot. Similarly, for DH-QMF, the TTS is computed as the overall number of Grover iterations required to ensure a target probability of success of observing an actual optimal solution,  multiplied by the time required to implement a single Grover iteration.

We have evaluated the wall-clock TTS of the three algorithms
introduced above for solving {\sc MaxCut} problems, and empirically
found exponential scaling laws for them already in the relatively small problem
size range of 4 to 800 spins.
In order to elucidate the ultimate performance limits of these solvers, we
assume no extrinsic noise, gate errors, or connectivity limitations exist in the hardware.
That is, we assume that phase decoherence ($T_2$) and energy
dissipation ($T_1$) times are infinite and gate errors are absent. 
Consequently, the overheads associated with performing quantum error correction and building fault-tolerant architectures and protocols are not included in our benchmarking study, as they would make the comparison less favourable for circuit model quantum algorithms against the MFB-CIM.
We also assume that all spins (represented by qubits in the circuit model)
can be coupled to each other via (non-local)
spin--spin interaction with a universal gate time of 10 nanoseconds.
Therefore, there is no need to implement expensive sequences
of swap gates or other bus techniques for
transferring quantum information across the hardware.
However, since energy dissipation and stochastic noise both constitute important computational
resources for the MFB-CIM, we allow a finite energy dissipation time $T_1$, as well as a finite
gate error limited by vacuum noise, for the MFB-CIM.

We emphasize that we compare optimistic lower bounds
on the TTS for the circuit-model quantum algorithms considered in this paper.
It is for this reason that we do not include the overhead
costs associated with quantum error correction and the realization of fault-tolerant
quantum computation schemes that become necessary for deep
circuits of DH-QMF and DAQC. The impact of such overhead costs, for instance,
when using topological surface code built of error-prone physical
qubits and gates for encoding logical qubits and logical operations,
is estimated more precisely in other recent works, for example, \mbox{in~\cite{babbush2021focus,gidney2021how,campbell2019applying}.}
The asymptotic overhead introduced
by fault-tolerant architectures can be inferred as follows.
For DAQC, the circuit depth of each Trotter layer scales linearly
with the problem size $n$; see~\cref{sec:res_qaoa}. Therefore, the  error rate of each logical gate
must scale inversely with $n$, necessitating a code distance logarithmic
in $n$. Fault-tolerant operations on an encoding scheme of distance $d$ introduce at
least a factor of $d$ in physical gate time overhead. Hence, we can expect
the TTS for the DAQC algorithm to increase by an $\Omega(\log n)$ factor.
Similarly, for DH-QMF, which is based on Grover's search requiring circuits
of depth  $\widetilde{\Theta}\left(\sqrt{2^n}\right)$  (see \cref{sec:res_qmf}),
the incurred overhead results in an increase in the TTS by a factor of $\Omega(n)$.
This rough estimate does not account for compilation overhead, which would typically further increase the TTS. In addition, it also does not account for overheads caused by decoding and active error correction.

From a fundamental viewpoint, such a comparative study is of interest
but the outcome is difficult to predict, because the three algorithms are based on
completely different computational principles,
as shown in \cref{table:1}. The DH-QMF algorithm iteratively deploys Grover's
search, which uses a unitary evolution of a superposition of
computation basis states in order to amplify the
amplitude of a target state by successive constructive interference, while the
amplitudes of all the other states are attenuated by destructive interference.
The DAQC algorithm attempts to prepare a pure state that has a large overlap with the ground
state of the optimization problem through an approximation of the adiabatic quantum evolution. Finally, the ground state search mechanism of the MFB-CIM employs a collective phase transition at the threshold of an optical parametric oscillator (OPO) network. The correlations formed among the squeezed vacuum states in OPOs below the threshold guide the network toward oscillating at a ground state.

It is worth noting that all the algorithms in our study in various ways rely on hybrid quantum--classical architectures  for computation.
In an MFB-CIM with self-diagnosis and dynamical feedback control, a
classical processor plays an important role by detecting when the OPO network is trapped in local minima, and destabilizes it out of those states.
The DH-QMF algorithm also relies on comparing the values of an objective function with a
(classical) threshold value. This threshold value is updated in a classical coprocessor  as DH-QMF proceeds. Finally, DAQC relies on tuning a set of parameters (e.g., the rotation
angles of quantum gates). These parameters can be treated as hyperparameters
of a predefined approximate adiabatic evolution and tuned for the problem type
solved by the algorithm. Alternatively, the quantum
circuit can be viewed as a variational ansatz, in which case the gate parameters
are optimized using a classical optimizer. In the latter case, the algorithm can be
considered as a variational quantum algorithm~\cite{cerezo2020variational}.
QAOA is commonly viewed as such an algorithm.
In previous studies, the contribution of the variational optimization of  DAQC 
parameters to the TTS has often been ignored. In fact, while both
approaches (i.e., hyperparameter tuning and variational optimization) have been
adopted for solving {\sc MaxCut} problems using QAOA~\cite{crooks2018performance,zhou2020quantum}, our investigation makes it clear
that the variational approach hurts the TTS scaling significantly. The
optimization landscape for such a variational quantum algorithm is ill-behaved,
which results in a poor and unstable scaling for TTS with respect to
the size of the {\sc MaxCut} instances (refer to \cref{app_qaoaopt}). As a
result, the TTS scalings reported in this paper rely on pre-tuned DAQC schedules rather than variational optimization.

\vspace*{5pt}

\section{Scaling of the MFB-CIM}
\label{sec:res_cim}
A CIM is a non-equilibrium, open-dissipative computing
system based on a network of degenerate OPOs to find a ground state of Ising problems~\cite{utsunomiya2011mapping,
wang2013coherent, marandi2014network, inagaki2016large, mcmahon2016fully}. The
Ising Hamiltonian is mapped to the loss landscape of the OPO network formed by
the dissipative coupling rather than the standard Hamiltonian coupling. By providing a sufficient gain to compensate for the overall network loss, a ground state of the target Hamiltonian is expected to build up spontaneously as a single oscillation mode~\cite{yamamoto2017coherent}.
However, the mapping of the cost function to the OPO network loss landscape
often fails in the case of a frustrated spin problem due to the OPO amplitude inhomogeneity~\cite{wang2013coherent,leleu2019destabilization}. In addition, with an increasing
number of local minima occurring as problem sizes become larger, the machine state is trapped in
those minima for a substantial amount of time, thereby causing the machine to report suboptimal
solutions~\cite{yamamoto2017coherent,kako2020coherent}.
Recently, self-diagnosis and dynamical feedback mechanisms have
been introduced by a measurement-feedback CIM (MFB-CIM) to overcome these
problems~\cite{leleu2019destabilization,kako2020coherent}. This is achieved by a mutual coupling field dynamically modulated for each OPO to suppress the amplitude inhomogeneity and simultaneously to destabilize the machine's state out of local minima.

\subsection{Principle of Operation}
\label{sec:MFB-CIM_A}

\begin{figure}[h!]
\centering
\includegraphics[width=1.0\linewidth]{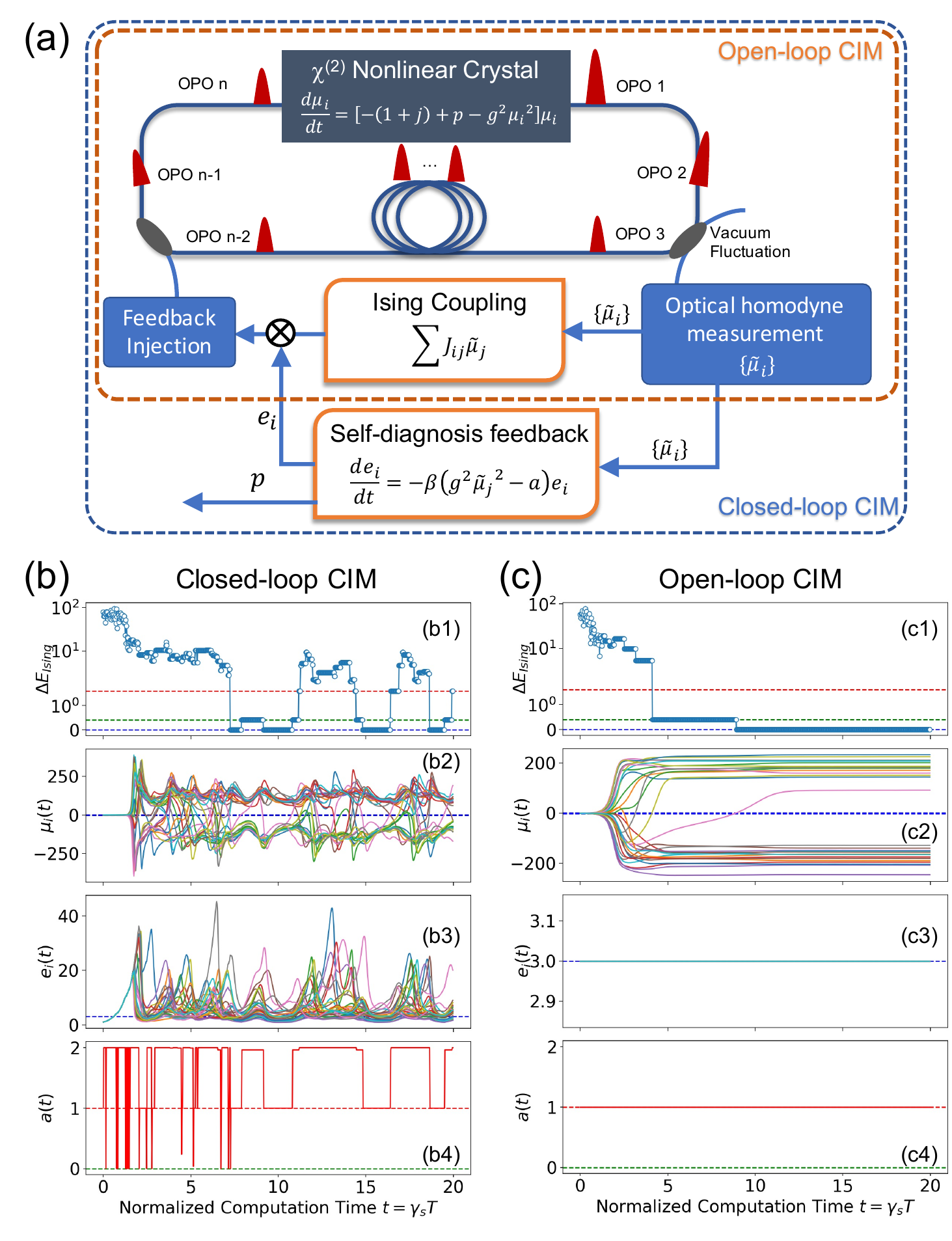}
\caption{(a) Schematic diagram of the measurement-feedback coupling CIMs with
and without the self-diagnosis and dynamic feedback control
(closed-loop and open-loop CIMs) indicated using dashed blue and orange lines, respectively. (b) and (c) Dynamical behaviour of the
closed-loop and open-loop CIMs, respectively. (b1) and (c1) Inferred Ising energy
(the dashed horizontal lines are the lowest three Ising eigen-energies). (b2) and (c2)
Mean-field amplitude $\mu_i(t)$. (b3) and (c3) Feedback-field amplitude $e_i(t)$.
(b4) Target squared amplitude $a(t)$. (c4) Pump rate $p(t)$.}
\label{fig:basic}
\end{figure}

A schematic diagram of two MFB-CIMs with predefined feedback control
(hereafter referred to as ``open-loop CIM'') and with self-diagnosis and dynamical
feedback control (hereafter referred to as ``closed-loop CIM''), is shown in~\cref{fig:basic}(a). If the fibre ring resonator has high finesse, both  CIMs are modelled via the Gaussian quantum theory
\cite{shoji2017quantum, inui2020noise}. The dynamics captured by the master equation for the density operator
(i.e., the Liouville--von Neumann equation) is driven by the parametric interaction
Hamiltonian,
\mbox{$\hat{\mathcal H} = i \hbar \frac{S}{2} \sum_{i} \left( \hat{a}^{\dagger 2}_i - \hat{a}^2_i \right)$}, the measurement-induced state reduction (the third term on the right-hand side in \cref{eq:master-equation}), the coherent injection (the fourth term on the right-hand side in \cref{eq:master-equation}),
as well as three Liouvillians. The Liouvillians pertain to the linear loss due to measurement and injection couplings, \mbox{$\hat{\mathcal L}_\text{c}^{(i)} = \sqrt{J} \hat{a}_{i}$},
two-photon absorption loss (i.e., parametric back conversion) in a degenerate
parametric amplifying device, $\hat{\mathcal L}_{2}^{(i)} = \sqrt{B/2}\,\hat{a}_{i}^{2}$,
and background linear losses, $\hat{\mathcal L}_{1}^{(i)} = \sqrt{\gamma_{\text{s}}} \hat{a}_{i}$, respectively~\cite{inui2020noise}. The master equation is thus
given by
\begin{widetext}
\begin{align}
\frac{d}{dt} \hat{\mathcal \rho}
= - \frac{i}{\hbar} \left[ \hat{\mathcal H}, \hat{\mathcal \rho} \right]
&+ \sum_{i=1}^{n} \sum_{k=1,2,\text{c}} \left( \left[ \hat{\mathcal L}_{k}^{(i)},
\hat{\mathcal \rho} {\hat{\mathcal L}_{k}^{(i)\dagger}} \right] + h.c. \right)\nonumber\\
{\label{eq:master-equation}}
&+ \sqrt{J}\sum_{i=1}^{n} \Bigl(\hat{a}_{i}\hat{\rho}+\hat{\rho}\hat{a}^{\dagger}_{i}-\langle \hat{a}_i+\hat{a}^{\dagger}_i\rangle\hat{\rho}\Bigr)w_{i}
+ \frac{J}{2} \sum_{i,k=1}^n e_{i}(t) J_{ik} \Bigl(\langle \hat{a}_{k}+\hat{a}^{\dagger}_{k}\rangle+\frac{w_{k}}{\sqrt{J}}\Bigr) [\hat{a}_i^{\dagger}-\hat{a}_i,\hat{\rho}].
\end{align}
\end{widetext}
In general, the numerical integration of~\cref{eq:master-equation} requires
exponentially growing resources as the problem size $n$ (i.e., the number of spins)
increases. Generally speaking, the size of the density matrix scales as
$\mathcal{O}({n_{0}}^{n} \times {n_{0}}^{n})$, where $n_{0} \gg 1$ is the
maximum number of photons possible for each OPO pulse. In MFB-CIMs, however, there is no entanglement
between the OPO pulses, that is, the OPO states are separable. Therefore, the
simulation's memory requirements reduce to $\mathcal O(n \times {n_{0}}^{2})$.
However, this reduction still yields too many c-number differential equations
due to the large upper bounds on the number of photons $n_{0} \lesssim
10^{7}$ and the number of spins $n \leq 1000$. The Gaussian quantum model has been
introduced to overcome this difficulty \cite{inui2020noise, kako2020coherent}.

In the case of a small saturation parameter, \mbox{$g^2 = B/\gamma_{\text{s}} \ll 1$}, we can
split the $i$-th OPO's pulse amplitude operator, $\hat{a}_i = \frac{1}{\sqrt{2}}
(\hat{X}_i + i\hat{P}_i)$, into the mean field and small fluctuation operators,
$\hat{X}_i = \langle \hat{X}_i\rangle + \Delta \hat{X}_i$ and $\hat{P}_i =
\langle \hat{P}_i\rangle + \Delta \hat{P}_i$. The saturation parameter $g^2$
corresponds to the inverse photon number at twice the threshold pump rate of a
solitary OPO. With an appropriate choice of the pump phase, each OPO mean-field is
generated only in an $\hat{X}$-quadrature, that is, $\langle \hat{P}_i\rangle = 0$.
The equation of motion for the mean field $\mu_i = \langle \hat{X}_i\rangle
/\sqrt{2}$ and the variances $\sigma_i = \langle \Delta {\hat{X}_i}^2 \rangle $
and $\eta_i = \langle \Delta {\hat{P}_i}^2 \rangle $ obey the following
equations \cite{inui2020noise}:

\begin{widetext}
\begin{align}
{\label{eq:mu-equation}}
\frac{d}{dt} \mu_i
&=\left[ - \left( 1 + j \right) + p - g^2 \mu^2_i \right] \mu_i
+ j\xi e_{i}(t) \sum_{k}{J_{i k} \tilde{\mu}_k }
+ \sqrt{j} \left( \sigma_i -1/2 \right) w_i\,, \\
{\label{eq:xval-equation}}
\frac{d}{dt} \sigma_i
&= 2 \left[ - \left( 1 + j \right) + p - 3 g^2 \mu^2_i \right] \sigma_i
- 2j {\left( \sigma_i -1/2 \right)}^2
+ \left[ \left( 1 + j \right) + 2 g^2 \mu^2_i \right],\\
{\label{eq:pval-equation}}
\frac{d}{dt} \eta_i
&= 2 \left[ - \left( 1 + j \right) - p -  g^2 \mu^2_i \right] \eta_i
+ \left[ \left( 1 + j \right) + 2 g^2 \mu^2_i \right].
\end{align}
\end{widetext}
Here, $t = \gamma_{\text{s}} T$ refers to   normalized and dimensionless time, where $T$ is
physical (or wall-clock) time, and $\gamma_{\text{s}}$ is the background loss rate of the cavity. The time $t$ is normalized so that the background
linear loss (with a signal amplitude decay rate of $1/e$) is $1$. The term $-(1+j)$ in
\cref{eq:mu-equation,eq:xval-equation,eq:pval-equation} represents a background
linear loss ($-1$) and an out-coupling loss ($-j$) for optical homodyne measurement and feedback injection, where $j = J/\gamma_{\text{s}}$ is a normalized out-coupling rate (see
\cref{fig:basic}(a)). The parameter $p = S/\gamma_{\text{s}}$ is a normalized linear gain coefficient
provided by the parametric device. The term $g^2 \mu^2_i$ represents two-photon
absorption loss (i.e., back conversion from signal to pump fields). The second and
third terms on the right-hand side of~\cref{eq:mu-equation}, respectively, represent
the Ising coupling term and the measurement-induced shift of the mean-field
$\mu_i$. The inferred mean-field amplitude, $\tilde{\mu}_k = \mu_k +
\sqrt{\frac{1}{4j}} w_k$, deviates from the internal mean-field amplitude
$\mu_k$ by a finite measurement uncertainty in the optical homodyne detection. The
random variable $w_k \sqrt{\Delta t}$ attains values drawn from the standard
normal distribution, where $\Delta t$ is a time step for the numerical
integration of \cref{eq:mu-equation,eq:xval-equation,eq:pval-equation}. The
$k$-th Ising spin $S_{k}=\pm 1$ is determined by the sign of the inferred
mean-field amplitude, $S_{k}=\tilde{\mu}_{k}/|\tilde{\mu}_{k}|$. $J_{ik}$ is the
Ising coupling coefficient and $e_{i}(t)$ is a dynamically modulated feedback-field amplitude, while $\xi = 1/\sqrt{{\frac{1}{N}} \sum_{i,j} \left| J_{ij} \right|}$ is a feedback-gain parameter. The second term on the right-hand side of
\cref{eq:xval-equation} represents the measurement-induced partial state
reduction of the OPO field. The last terms of
\cref{eq:xval-equation,eq:pval-equation}, respectively, represent the variance
increase by the incident (fresh) vacuum field fluctuations via linear loss and
the pump noise coupled to the OPO field via gain saturation.

The dynamically modulated feedback-field amplitude $e_i (t)$ is introduced
to reduce the amplitude inhomogeneity~\cite{leleu2019destabilization}, which is
determined by the inferred signal amplitude $\tilde{\mu}_i$:
\begin{equation}{\label{eq:error-equation}}
\frac{d}{dt} e_i (t) = - \beta \left[ g^2 {\tilde{\mu}_i}^2 - a \right] e_i (t).
\end{equation}
Here, $\beta$ is a positive constant representing the rate of change for the exponentially growing or attenuating
feedback amplitude $e_i (t)$, and $a$ is a target squared amplitude. Both $a$ and the pump rate $p$ are
dynamically determined by the difference of the current Ising energy
$\mathcal{E}(t)=-\sum_{i<k} J_{ik} S_{i} S_{k}$ and the lowest Ising energy
$\mathcal{E}_{\text{opt}}$ visited previously:
\begin{align}
{\label{eq:a}}
a(t) &= \alpha + \rho_{a} \tanh
\left( \frac{\mathcal{E}(t) - \mathcal{E}_{\text{opt}}}{\Delta} \right),\\
{\label{eq:p}}
p(t) &= \pi - \rho_{p} \tanh
\left( \frac{\mathcal{E}(t) - \mathcal{E}_{\text{opt}}}{\Delta} \right).
\end{align}
Here, $\pi$, $\alpha$, $\rho_{a}$, $\rho_{p}$, and $\Delta$ are predetermined
positive parameters which characterize the self-diagnosis and dynamic feedback
control.

The machine can distinguish the following three modes of operation from the
energy measurements. When \mbox{$\mathcal{E}(t) - \mathcal{E}_{\text{opt}} < -\Delta$}, the
machine is in a gradient descent mode and moving toward a local minimum, in which
case the pump is set to a positive value of $\pi + \rho_{p}$ (leading to parametric
amplification). When $|\mathcal{E}(t) - \mathcal{E}_{\text{opt}}| \ll \Delta$, the
machine is close to, or trapped in, a local minimum, in which case the pump is
switched off (i.e., there is no parametric amplification) so as to destabilize the current
spin configuration. When $\mathcal{E}(t) - \mathcal{E}_{\text{opt}} > \Delta$,
the machine is attempting to escape
from a previously visited local minimum, in which case the pump is set to a negative value of
$\pi - \rho_{p}$ (i.e., there is parametric de-amplification) to increase the rate of spin
flips.

\cref{fig:basic}(b) shows the time evolution of a closed-loop CIM to demonstrate
its inherent exploratory behaviour from one local minimum to another. We solve a \mbox{{\sc
MaxCut}} problem with randomly generated discrete edge-weights $J_{ij} \in \{-1, -0.9, \ldots, 0.9, 1\}$ over $n=30$ vertices, for which an exact solution is obtained
by performing an exhaustive search. The dynamical behaviour of the inferred Ising energy
measured from the ground state energy, $\Delta \mathcal{E}(t) = \mathcal{E}(t) -
\mathcal{E}_{\text{G}}$, the mean amplitude, $\mu(t)$, the feedback-field
amplitude, $e(t)$, and the target squared amplitude, $a(t)$, are shown in
\cref{fig:basic} (b) and (c). The results shown in \cref{fig:basic}(b) are
taken from a single trial for one particular problem instance
and a particular set of noise amplitudes $w_i \sqrt{\Delta t}$. The
feedback parameters are set to $\alpha = 1.0$, $\pi = 0.2$, $\rho_a = \rho_p =
1.0$, $\Delta = 1/5$, and $\beta = 1.0$~\cite{kako2020coherent}. The saturation
parameter and the out-coupling loss are chosen as $g^2 = 10^{-4}$ and $j=1$, respectively. The time step $\Delta t$ for the numerical
integration of \cref{eq:mu-equation,eq:xval-equation,eq:pval-equation} is
identical to the normalized round-trip time
$\Delta t_{\text{c}} = \gamma_{\text{s}} \Delta T_{\text{c}} = 0.025$. This means the signal-field lifetime $1/\gamma_\text{s}$ is 40 times greater than the round-trip time.

As shown in \cref{fig:basic}(b1), the inferred Ising energy $\mathcal{E}(t)$
fluctuates up and down during the search for a solution even after the machine finds
one of the degenerate ground states.
As shown in \cref{fig:basic}(b2), the measured squared
amplitude $g^2 {\tilde{\mu_i}}^2$ is stabilized to the target squared amplitude
$a$ through the dynamically modulated feedback mean-field $e_i (t)$. Several OPO
amplitudes, however, flipped their signs
followed by an exponential increase in $e_i(t)$, while most
other OPOs maintained a target amplitude. During this spin-flip
process, the feedback-field amplitude $e_i(t)$ increases exponentially and then decreases
exponentially after the OPO's squared amplitude $g^2 \tilde{\mu}^2_{i}$ exceeds
the target squared amplitude $a(t)$. The mutual coupling strength
$\sum_k J_{ik} \tilde \mu_k$ is adjusted in order to decrease the energy
continuously by flipping the ``wrong'' spins and preserving the ``correct'' ones.
If the machine reaches local minima, which may also include global minima (in which case there are degenerate ground states), the current Ising energy $\mathcal{E}(t)=-\sum_{i<k} J_{ik} S_{i}
S_{k}$ is roughly equal to the minimum Ising energy $\mathcal{E}_{\text{opt}}$
previously visited ($\mathcal{E}(t) \simeq \mathcal{E}_{\text{opt}}$). The machine then
decreases the target squared amplitude $a$, which helps it to escape from the
local minimum. During this escape, the current Ising energy $\mathcal{E}(t)$
becomes greater than the minimum Ising energy $\mathcal{E}_{\text{opt}}$. The machine
then switches the pump rate $p$ to a negative value and deamplifies the signal
amplitude, which results in further destabilization of the local minimum. As a
consequence of such dynamical modulation of the pump rate $p$ and the target
squared amplitude $a$, the machine continually escapes local minima, migrating
from one local minimum to another as the computation carries on.

\begin{figure}[h!]
\centering
\includegraphics[width=1.0\linewidth]{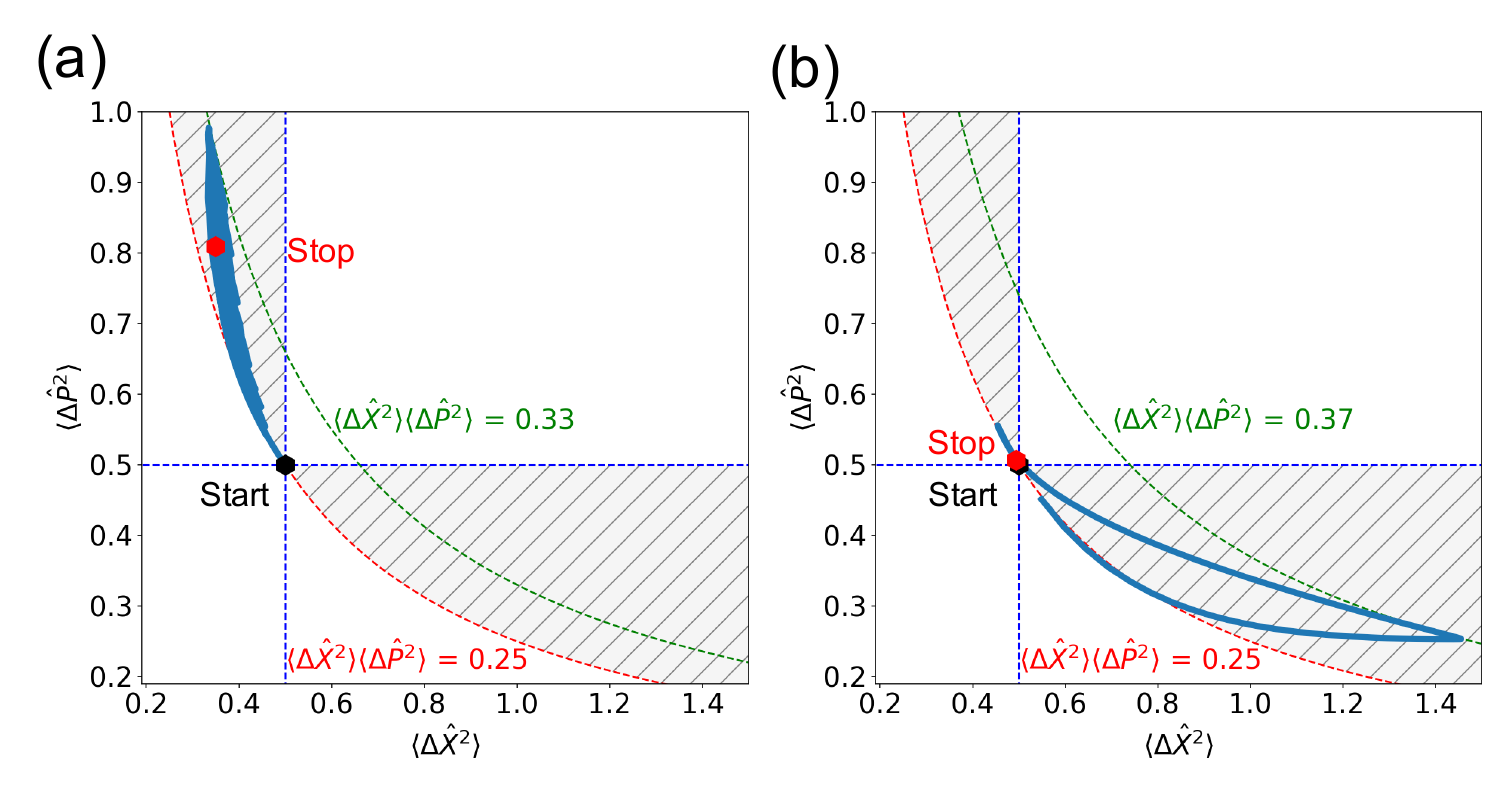}
\caption{Variances $\langle \Delta \hat{X}^2 \rangle$ and $\langle \Delta
\hat{P}^2 \rangle$ for (a) a closed-loop CIM and (b) an open-loop CIM. The shaded areas
show the quantum domains ($\langle \Delta \hat{X}^2 \rangle < 1/2$ or $\langle
\Delta \hat{P}^2 \rangle < 1/2$). Note that these are the results for one particular OPO, i.e., for one of the trajectories shown in \cref{fig:basic}(b) and (c).}
\label{fig:dx2-dp2-plot}
\end{figure}

\cref{fig:basic}(c) shows the time evolution of an open-loop CIM, in which both the
pump rate $p$ and the feedback-field amplitude $e_i (t)$ are predetermined constants.

As shown in \cref{fig:dx2-dp2-plot}(a) and (b), the quantum states of the OPO
fields satisfy the minimum uncertainty product, $\langle \Delta \hat{X}^2
\rangle \langle \Delta \hat{P}^2 \rangle = 1/4$, with a small excess factor of
$\sim 30\%$ despite the open-dissipative nature of the machine. We note that
each OPO state is in a quantum domain ($\langle \Delta \hat{X}^2 \rangle < 1/2$
or $\langle \Delta \hat{P}^2 \rangle < 1/2$), which is shown by the shaded area
in \cref{fig:dx2-dp2-plot}. This is a consequence of the repeated homodyne
measurements performed during the computation, which iteratively reduces the entropy in the machine and
partially collapses the OPO state such that it comes close to being a minimum-uncertainty state. In a closed-loop CIM, parametric amplification with a positive pump
rate ($p > 0$) is employed only in the initial stage, but parametric
deamplification with a negative pump rate ($p<0$) is used later on. The
resulting squeezing ($\langle \Delta \hat{X}^2 \rangle < 1/2$) rather than
anti-squeezing ($\langle \Delta \hat{X}^2 \rangle > 1/2$) is favourable for
exploration when using repetitive spin flips. In contrast, parametric amplification
with a positive pump rate is used in an open-loop CIM throughout the
computation.

\subsection{Time-to-Solution}
\label{sec:MFB-CIM_B}

\begin{figure}[h!]
\centering
\includegraphics[width=.8\linewidth]{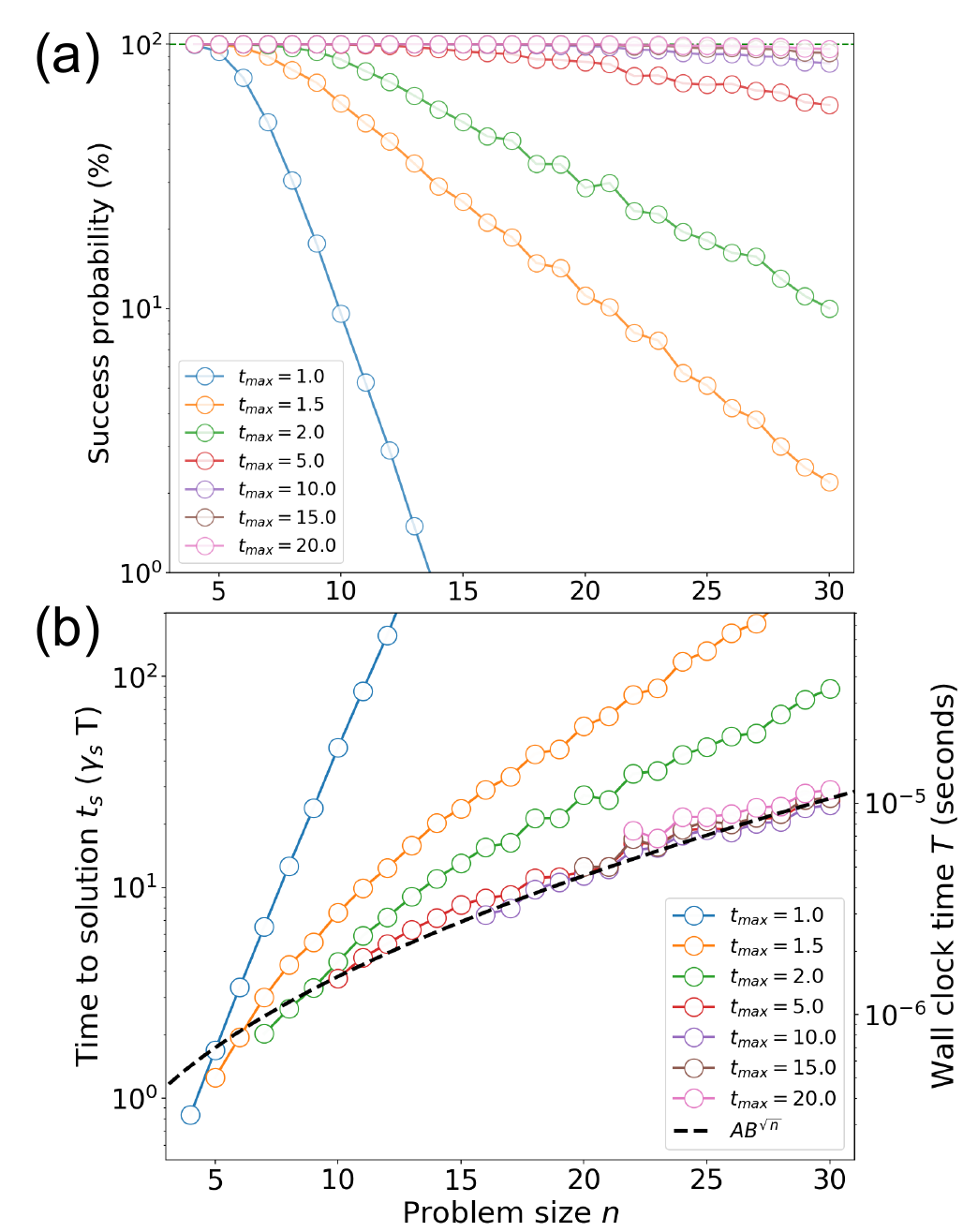}
\caption{(a) Success probability $P_{\text{s}}$ and (b) time-to-solution (in units of
signal field decay time $1/\gamma_{\text{s}}$) as a function of problem size $n$ for
various runtimes $t_{\text{max}}$. The black dotted line shows the best-fit  TTS curve of
the form $A B^{\sqrt{n}}$.}
\label{fig:Ps-TTS}
\end{figure}

Figures \ref{fig:Ps-TTS}(a) and (b) show the median of the success probability $P_{\text{s}}$ and the TTS $t_{\text{s}}$ of the closed-loop CIM as a function of problem
size \mbox{$n=4, 5, \ldots, 30$} with varying runtime $t_{\text{max}}$. We perform 1000 trials,
with a trial considered successful if the machine finds an exact solution within
$t_{\text{max}}$. The success probability $P_{\text{s}}$ decreases exponentially with respect
to $n$, especially for $t_{\text{max}}\leq 5$. For a greater value of $t_{\text{max}}$, the
slope of the decay improves as shown in \cref{fig:Ps-TTS}(a). The TTS is defined as the
expected computation time required to find a ground state for a particular
problem instance with 99\% confidence. As such, it is defined via
\begin{equation}
{\label{eq:TTS-definition}}
t_{\text{s}} = R_{99} \cdot t_{\text{max}}\,,
\end{equation}
where $R_{99}= \frac{\log(0.01)}{\log(1-P_{\text{s}})}$ is the number of trials required
to achieve a 99\% probability of success. We solve 1000 instances for each
problem size ($n=4, \dots, 30$) to evaluate the median $P_{\text{s}}$ and TTS.
Note that $t_\text{s}$ refers to the {\em normalized} and dimensionless TTS, while the actual wall-clock TTS (in seconds) is denoted by $T$. These two notions of TTS are related via the equation \mbox{$t_\text{s}=\gamma_\text{s}T$.}
The wall-clock time $T$ is estimated by assuming a cavity round-trip time of
$\Delta T_\text{c}=10$ nanoseconds (all-to-all spin coupling is implemented in 10 nanoseconds), and a $1/e$
signal amplitude decay time of 400 nanoseconds ($\gamma_\text{s} \Delta T_\text{c} = 0.025$).
An important observation from~\cref{fig:Ps-TTS}(b) is that the optimal median TTS
scales as an exponential function of the square root of the problem size, that is,
an exponential of $\sqrt{n}$ rather than $n$. This unique trend was first noticed in
\cite{hamerly2019experimental}.

\begin{figure*}[t]
\centering
\begin{tabular}{c @{\hskip 0.2in} c}
    \text{(a) 21-weight Problem Instances}
    & \text{(b) SK Model Instances} \\
    \includegraphics[width=0.4\linewidth]{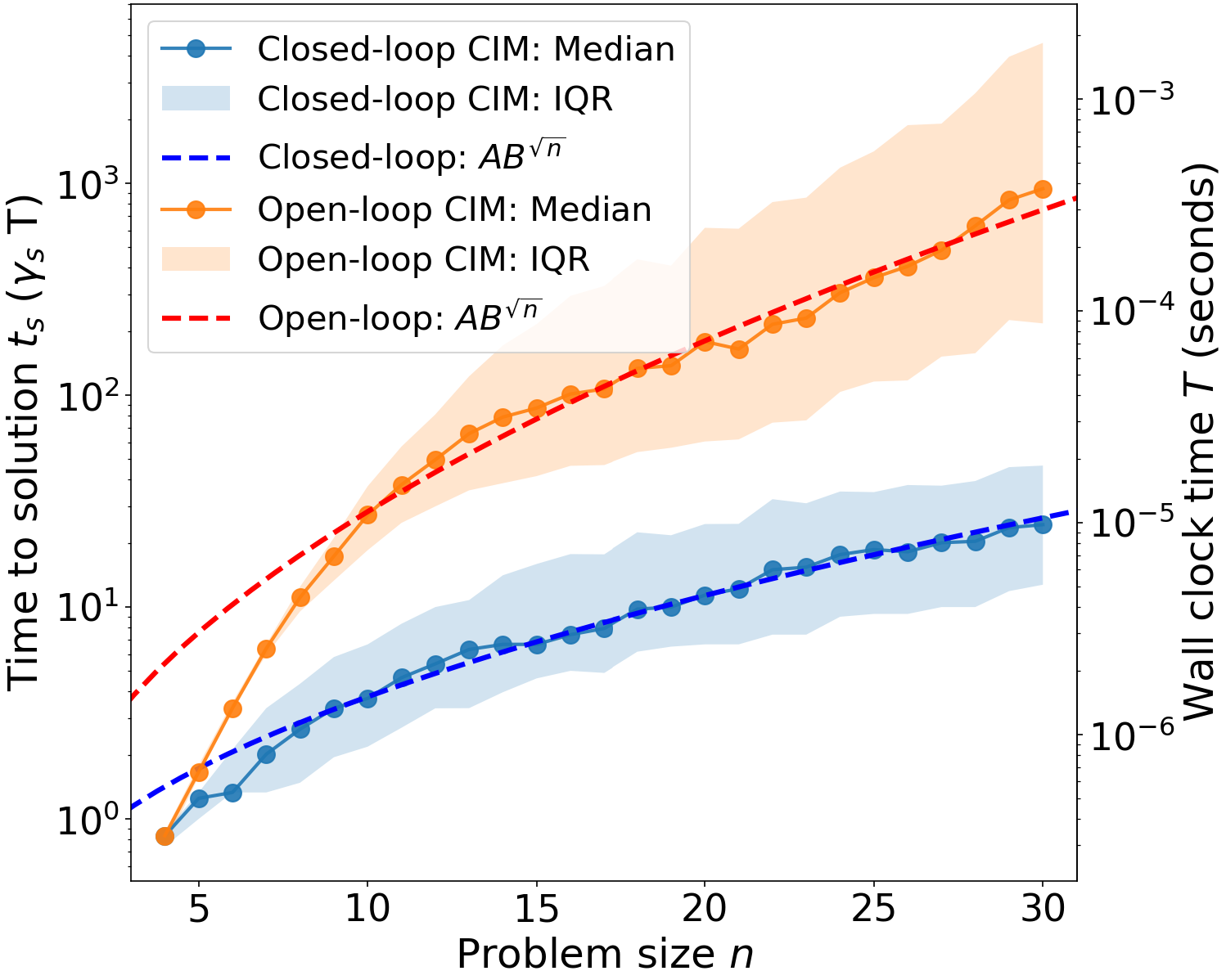}
    & \includegraphics[width=0.4\linewidth]{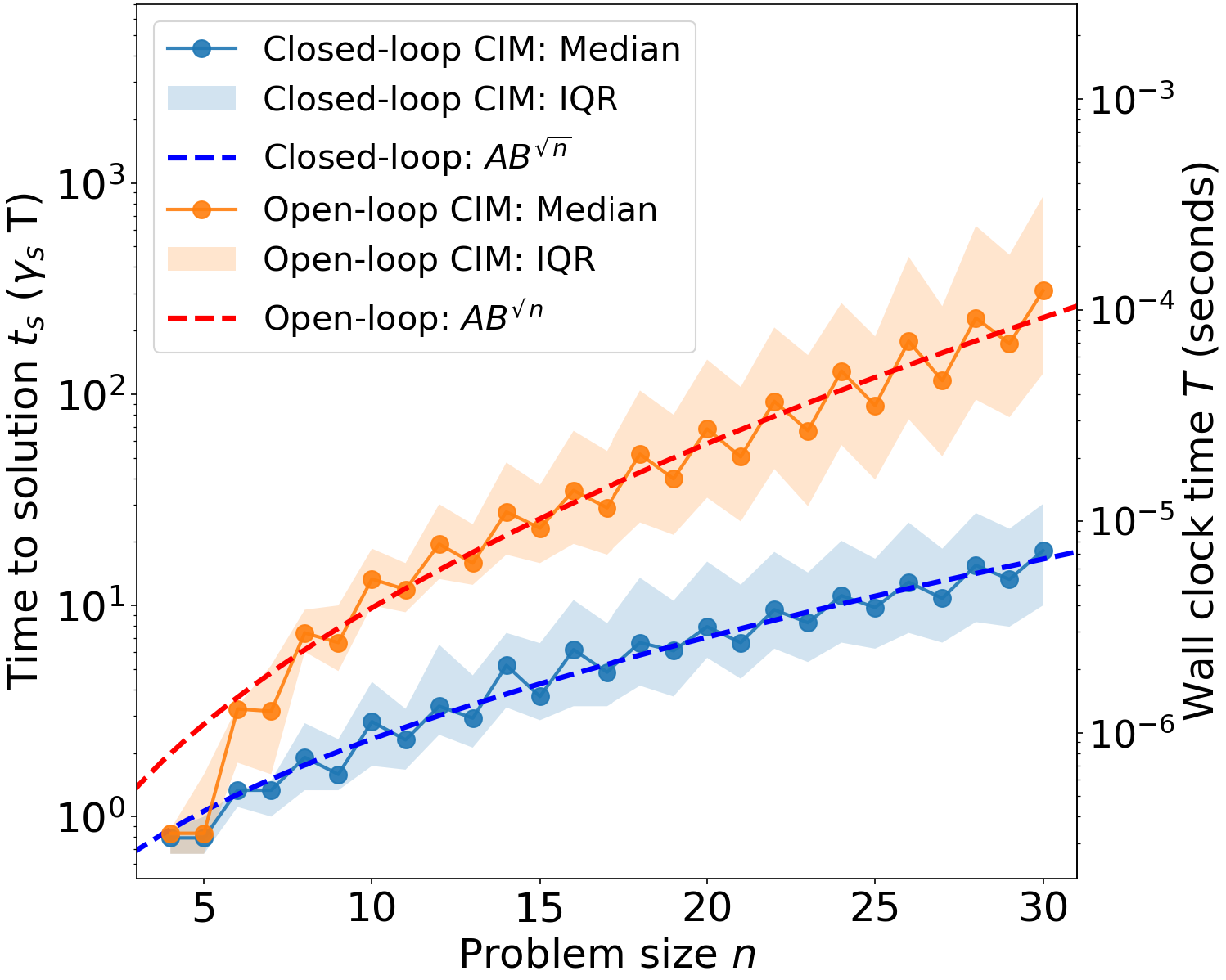}
\end{tabular}
\caption{The optimal (median) time-to-solution of the closed-loop CIM
and open-loop CIM on (a) 21-weight randomly generated $J_{ij}$ and (b) binary-weight randomly generated instances ($J_{ij}= \pm 1$, SK model).
The shaded regions represent the interquartile range (IQR), showing the region between the 25th and 75th percentiles obtained from the 1000 instances. The dashed blue and red lines are fitted curves of the form $AB^{\sqrt{n}}$.}
\label{fig:TTS-comp}
\end{figure*}
\cref{fig:TTS-comp}(a) and (b) show the optimum TTS of the closed-loop CIM and
the open-loop CIM with respect to the problem size $n$. We solve two types of {\sc
MaxCut} problems. The first type are randomly generated instances with
edge weights $J_{ij} \in \{-1, -0.9, \ldots, 0.9, 1\}$. We  refer to these instances
as 21-weight {\sc MaxCut} problem instances. The second type are randomly generated
Sherrington--Kirkpatrick (SK) spin glass instances with $J_{ij}= \pm 1$.
We study the open-loop CIM with the same Gaussian quantum model without
dynamical modulation of $e_{i}(t)$, $a_{i}(t)$, and $p_{i}(t)$, but with
measurement-induced state reduction (the third term of \cref{eq:mu-equation} and
the second term of \cref{eq:xval-equation})~\cite{inui2020noise}. We set the
feedback parameters $\beta=0$, $\rho_a = 0$, and $\rho_p = 0$ for the open-loop
CIM in order to have a constant feedback field strength $e_i(t)= e_i(0)
=1.0$. The pump rate $p$ is linearly increased from $p=0.5$ at $t=0$ (below
threshold) to $p=1.0$
at $t_{\text{max}}$ (above threshold). As shown in \cref{fig:TTS-comp}(a) and (b), the
performance of the closed-loop CIM is superior to that of the open-loop CIM for
both types of {\sc MaxCut} problems.

\cref{tab:TTS-AB} summarizes the best-fitting parameters $A$ and $B$ for a
function of the form $t_{\text{s}}= A B^{\sqrt{n}}$ in both the closed-loop and open-loop CIMs.
The smaller coefficient values for $B$ for the closed-loop CIM than those for the open-loop
CIM highlight the superior scaling of the closed-loop CIM compared
to the open-loop variant. We note that $A$ is expressed in units of a normalized
time $t_{\text{s}} = \gamma_\text{s} T$, where $T$ is the wall-clock time. 
It is worth noticing that the scaling law of the sub-exponential function 
is not necessarily optimal for fitting the data within the problem size range  $n \leq 30$.
In \cref{sec:comp}, we present results for a much wider range for the SK model. 
Additional considerations with respect to inferring the true scaling law are discussed in~\cref{app_TTS_scaling}.

\begin{table}[hb]
\centering
\caption{Parameters $A$ and $B$ found by regression of a function of the form
$A B^{\sqrt{n}}$ to the TTS curves of the closed-loop and open-loop CIMs for
the two types of {\sc MaxCut} instances.}
\label{tab:TTS-AB}
\small
\begin{tabular}{|l|
c@{\hskip 0.1in}|
c@{\hskip 0.1in}|
c@{\hskip 0.1in}|
c@{\hskip 0.1in}|
c@{\hskip 0.1in}|
c@{\hskip 0.1in}|}
\hline
\multirow{3}{*}{}
& \multicolumn{2}{c|}{21-weight random $J_{ij}$}
& \multicolumn{2}{c|}{Binary random $J_{ij}$} \\ \cline{2-5}
& \multicolumn{2}{c|}{$n=4, \ldots, 30$}
& \multicolumn{2}{c|}{$n=4, \ldots, 30$} \\ \cline{2-5}
& \, Closed loop
& \, Open loop
& \, Closed loop
& \, Open loop  \\ \hline
A & 0.26 & 0.32 & 0.16 & 0.13 \\
B & 2.32 & 4.12 & 2.33 & 3.92 \\ \hline
\end{tabular}
\end{table}

\subsection{Discrete-Time Model}
\label{sec:cim-DTmodel}

\cref{sec:MFB-CIM_B} presented the results of our study of the performance of closed-loop and
open-loop CIMs with a high-finesse cavity. Nevertheless, it is obvious that a low-finesse cavity with a larger signal decay rate $\gamma_{\text{s}}$ is favourable
in terms of the runtime of the algorithm. This is because the
wall-clock time $T$ scales as \mbox{$T=t_\text{s}/\gamma_\text{s}$.}
However, it appears that the continuous-time
Gaussian quantum theory based on the master equation [\cref{eq:master-equation}]
breaks down in the case of a low-finesse cavity. Here, we describe a new discrete-time
Gaussian quantum model~\cite{ng2021efficient}.

We treat the MFB-CIM as an $n$-mode bosonic system with $2n$ quadrature
operators, $\hat{X}_1,\hat{P}_1, \ldots, \hat{X}_n,\hat{P}_n$, satisfying
$[\hat{X}_{k},\hat{P}_{k'}] = i \delta_{kk'}$. If the system is in a Gaussian
state, it is fully characterized by a mean-field vector $\mu$ and a covariance
matrix $\Sigma$. In other words, the density operator of each OPO pulse can be
written as ${\hat{\rho}}_{i} (\mu_{i}, \Sigma_{i})$, where
\begin{align}
{\label{eq:disc-mu}}
\mu_i & = (\langle \hat{X}_i \rangle,\langle \hat{P}_i \rangle), \\
{\label{eq:disc-sigma}}
\Sigma_i & =
{\scriptsize
\begin{pmatrix}
\langle {\hat{X}_i}^2 \rangle & \frac{1}{2} \langle \Delta \hat{X}_i \Delta
\hat{P}_i + \Delta \hat{P}_i \Delta \hat{X}_i \rangle\\
\frac{1}{2} \langle \Delta \hat{X}_i \Delta \hat{P}_i + \Delta \hat{P}_i \Delta
\hat{X}_i \rangle & \langle {\hat{P}_i}^2 \rangle
\end{pmatrix}}.
\end{align}

We let $\hat{\rho} \left(\mu_{i}(\ell), \Sigma_{i}(\ell) \right)$ denote the
state of the $i$-th OPO pulse just before it starts its $\ell$-th round trip
through the cavity. To propagate the state of the $i$-th signal pulse from
$\hat{\rho} (\mu_{i}(\ell), \Sigma_{i}(\ell))$ to
$\hat{\rho} (\mu_{i}(\ell+ 1), \Sigma_{i}(\ell+ 1))$, we
perform the following five discrete maps iteratively: the background linear-loss
map $\mathcal{B}$, the OPO crystal propagation map $\chi$, the out-coupling loss
map $\mathcal{B}_{\text{out}}$, the homodyne detection map $H$, and the feedback
injection map $\mathcal{D}$. These discrete maps are defined in
\cref{app_discmodel}.

\begin{figure*}[tb]
\centering
\begin{tabular}{c @{\hskip 0.2in} c}
    \text{(a) 21-Weight Problem Instances}
    & \text{(b) SK Model Instances} \\
    \includegraphics[width=0.4\linewidth]{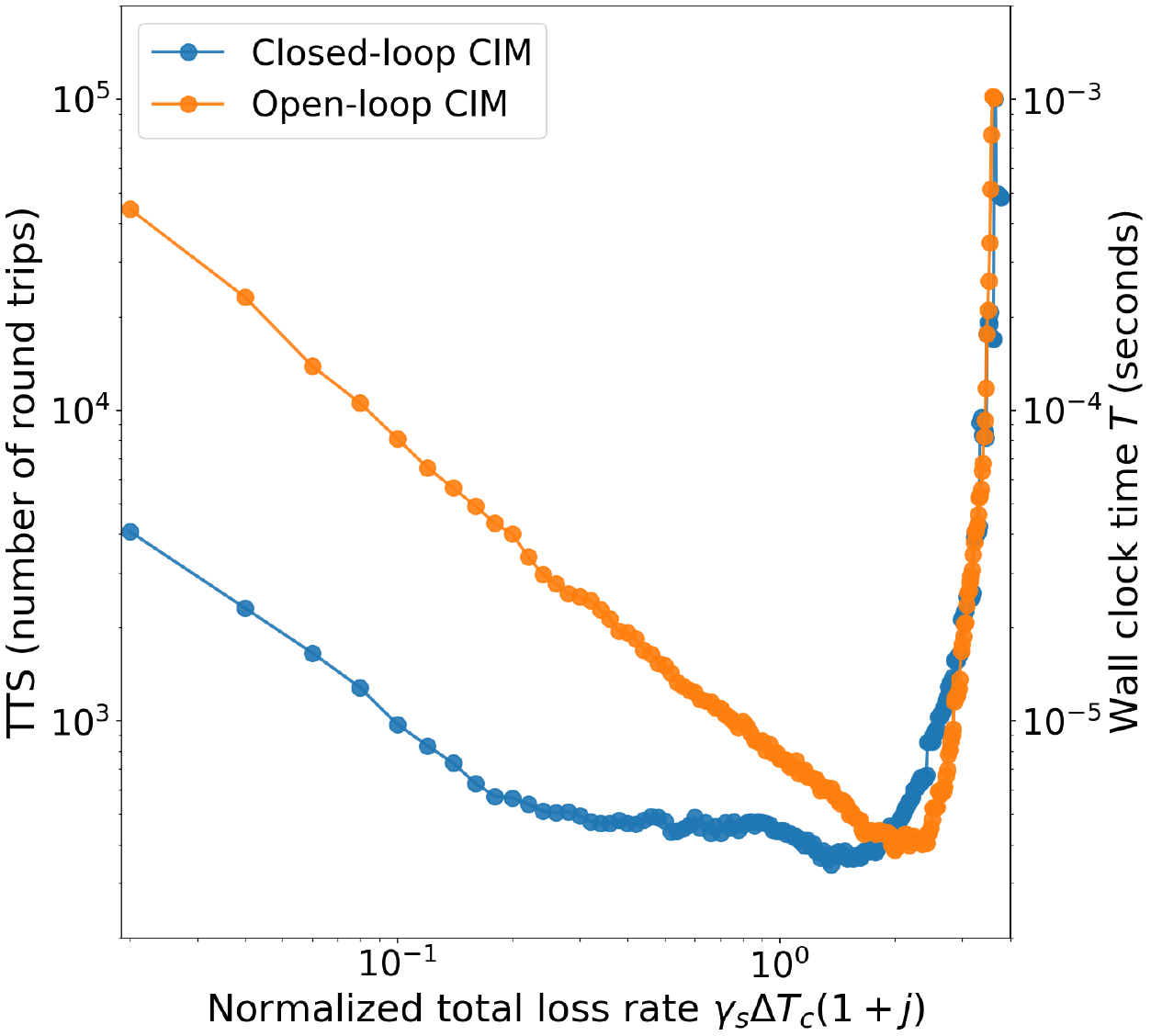}
    & \includegraphics[width=0.4\linewidth]{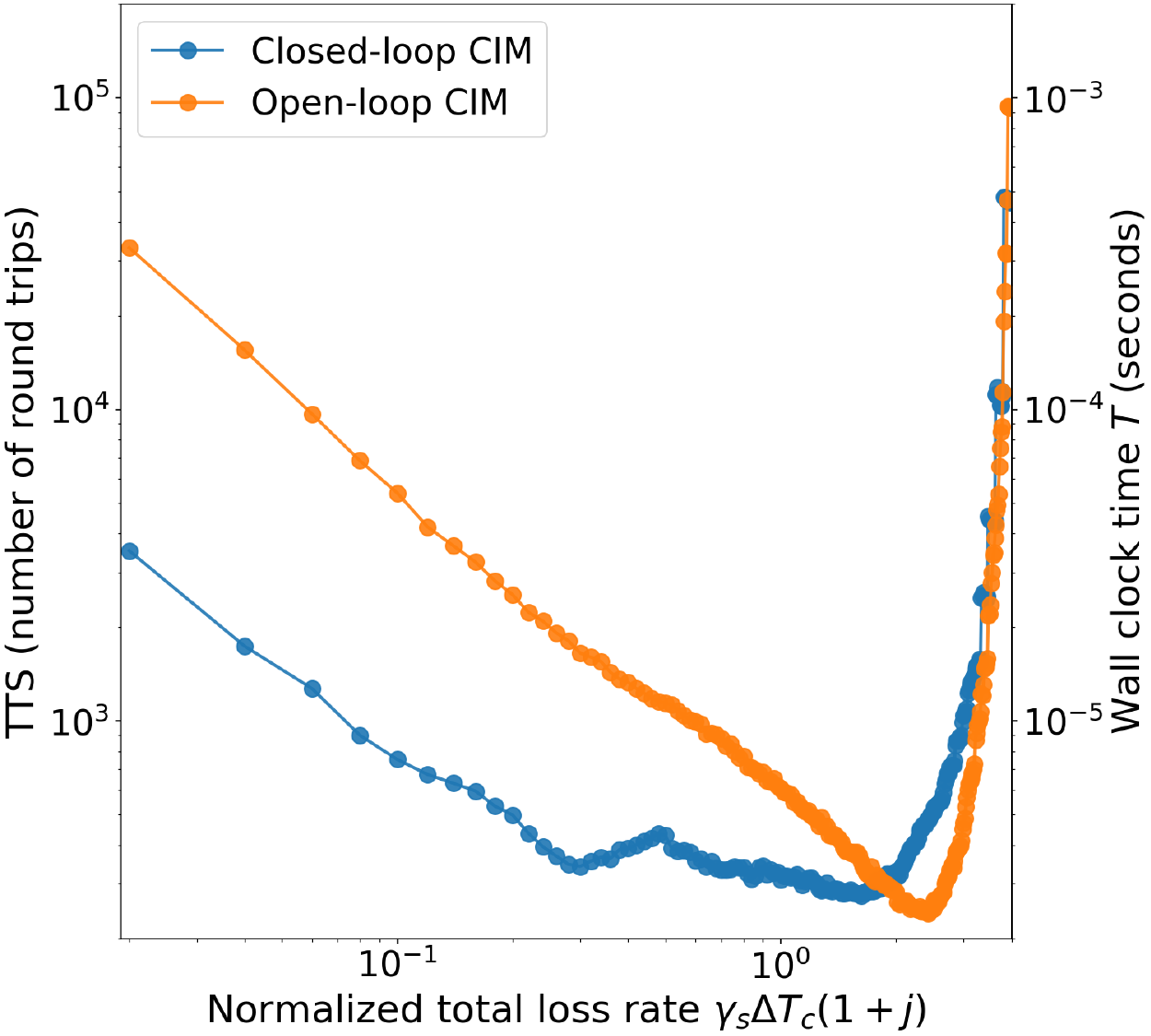}
\end{tabular}
\caption{Median TTS expressed in units of round trips (left y-axis) and the corresponding wall-clock time (right y-axis) of the closed-loop CIM and the
open-loop CIM versus the normalized total loss rate $\gamma_{\text{s}} {\Delta T}_{\text{c}}(1+j)$
for (a) 21-weight problem instances and (b) SK model instances, for size $n=30$ and $j$ kept constant at the value 1.}
\label{fig:TTS-finesse}
\end{figure*}

In order to see how the wall-clock TTS of the closed-loop and open-loop CIMs
is decreased by increasing the total cavity loss rate $\gamma_{\text{s}}(1+j)$, we solve the
21-weight {\sc MaxCut} instances and the SK model instances for \mbox{$n=30$} to explore
the TTS as a function of the normalized total loss rate $\gamma_{\text{s}} {\Delta T}_{\text{c}}(1+j)$. 
The results are shown in \cref{fig:TTS-finesse}. 
The saturation parameter and the out-coupling loss are chosen as $g^2=10^{-4}$ and $j=1$, respectively. The feedback parameters are set to $\alpha = 0.5$, $\pi = 0.2$, $\rho_a = \rho_p = 0$ (while we keep $a$ and $p$ constant), and $\beta = 0.2$.

As expected, the TTS (expressed in terms of the number of round trips) decreases monotonically
for both problem types and for both the closed-loop and open-loop CIMs as long as $\gamma_{\text{s}} {\Delta T}_{\text{c}}(1+j) \lesssim 0.1$ (i.e., in the case of a high-finesse cavity). However, if $\gamma_{\text{s}} {\Delta T}_{\text{c}}(1+j) \gtrsim 1$ (i.e., in the case of a very-low-finesse cavity), the TTS increases for both the
closed-loop and the open-loop CIMs.  
This is because one homodyne measurement per
round-trip loss does not provide sufficiently accurate information about the internal
OPO pulse state and, therefore, the measurement-feedback circuit fails to
implement the Ising Hamiltonian and self-diagnosis feedback properly.  At $n=30$,
the optimum normalized loss rate is $\gamma_{\text{s}} {\Delta T}_{\text{c}}(1+j) \approx 1$ for both
the closed-loop and the open-loop CIMs.

\section{Scaling of DAQC}
\label{sec:res_qaoa}
\begin{figure}[b]
\centering
\begin{tikzpicture}
\node[align=center] at (-0.3,0.0) {$\ket{+}$};
\node[align=center] at (-0.3,0.5) {$\ket{+}$};
\node[align=center] at (-0.3,1.0) {$\ket{+}$};
\node[align=center] at (-0.3,1.5) {$\ket{+}$};

\draw (0,0.0) -- (0.15,0.0);
\draw (0,0.5) -- (0.15,0.5);
\draw (0,1.0) -- (0.15,1.0);
\draw (0,1.5) -- (0.15,1.5);
\draw (0.15,-0.3) -- (0.15,1.8) -- (1.45,1.8) -- (1.45,-0.3) -- (0.15,-0.3);
\node[align=center] at (0.8,0.75) {$e^{-i\gamma_1H_\text{P}}$};
\draw (1.45,0.0) -- (1.6,0.0);
\draw (1.45,0.5) -- (1.6,0.5);
\draw (1.45,1.0) -- (1.6,1.0);
\draw (1.45,1.5) -- (1.6,1.5);
\draw (1.6,-0.3) -- (1.6,1.8) -- (2.9,1.8) -- (2.9,-0.3) -- (1.6,-0.3);
\node[align=center] at (2.25,0.75) {$e^{-i\beta_1H_\text{M}}$};
\draw (2.9,0.0) -- (3.05,0.0);
\draw (2.9,0.5) -- (3.05,0.5);
\draw (2.9,1.0) -- (3.05,1.0);
\draw (2.9,1.5) -- (3.05,1.5);
\node[align=center] at (3.3,0.75) {$\cdots$};
\draw (3.55,0.0) -- (3.7,0.0);
\draw (3.55,0.5) -- (3.7,0.5);
\draw (3.55,1.0) -- (3.7,1.0);
\draw (3.55,1.5) -- (3.7,1.5);
\draw (3.7,-0.3) -- (3.7,1.8) -- (5.0,1.8) -- (5.0,-0.3) -- (3.7,-0.3);
\node[align=center] at (4.35,0.75) {$e^{-i\gamma_pH_\text{P}}$};
\draw (5.0,0.0) -- (5.15,0.0);
\draw (5.0,0.5) -- (5.15,0.5);
\draw (5.0,1.0) -- (5.15,1.0);
\draw (5.0,1.5) -- (5.15,1.5);
\draw (5.15,-0.3) -- (5.15,1.8) -- (6.45,1.8) -- (6.45,-0.3) -- (5.15,-0.3);
\node[align=center] at (5.8,0.75) {$e^{-i\beta_pH_\text{M}}$};
\draw (6.45,0.0) -- (6.6,0.0);
\draw (6.45,0.5) -- (6.6,0.5);
\draw (6.45,1.0) -- (6.6,1.0);
\draw (6.45,1.5) -- (6.6,1.5);
\draw[decorate,decoration={brace},thick] (6.7, 1.7) to
	node[midway,right] (bracket) {$\ket{\psi(\gamma, \beta)}$}
	(6.7,-0.2);

\end{tikzpicture}
\caption{DAQC circuit with $p$ layers. The rotation parameters satisfy
$\gamma_i \in [0, \pi)$ and $\beta_i \in [0, \pi/2)$.
This circuit ansatz results from Hamiltonian simulation implementing a discretized adiabatic evolution in terms of a first-order Suzuki--Trotter expansion.}
\label{fig_qaoaansatz}
\end{figure}
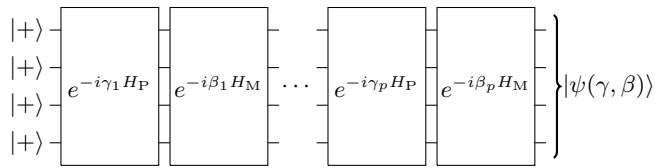

We now analyze the efficacy of the DAQC algorithm in solving
\mbox{\textsc{MaxCut}} problems. In this paper, DAQC is associated with the first-order Suzuki--Trotter
expansion of the adiabatic Hamiltonian evolution. This algorithm attempts to prepare the ground state of a target
Hamiltonian $H_\text{P}$.
A typical circuit for DAQC is shown in \cref{fig_qaoaansatz}.
The state $\ket{+}^{\otimes n}$ is prepared on $n$ qubits,
and is evolved through a sequence of $p$ ``layers''.
Each layer consists of an evolution
according to $H_\text{P}$
along a computational basis, here chosen to
be the Pauli-$Z$ eigenbasis, followed by an
evolution under a mixing Hamiltonian
$H_\text{M} =\sum_{i} X_i$.
A vector of tunable parameters
$\gamma= (\gamma_1, \ldots,\gamma_p)$ is chosen,
where each entry $\gamma_i$ corresponds to the angle of
rotation along $H_\text{P}$ in the $i$-th layer.
Similarly, a vector $\beta= (\beta_1,\ldots, \beta_p)$
is chosen for the $H_\text{M}$ evolutions. Finally,
the qubits undergo projective measurements in the
computational basis, and the measurement results
are used to compute the energy values of
$H_\text{P}$.

A ``shot'' of the circuit with parameters $(\gamma, \beta)$ is defined as a single execution of the circuit from preparation to measurement, and returns a single energy measurement. Multiple shots performed with the same parameters
$(\gamma, \beta)$ can return different results, as they are taken from
independent copies of the same prepared state $\ket{\psi(\gamma, \beta)}$. For
the weighted {\sc MaxCut} problem, we use the target Hamiltonian $H_\text{P} = \sum_{i,
j}J_{ij} Z_iZ_j$, which is diagonal in the computational basis and whose ground
states correspond to the largest cuts of the complete \mbox{$n$-vertex} graph with
edge weights $J_{ij}$.

We study two schemes for optimizing the gate parameters
of the DAQC algorithm. The first scheme treats gate parameters as hyperparameters that follow a tuned DAQC schedule. The second scheme uses a variational hybrid \mbox{quantum--classical} protocol to optimize the gate parameters, similar to the method typically used for QAOA. In our numerical experiments, we observed a better TTS scaling for the first scheme compared to the second scheme  (see~\cref{app_hptuning,app_qaoaopt}); therefore, we use the first scheme to conduct our scaling analysis.

\subsection{Time-to-Solution Scaling of DAQC}
\label{sec:tts-qaoa}

To study the time-to-solution of the DAQC algorithm in  solving {\sc MaxCut} problems,
we analyze the algorithm using pre-tuned Trotterized adiabatic scheduling.
We use randomly generated graphs of size $n \in \{10, \ldots, 20\}$. Our test
set consists of $1000$ graphs of each size, with edge weights
$J_{k\ell}=\pm 0.1 j$, where $j\in\{0,1,\dots, 10\}$.

Given a parameter vector $(\gamma, \beta)$, 
we evaluate the TTS of DAQC 
as a product of two terms~\cite{aramon2019physics},
\begin{equation}
\mathrm{TTS}(\gamma, \beta) =
R_{99}(\gamma, \beta) \cdot t_{\text {ss}}\,,
\end{equation}
where $t_{\text{ss}}$ is the time taken for a single shot.

The $R_{99}$ is the number of shots that must be performed to ensure a 99\%
probability of observing the ground state of $H_\text{P}$. It is a metric
commonly used to benchmark the success of heuristic optimization
algorithms. If the state $\ket{\psi(\gamma, \beta)}$ has a probability $p$ of
being projected onto the ground state, then
\begin{equation}
    R_{99}(\gamma, \beta) = \frac{\log(0.01)}{\log(1-p)}.
\end{equation}

We estimated the time required for a single shot using the following assumptions
for an ideal, highly performant quantum computer with access to arbitrary-angle, single-qubit
$X$-rotations and two-qubit $ZZ$-rotations.

\begin{assumption}
\label{ass:ideal-qpu}
The preparation and measurements of qubits collectively take 1.0 microseconds. The
processor performs any single-qubit or two-qubit gate operations in 10 nanoseconds. Gate operations
may be performed simultaneously if they do not act on the same qubit. In addition, all
components of the circuit are noise-free and, therefore, there is no overhead for quantum
error correction or fault-tolerant quantum computation.
\end{assumption}

\begin{figure*}[t]
\centering
\begin{tabular}{c @{\hskip 0.17in} c @{\hskip 0.17in} c}
    \text{(a) TTS Scaling for 21-Weight Graphs}
    & \text{b) TTS Scaling versus Number}
    & \text{(c) TTS Scaling for  the SK Model} \\
     \text{for Selected Numbers of DAQC Layers}
    & \text{of DAQC Layers}
    & \text{for a 20-Layer DAQC}
    \\
    \includegraphics[clip, width=0.317\linewidth]{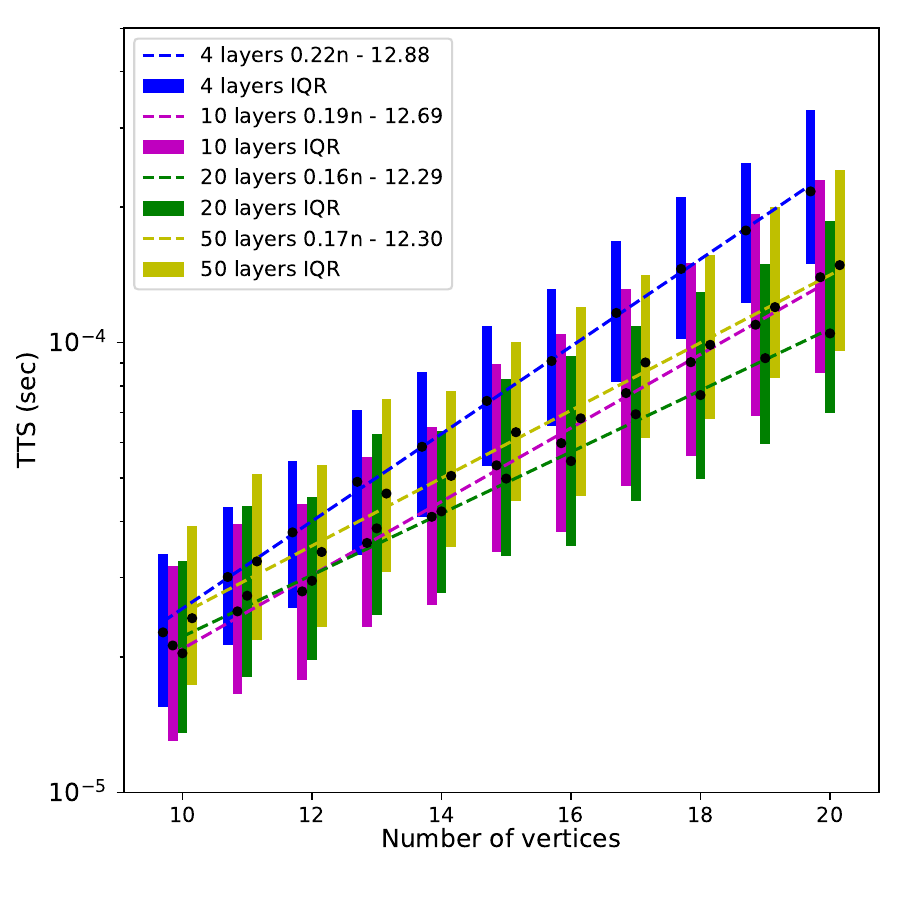}
    & \includegraphics[clip, width=0.317\linewidth]{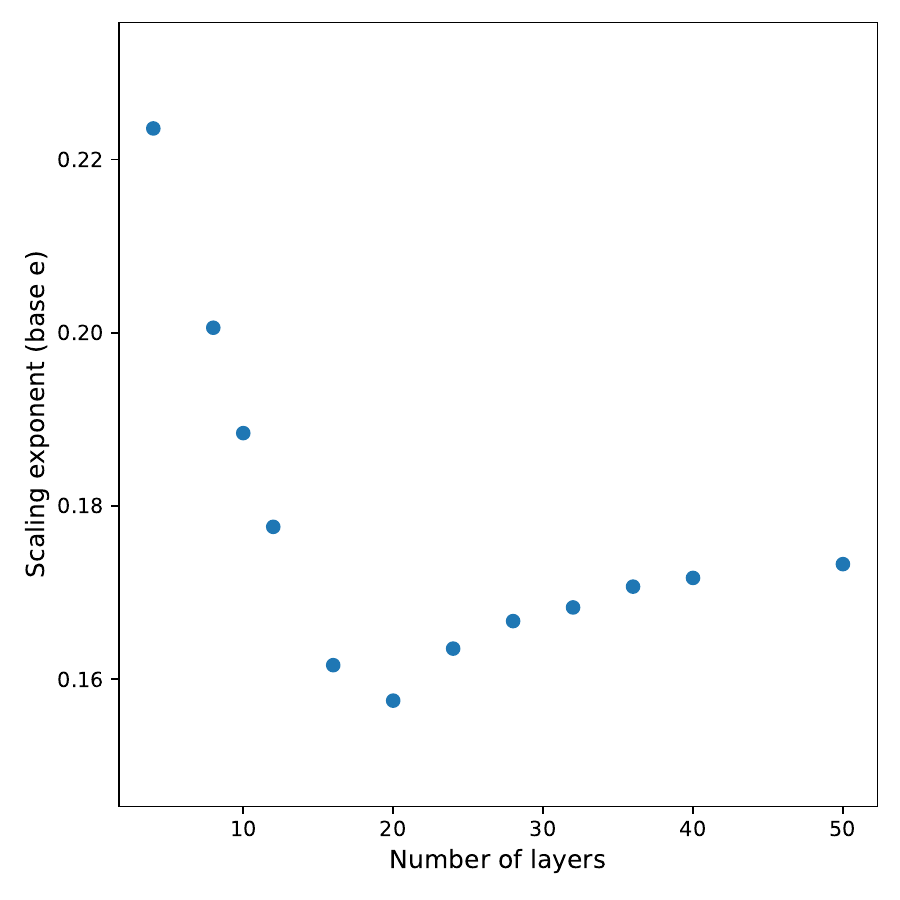}
    &  \includegraphics[clip, width=0.317\linewidth]{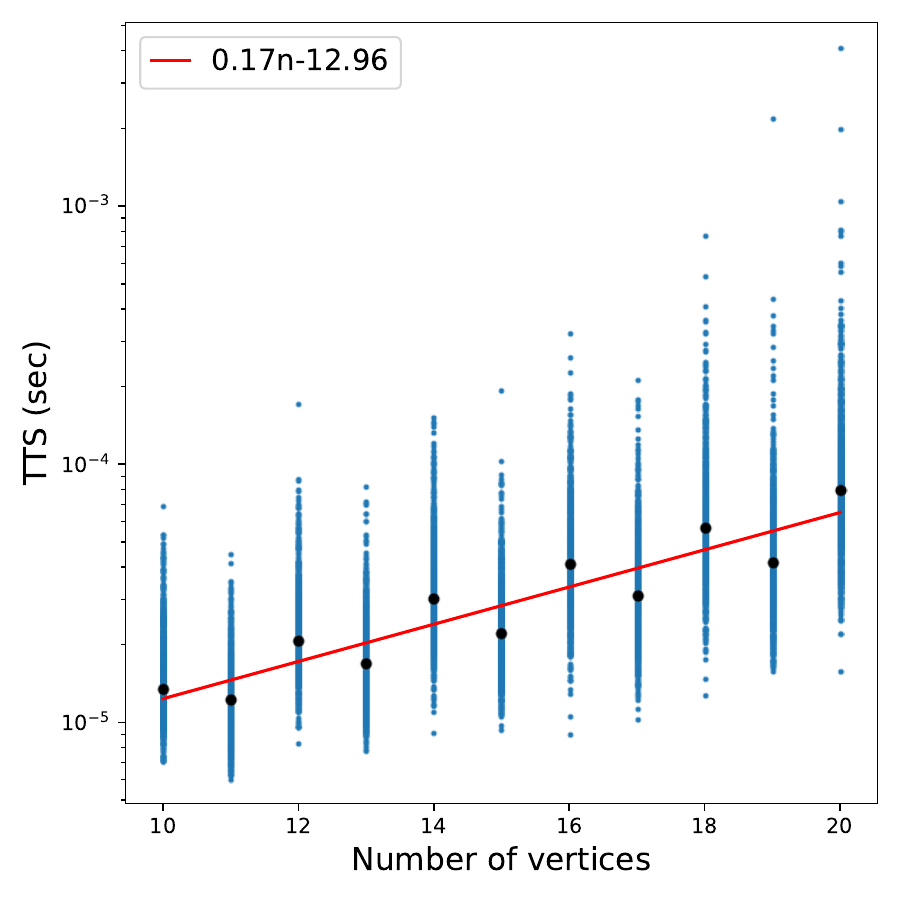}
\end{tabular}

\caption{Scaling of the DAQC algorithm in solving {\sc MaxCut} problems. The TTS results are obtained by simulating DAQC, using pre-tuned adiabatic scheduling rather than optimizing its parameters variationally. 
The number of qubits required to implement the algorithm is $n$. \newline
(a) TTS scaling for a 4-, 10-, 20-, and 50-layer DAQC algorithm as the problem size
grows from 10 to 20 vertices.
A best-fit line (dashed) is drawn to the median of the TTSs of the 1000 instances of each size, whose IQR ranges are represented using coloured bars.
The equation of this linear regression is given by
$\ln(\text{TTS}) = mn+b$, where $n$ is the problem size. In \cref{app_TTS_scaling}, we present the results of additional regression 
analysis for more-general scaling laws of the 
form $\log(\text{TTS}) = mn^c+b$. The highest confidence with respect to the quality of the regression fit is indeed obtained at an exponent value close to $c = 1$, which supports our conjecture that DAQC scales exponentially. In actuality, the scaling is found to be slightly sub-exponential, at the value $c \approx 0.9$. 
(b) Slope of the linear regression for a range of layers. The
best scaling for DAQC on these 21-weight {\sc MaxCut} instances is observed
at 20 layers.  (c) TTS scaling for the SK model, when using a 20-layer DAQC. A best-fit linear-regression is drawn to the median
of the TTSs of the 1000 instances for each size.}
\label{fig_qaoascaling}
\end{figure*}

For each problem size varying from 10 to 20 vertices, \cref{fig_qaoascaling} shows a plot of the median TTS, suggesting that the TTS scales exponentially with respect to problem size. With more layers, DAQC has a lower potential $R_{99}$, but a single shot takes more time. We found the best scaling was achieved with $p \approx 20$ layers. However, near-term hardware will suffer from various sources of noise, such as decoherence and control noise, which will restrict us to employing
shallow DAQC circuits with only a few layers, for example, $p=4$.

The DAQC parameters $(\gamma, \beta)$ used in
\cref{fig_qaoascaling} were produced using the formula explained in what follows.
Recall the setup for quantum adiabatic evolution \cite{farhi2000quantum}. Given
 an initial Hamiltonian $H_0$ and a target Hamiltonian $H_1$, we consider the
 time-dependent Hamiltonian
$$H(t) = s(t)H_1 + (1-s(t))H_0, \hskip 0.1in  t \in [0, T]$$
over a total annealing time $T$, where the function $s(t)$ is an increasing
schedule satisfying $s(0)=0$ and $s(T)=1$. The time-dependent Hamiltonian
$H(t)$ is then applied to the ground state of $H_0$. Let
$\psi(t)$ denote the wavefunction at time $t$, so that $\psi(0)$ is the ground
state of $H_0$ and $\psi$ evolves according to the Schr\"{o}dinger
equation $$\dot{\psi} = -i\Big(s(t)H_1
+(1-s(t))H_0\Big)\psi.$$
We use Trotterization to approximate the prepared state $\psi(T)$. Let
$$c_k:=
\int_{(k-1)T/p}^{kT/p}s(t)\,dt \quad\! \text{and} \quad\!
b_k:=\int_{(k-1)T/p}^{kT/p}(1-s(t))\,dt.$$
Then,
\begin{equation}\psi(T) \approx
e^{-ib_pH_0}e^{-ic_pH_1}\cdots
e^{-ib_1H_0}e^{-ic_1H_1}\psi(0),
\end{equation}and this approximation
becomes exact in the limit as $p\to\infty$.

\begin{figure}[ht]
\centering
\includegraphics[trim=0 0 0 30, width=0.4\textwidth]{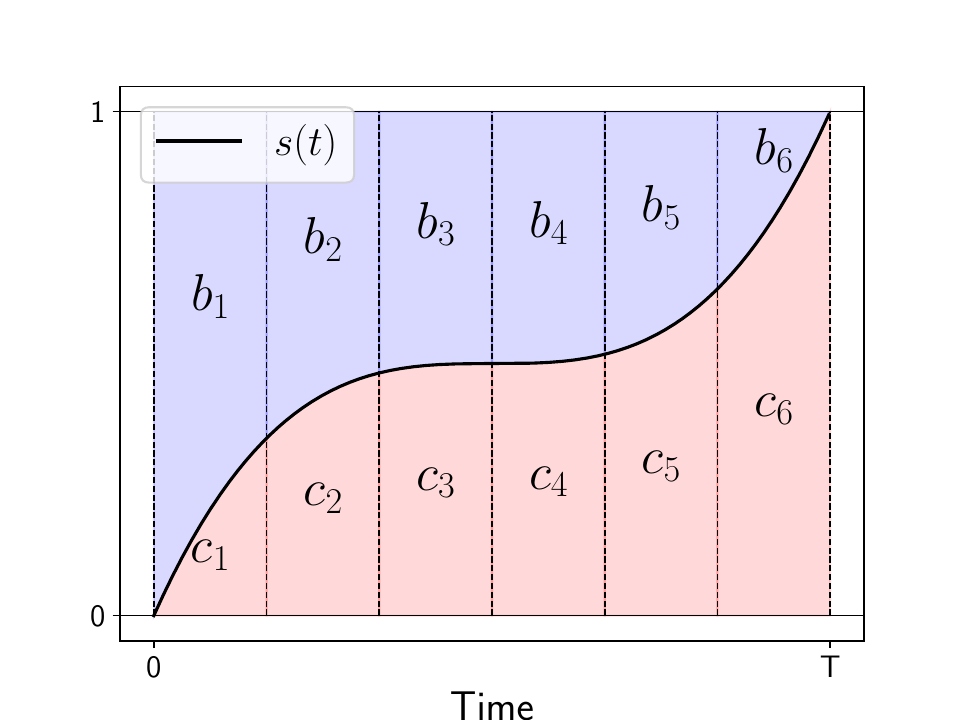}
\caption{Trotterization of adiabatic evolution into $p=6$ layers. The integrals computing
$b_k$ and $c_k$ yield the coefficients for
$H_0$ and $H_1$, respectively.}\label{adiabatic}
\end{figure}

The Hamiltonians $H_0$ and $H_1$ are both chosen to
have a Frobenius norm equal to 1. We divide both $H_\text{M}$ and
$H_\text{P}$ by their corresponding norms, which can easily be calculated, as each Hamiltonian
is a sum of the orthogonal Pauli terms
$$H_0 = \frac{1}{\lVert H_\text{M}\rVert}H_\text{M} = -\frac{1}{\sqrt{n}}\sum_{i}X_i$$
and
$$H_1 = \frac{1}{\lVert{H_\text{P}\rVert}}H_\text{P} = \frac{1}{\sqrt{\sum_{i, j}J_{ij}^2}}\sum_{i, j}J_{ij}Z_iZ_j.$$
Thus,
$$\gamma_k = \int_{(k-1)T/p}^{kT/p}\frac{s(t)}{\lVert H_\text{P}\rVert}\,dt
\:\:\: \text{and} \:\:\:
\beta_k = \int_{(k-1)T/p}^{kT/p}\frac{1-s(t)}{\lVert H_\text{M}\rVert}\,dt.$$
Empirically, we found that enforcing this Frobenius normalization has yielded a
very well-performing schedule for DAQC for multiple problem types. The theoretical
basis for this is yet to be fully understood.

The schedule $s(t)$ should have an ``inverted S'' shape~\cite{mbeng2019optimal,roland2002quantum} in order to handle the squeezed energy gap in
the middle. We take $s(t)$ to be a cubic function with the general form
\begin{equation}\label{cubicform}
s(t) = \frac{t}{T} + a \cdot \frac{t}{T}\left(\frac{t}{T} - \frac{1}{2}\right)\left(\frac{t}{T} - 1\right)
\end{equation}
for a free parameter $a$. When $a=0$, $s(t)$ is a straight linear path. When
$a=4$, $s(t)$ is a curved path with a slope of $0$ at $t=T/2$. We found by empirical
means that $a=4$ and $T=p(1.6 + 0.1n)$ are the best hyperparameters. See \cref{app_hptuning} for more details.

\begin{figure}[ht]
\centering
\includegraphics[width=0.49\textwidth]{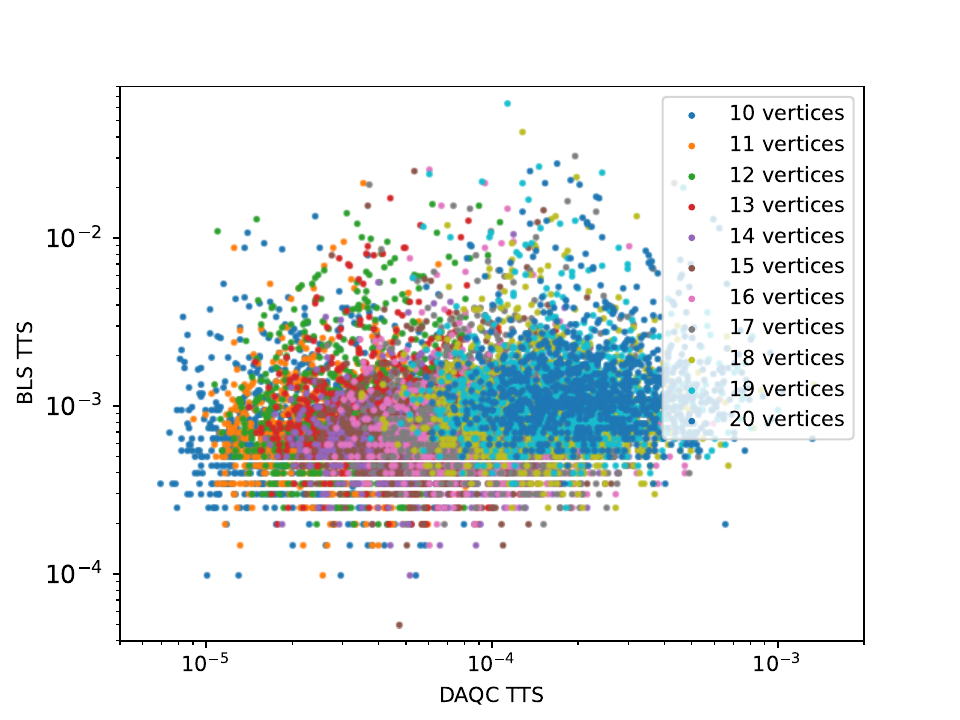}
\caption{
Scatter plot of DAQC-TTS versus BLS-TTS indicates there is no significant
correlation between the difficulty of an instance for DAQC  versus the difficulty of an
instance for Breakout-Local Search.}
\label{fig_qaoahistogram}
\end{figure}

We also compare the TTS for DAQC to the TTS for Breakout-Local Search (BLS), a
classical search algorithm. For each graph instance, 20 runs of BLS were
performed, and runtimes were averaged rto obtain the TTS. The algorithm's runtime for
each run was capped at 0.1 seconds, although the minimum value was almost always found
within that time. \cref{fig_qaoahistogram} demonstrates that the TTS for DAQC
shows no significant correlation with the TTS for BLS.

\subsection{Challenges Encountered when Using the Variational Approach}
\label{sec:tts-qaoa-variationally}
We have also explored using the variational
\mbox{quantum--classical} protocol, which is typical for
the approach known as QAOA. This protocol includes 
an optimization loop which learns better parameters $(\gamma, \beta)$
by using the data from  already-performed shots.
However, we found that including an optimization step did not improve
the total TTS for the following reasons, and therefore did not
include the step in our analysis. The $R_{99}$ is impossible to measure
without knowledge of the ground state, and
therefore any optimization routine must instead rely on energy measurements.
A common approach is to use the expected energy, $\bra{ \psi(\gamma, \beta)
}H_\text{P} \ket{\psi(\gamma, \beta)}$, which is estimated by averaging over
the multiple  shots taken with the parameters $(\gamma, \beta)$.
This approach suffers from two limitations.
First, we must use a large number of shots to accurately estimate
the expected energy, which makes the optimization step costly. This is consistent with the challenges encountered in overcoming the problem known as barren plateau phenomenon~\cite{holmes2021connecting,mcclean2018barren}.  
Second, the
expected energy is an imperfect stand-in for $R_{99}$, and therefore optimization
typically offers little to no improvement upon the annealing-inspired parameter
schedule. See \cref{app_qaoaopt} for more details.

\section{Scaling of DH-QMF}
\label{sec:res_qmf}

\begin{figure}[b]
 \centering
 \includegraphics[width=1.0\linewidth]{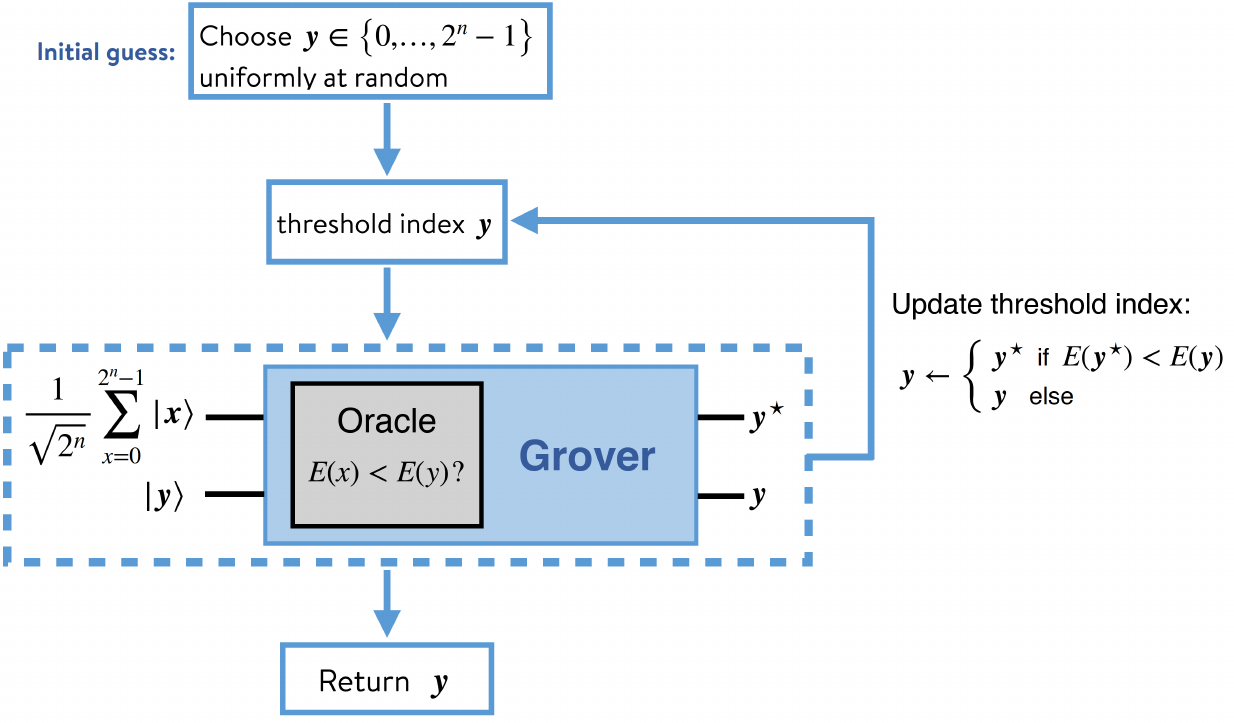}
\caption{
Schematic illustration of the
\mbox{D\"urr--H{\o}yer} algorithm for quantum minimum finding /mbox{(DH-QMF)} applied to searching for a
spin configuration corresponding to the energy minimum (ground state). The
possible spin configurations are labelled by the indices
$y\in\left\{0,\dots, 2^{n}-1\right\}$. The algorithm starts by choosing
uniformly at random an {\em initial guess} for the ``threshold index''
$y$, whose energy $E(y)$ serves as a threshold: solutions to the problem cannot
have an energy value larger than this threshold. The main step of the algorithm is a
loop consisting of Grover's search for a spin configuration with an energy value
strictly smaller than the threshold energy, followed by a threshold-index
update. This loop needs to be repeated many times until the threshold index
eventually holds the solution with a probability of success higher than
a given target lower bound, say, e.g.,\ $p_{\text{\tiny succ}}=0.99$. The final
step returns the threshold index as output. A key element of the Grover's search
subroutine is an oracle which marks all states whose energies are strictly
smaller than the threshold energy.  Note that Grover's search may fail to output
a marked state.}
\label{fig:QMF-scheme}
\end{figure}

We now consider using
{D\"urr} and H{\o}yer's algorithm for quantum minimum finding (\mbox{DH-QMF})~\cite{durr1996quantum} to find the ground state of an Ising
Hamiltonian corresponding to a \mbox{\textsc{MaxCut}} problem. Given a real-valued
function \mbox{$E: S \to \mathbb R$} on a discrete domain $S$ of size $N= |S|$, \mbox{DH-QMF}
finds a minimizer of $E$ (out of the possibly many) using
$\mathcal{O}(\sqrt{N})$ queries to $E$. In our case, the domain
$S$ is the set of all spin configurations of a classical Ising Hamiltonian on
$n$ sites ($N= 2^n$), and the function $E$ maps each spin configuration to its
energy. The \mbox{DH-QMF}  algorithm is a randomized algorithm, that is, it succeeds in finding the optimal
solution only up to a (high) probability. The probability of failure of \mbox{DH-QMF} can
be made arbitrarily small without changing the mentioned complexity. A schematic
illustration of \mbox{DH-QMF} is shown in \cref{fig:QMF-scheme}, and additional technical
details can be found in \cref{app:app_qmf}.

Given an $n$-spin Ising Hamiltonian
\begin{equation}
\label{Eq:Ising-Hamiltonian}
H=-\sum_{0\le i< j \le n-1}J_{ij}Z_iZ_j
\end{equation}
corresponding to an undirected weighted graph of size $n$, its $N=2^n$ energy
eigenstates can be labelled by the integer indices $0\le y\le N-1$, with
the corresponding energy eigenvalues $E(y)$. The index $y$ associated with a computational basis state
$\ket{y}=\ket{\eta_{0}}\otimes\dots\otimes\ket{\eta_{n-1}}$ represented by the
classical bits $\eta_j\in\{0,1\}$ is the binary representation
$y=\sum_{j=0}^{n-1}\eta_j2^j$ of the bit string $(\eta_0, \ldots, \eta_{n-1})$.

The algorithm starts by choosing uniformly at random an index $y\in
\{0,\dots,N-1\}$ as the initial ``threshold index''. The threshold index is
used to initiate a Grover's search~\cite{grover1997quantum,boyer1998tight}. The
Grover subroutine searches for a label $y^\star$ whose energy is strictly
smaller than the threshold value $E(y)$. We measure the output of Grover's search and (classically) ascertain whether the
search has been successful, $E(y^\star)<E(y)$, in which case we (classically) update the
threshold index from $y$ to $y^\star$, and then continue by performing the next Grover's search using
the new threshold. The threshold is not updated if Grover's search fails to find
a better threshold.

In this paper, we assume a priori knowledge of a hyperparameter we call the
number of ``Grover iterations'' (see \cref{sec:meth_qmf}) inside every
Grover's search subroutine that guarantees a sufficiently small failure
probability. However, the practical scheme for using \mbox{DH-QMF} consists of multiple
trials of Grover's search and iterative updates to the threshold index. We
terminate this loop when the Grover subroutine repeatedly fails to provide any
further improvement to $y$ and the probability of the existence of undetected
improvements drops below a sufficiently small value. Finally, we return the last
threshold index as the solution. As shown in~\cite{durr1996quantum}, the overall
required number of Grover iterations needed to find the ground state with
sufficiently high probability, say $1/2$, is in $\mathcal{O}(\sqrt{N})$.

\subsection{Time-to-Solution Benchmark for \mbox{DH-QMF}}
\label{sec:res-qmf}

We investigate the scaling of the time required by \mbox{DH-QMF} to find a solution of
weighted {\sc MaxCut} instances with a $0.99$ success probability, assuming an
optimistic scenario that is explained in \cref{sec:meth_qmf}. This runtime is
analogous to the TTS measure defined in previous sections for
the heuristic algorithms of the \mbox{MFB-CIM} and DAQC and we therefore call this runtime a TTS
as well. For each instance of the problem we have estimated an \emph{optimistic
lower bound} on the runtime of the quantum algorithm with numbers of Grover's
iterations in \mbox{DH-QMF} set (ahead of any trials) to achieve an at least 0.99 success
probability. As this optimal number of Grover's iterations is dependent on
the specific \mbox{{\sc MaxCut}} instance, we consider this an optimistic bound on
performance of \mbox{DH-QMF}. We use the same test set of randomly generated 21-weight
{\sc MaxCut} instances as in  previous sections.

\begin{figure*}[t]
\centering
\begin{tabular}{c @{\hskip 0.2in} c}
    \text{(a) TTS for 21-Weight Problem Instances}
    & \text{(b) TTS for SK Model Instances} \\
    \includegraphics[width=0.4\linewidth]{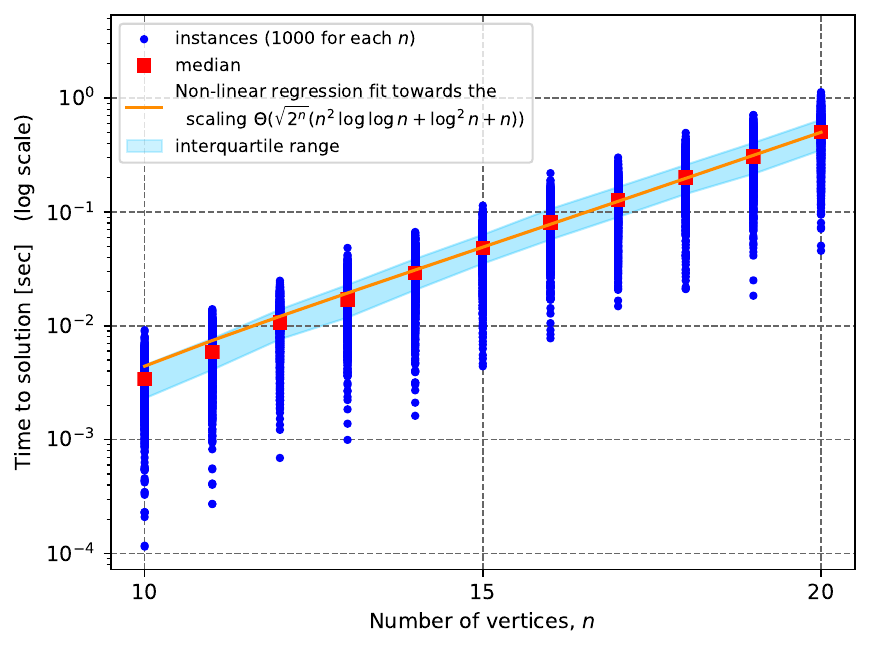}
    & \includegraphics[width=0.4\linewidth]{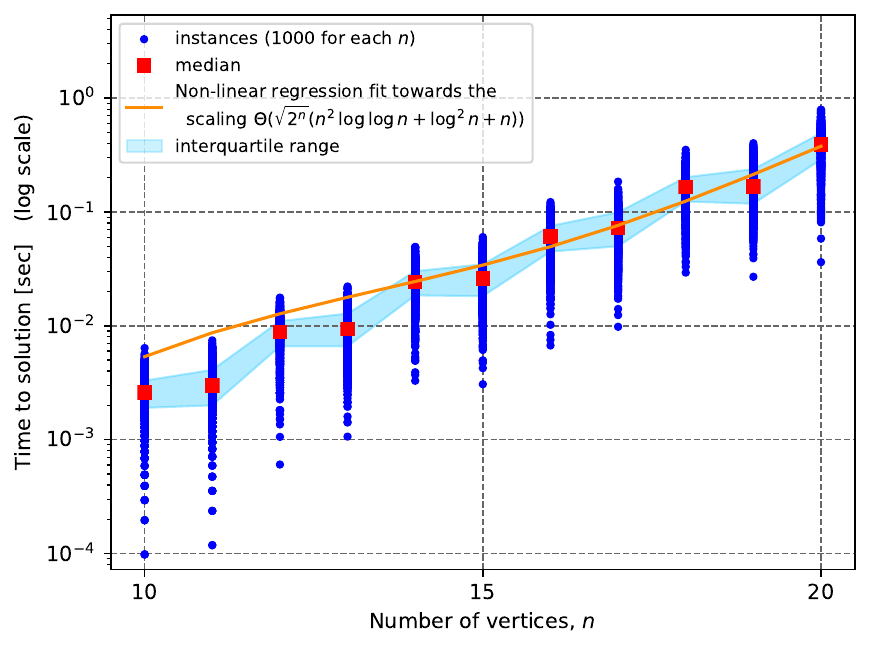}
\end{tabular}
\caption{
Scaling of {D\"urr} and H{\o}yer's algorithm for quantum minimum finding \mbox{(DH-QMF)} in solving {\sc MaxCut}.
(a) Time-to-solution (TTS) for 21-weight problem instances.
(b) TTS for the SK model instances. In both cases, for each value of
the number of vertices in the range $10\le n\le 20$, \mbox{DH-QMF} has been
emulated for 1000 (dark blue data) {\sc MaxCut} instances (see main text). A non-linear least-squares regression (orange curve) has been performed
to fit the expected runtime scaling in \cref{eq:QMF-circuit-depth},
respectively, resulting in
a sum of squared residuals approximately\ $1.2\times 10^{-4}$ $\text{seconds}^2$ for 21-weighted instances and $3.30\times 10^{-3}$ $\text{seconds}^2$ for the SK model instances.
A logarithmic scale has been used to display the data and the regression fits.
Note that the contributions from the logarithmic factors become more (less) significant for smaller (larger) problem sizes.}
\label{fig:QMF-Regression}
\end{figure*}

Our results are illustrated in~\cref{fig:QMF-Regression}. The optimistic
values for the TTS are in the range of orders of magnitude of $1.0$ milliseconds -- $1.0$ seconds for
the considered range of the number of vertices, $10\le n\le 20$, using the same
set of assumptions for the quantum processor as in \cref{ass:ideal-qpu}.

Our estimates for the runtime of the quantum algorithm are obtained as follows.
We note that \mbox{DH-QMF} consists of a sequence of Grover's search algorithms. The total
runtime of \mbox{DH-QMF} is therefore the sum of the runtimes of the quantum circuits, each of which
corresponds to a Grover's search. The runtime of each such circuit is
calculated using the \emph{depth} of that circuit, which is the length of the
longest sequence of native operations on the quantum processor (i.e., qubit
preparations, single-qubit and two-qubit gates, and qubit measurements) in that
circuit, assuming maximum parallelism between independent operations. This path
is also known as the ``critical path'' of a circuit. The runtime of the
circuit is therefore identical to the sum of the runtimes of the operations along
the critical path, with a contribution of 1.0 microsecond in total for both qubit initialization  and measurement, and 10 nanoseconds for any quantum gate operation along the critical
path.

The asymptotic scaling of the TTS is identical to the scaling of the circuit
depth, which is
\begin{equation}
\Theta\left(\sqrt{2^n} \left(n^2\log\log n+ (\log n)^2 +n\right) \right),
\label{eq:QMF-circuit-depth}
\end{equation}
as shown in~\cref{app:app_qmf}. Here the
$\Theta\left(\sqrt{2^n}\right)$ contribution is that of the number of
Grover iterations (identical to the query complexity of Grover's search), while
the $\text{poly}(n,\log n, \log\log n )$ factors are the contribution of each
single Grover iteration consisting of an oracle query with implementation cost
$\Theta\left(n^2\log\log n+ (\log n)^2 \right)$ and the Grover diffusion with
cost $\Theta\left(n\right)$. A nonlinear least-squares regression toward
this scaling is shown in~\cref{fig:QMF-Regression} for both the 21-weight and the SK model problem instances, respectively. Note that the contributions of logarithmic terms are significant only for small problem sizes.

\begin{figure*}[t]
\centering
\begin{tabular}{c @{\hskip 0.2in} c}
    \text{(a) Single-Qubit Gates Count}
    & \text{(b) \textsc{Cnot} Gate Count} \\
    \includegraphics[width=0.4\linewidth]{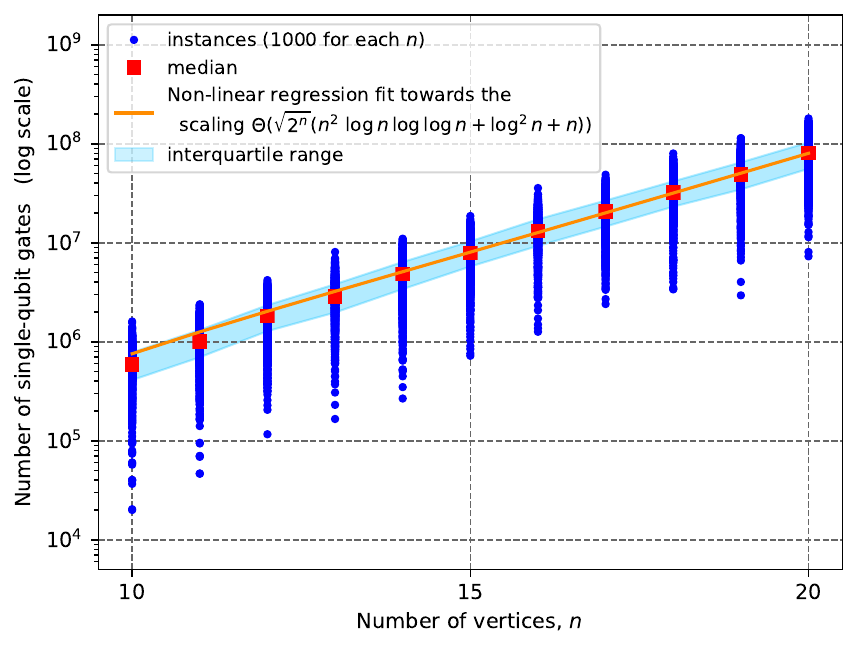}
    & \includegraphics[width=0.4\linewidth]{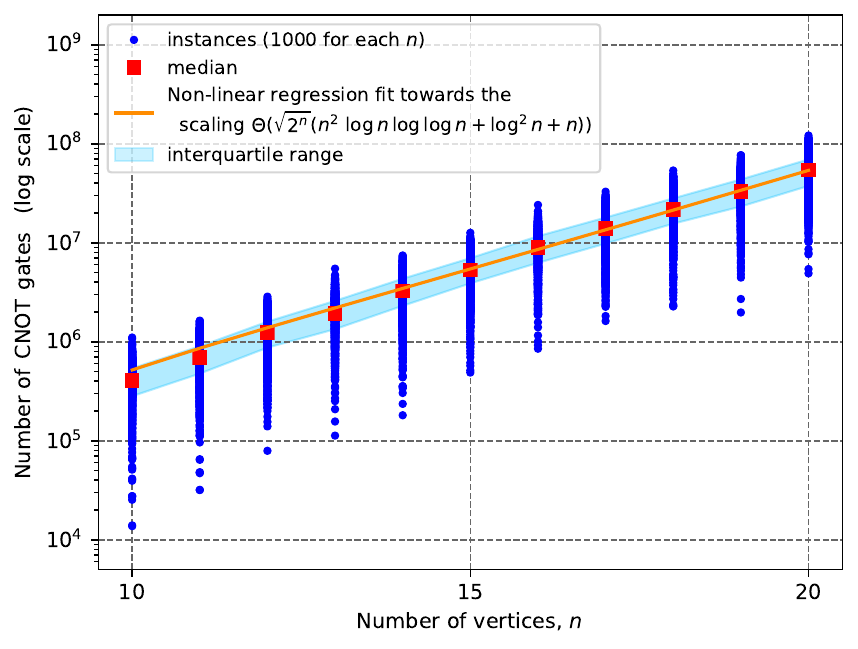}
\end{tabular}
\caption{Optimistic gate counts for DH-QMF in solving the {\sc MaxCut} problem.
For each value of the number of vertices in the range $10\le n\le 20$, the \mbox{DH-QMF}
algorithm was emulated for 1000 (blue data) 21-weight {\sc MaxCut} instances,
see main text. Concrete counts were conducted for the (a) overall number of
single-qubit gates, and (b) two-qubit CNOT gates.
A non-linear least-squares regression (orange curve) has been performed to fit
the expected gate complexity given in~\cref{Eq:QMF-gate-complexity},
respectively. A logarithmic scale has been used to display the data and the regression fits. 
The number of qubits required to implement the algorithm scales as  $\mathcal{O}\left(n+\log n\right)$.}
\label{fig:QMF-gate-count}
\end{figure*}

Alongside the optimistic runtime, we have also computed lower bounds on the
number of quantum gates, including concrete counts for the overall number of
single-qubit gates, two-qubit \textsc{Cnot} gates, and $T$ gates (see
\cref{fig:QMF-gate-count}). Our circuit analysis in~\cref{app:app_qmf}
yields the gate complexity
\begin{equation}
\label{Eq:QMF-gate-complexity}
\Theta\left(\sqrt{2^n}\left( n^2\log n\log\log n+ (\log n)^2 +n\right)\right).
\end{equation}
Our resource estimates have been generated using ProjectQ
\cite{steiger2018projectq}.

\subsection{Optimal Number of Grover Iterations}
\label{sec:meth_qmf}

In what follows, we explain how the algorithm can always be designed such that
the output is indeed a ground state with a probability higher than any
target lower bound for the probability of success, for example, 0.99.

A key component of Grover's search as part of QMF is an oracle that marks every
input state $\ket{x}$ whose energy is strictly smaller than the energy
corresponding to the threshold index $y$ (see \cref{fig:QMF-scheme}). We call it
the ``QMF oracle'' and denote it by $O_{\text{\tiny QMF}}$ to distinguish it
from the ``energy oracle'' $O_E$ which computes the energy of a state under
the problem Hamiltonian. The oracle $O_{\text{\tiny QMF}}$ uses an ancilla
qubit initialized in the state $\ket{z}$ to store its outcome
\begin{equation}
O_{\text{\tiny QMF}}: \,\ket{x}\ket{z}\longmapsto \ket{x}\ket{z\oplus f(x)},
\label{Eq:Oracle_QMF}
\end{equation}
where $f(x)=1$ if, and only if, $E(x)<E(y)$, and $f(x)=0$ otherwise. Here,
$\oplus$ represents a bitwise XOR. The QMF oracle is constructed from multiple
uses of the energy oracle $O_E$ and an operation that compares the values held
by two registers. Details of this construction can be found
in~\cref{app:QMF-oracle}. The combined effect of querying $O_{\text{\tiny
QMF}}$ followed by the Grover diffusion (together forming the \emph{Grover
iteration} to be repeated $\mathcal{O}\left(\sqrt{2^n}\right)$ times) results in
constructively amplifying the amplitudes of the marked items while diminishing
the amplitudes of the unmarked ones.

When there are multiple solutions to a search problem, as is frequently the case
in the Grover subroutine of QMF, the {\em optimal number of Grover iterations} needed
to maximize the success probability depends on the number of marked items as
well. Indeed, suppose we were to have knowledge of the number of marked items $t$ ahead
of time. Then, the optimal number of Grover iterations could be obtained from the
closed formulae provided in \cite{boyer1998tight}:
\begin{align}
\wp_{\text{\tiny succ}} &= \sin^2\left((2m+1)\theta\right),\nonumber \\
\wp_{\text{\tiny fail}}& = \cos^2\left((2m+1)\theta\right).
\label{Eq:Grover-prob_succ_fail}
\end{align}
Here, $m$ is the number of Grover iterations, and $\theta$ is defined by
$\sin^2\theta=t/N$. Hence, the success probability is maximized for the optimal
number of Grover iterations $m_{\text{\tiny opt }} = \left\lfloor \pi/{4\theta}
\right\rfloor$. We also observe that after exactly $m_{\text{\tiny opt }}$
iterations the failure probability obeys
$$\wp_{\text{\tiny fail}} \le \sin^2\theta=t/N,$$
which is negligible when $t\ll N$.

In practice, $t$ and, consequently, $m_{\text{\tiny opt}}$ are often unknown.
Nevertheless, \cite[Sec.~4 and Theorem 3]{boyer1998tight} propose a method to
find a marked item with query complexity $\mathcal{O}\left(\sqrt{N/t}\right)$
even when no knowledge of the number of solutions is assumed.

To simplify the analysis for our benchmark in this paper, we examine each
{\sc MaxCut} instance and assume $t$ is known every time Grover's search is
invoked. This assumption provides a lower bound on the performance of \mbox{DH-QMF}. In
view of the previous discussion, having knowledge of $t$ allows us to compute
$m_{\text{\tiny opt}}$, $\wp_{\text{\tiny succ}}$, and $\wp_{\text{\tiny
fail}}$.

We then boost the overall success probability of Grover's search to any target
success probability $p_{\text{\tiny G}}$ by repeating it $K$ times, where $K$
satisfies
\begin{equation}
p_{\text{\tiny G}} \le 1-\wp_{\text{\tiny fail}}^{\, K}\,.
\label{Eq:Grover-step_success_threshold}
\end{equation}
Moreover, if \mbox{DH-QMF} requires $J$ non-trivial threshold index updates in total,
we must succeed in every boosted Grover search (each including $K$ Grover
searches). The probability of this event is thus at least $p_{\text{\tiny
G}}^{\,J}$. Finally, let us denote the target lower bound for the probability
of success of the overall \mbox{DH-QMF} algorithm by $p_{\text{\tiny succ}}$. We then
must have
\begin{equation}
p_{\text{\tiny succ}}\le
p_{\text{\tiny G}}^{\,J}\,.
\label{Eq:Grover_overall_success}
\end{equation}
We achieve a lower bound for $K$ using
\cref{Eq:Grover-step_success_threshold,Eq:Grover_overall_success}:
\begin{equation}
K\ge \frac{\log\left(1- p_{\text{\tiny succ}}^{\;\,\frac{1}{J}}\right)}
{\log \wp_{\text{\tiny fail}}}\;.
\label{Eq:Grover_boosting_steps}
\end{equation}
Note that this number still depends on the optimal number $m_{\text{\tiny opt}}$
of Grover iterations. The remainder of this section explains how the latter number
is sampled for each {\sc MaxCut} instance via Monte Carlo simulation.

\begin{figure}[ht]
  \includegraphics[width=0.99\linewidth]{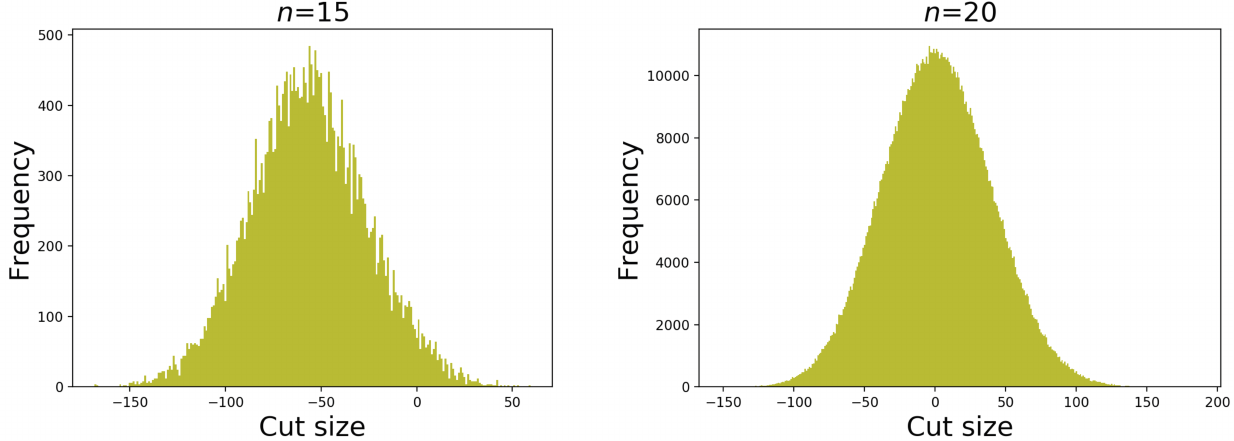}
  \centering
\caption{
Typical cut-size histograms of undirected random weighted
graphs with weights $w_{k\ell}=\pm 0.1 j$, where $j\in\{0,1,\dots, 10\}$.
Two instances are shown for random graphs with $n=15$ (left) and $n=20$ (right)
vertices. Note that, for a fully connected graph with $n$ vertices, the overall
number of edges is $n(n-1)/2$. }
\label{fig:cut_frequency_distribution}
\end{figure}

Given a weighted graph, we first generate the histogram of the sizes of all cuts
in the graph. Examples of such histograms are provided in
\cref{fig:cut_frequency_distribution}. This cut-size histogram allows us to
perform a Monte Carlo simulation of the progression of \mbox{DH-QMF} as follows. The \mbox{DH-QMF} algorithm
starts by choosing uniformly at random an initial cut $C$ as the threshold
index. The resulting energy threshold is therefore sampled according to the
cut-size histogram. Grover search then attempts to find a larger cut. The
number of these cuts is $t$ in the notation above, and can be found if 
the cut-size histogram is known. Using \cref{Eq:Grover-prob_succ_fail}, we can also
compute the optimal number $m_{\text{\tiny opt}}$ of Grover iterations needed to
achieve the highest possible success rate $\wp_{\text{\tiny succ}}$ in that
search. We furthermore can now use \cref{Eq:Grover_boosting_steps} to predict
the number $K$ of Grover searches needed to boost the success probability to at
least $p_{\text{\tiny G}}$. The cut $C$ is now replaced with a larger cut also
selected at random using the cut-size histogram, and this simulation is repeated
for the next iteration in \mbox{DH-QMF}.

We repeatedly sample and update the threshold until we find a maximum cut (i.e.,
at an iteration where \mbox{$t=0$}). At this point, we stop our Monte Carlo simulation
(even though in practice it will not be known that $t$ has become zero). For
each sampling step $j$, we count the total number $t_j$ of states contributing
to strictly greater cuts and use it to calculate the optimal number
$m_{\text{\tiny opt}}^{[j]}$ of Grover iterations as well as the number of
boosting iterations $K_j$ via \cref{Eq:Grover_boosting_steps}.

We now obtain an optimistic TTS as well
as an optimistic gate count estimate using the formulae
\begin{align}
\hspace{-4mm}\text{TTS}&=\sum_{j=1}^J K_jm_{\text{\tiny opt}}^{[j]}\times
\textsc{Runtime}, \label{Eq:TTS_Grover}\\
\hspace{-4mm}\#\,\text{gates}&=\sum_{j=1}^J K_jm_{\text{\tiny opt}}^{[j]}\times
\textsc{\small GateCount}.\label{Eq:TTS_Gate_count}
\end{align}
Here, $\textsc{\small Runtime}$ denotes the running time and $\textsc{\small
GateCount}$ indicates the gate count for a single Grover iteration. In
\cref{sec:res-qmf} we provided optimistic estimates for the number of
single-qubit gates, $\textsc{Cnot}$ gates, and $T$ gates. The quantum circuit
implementation of a single Grover iteration is presented in
\cref{app:app_qmf}.

\section{Comparison of the Three Algorithms}
\label{sec:comp}

\begin{figure*}[t]
 \centering
\begin{tabular}{c @{\hskip 0.2in} c}
    \text{(a) TTS Scalings for 21-Weight Graphs}
    & \text{(b) TTS Scalings as Functions of $\sqrt{n}$ for the SK Model} \\
  \includegraphics[width=0.47\linewidth]{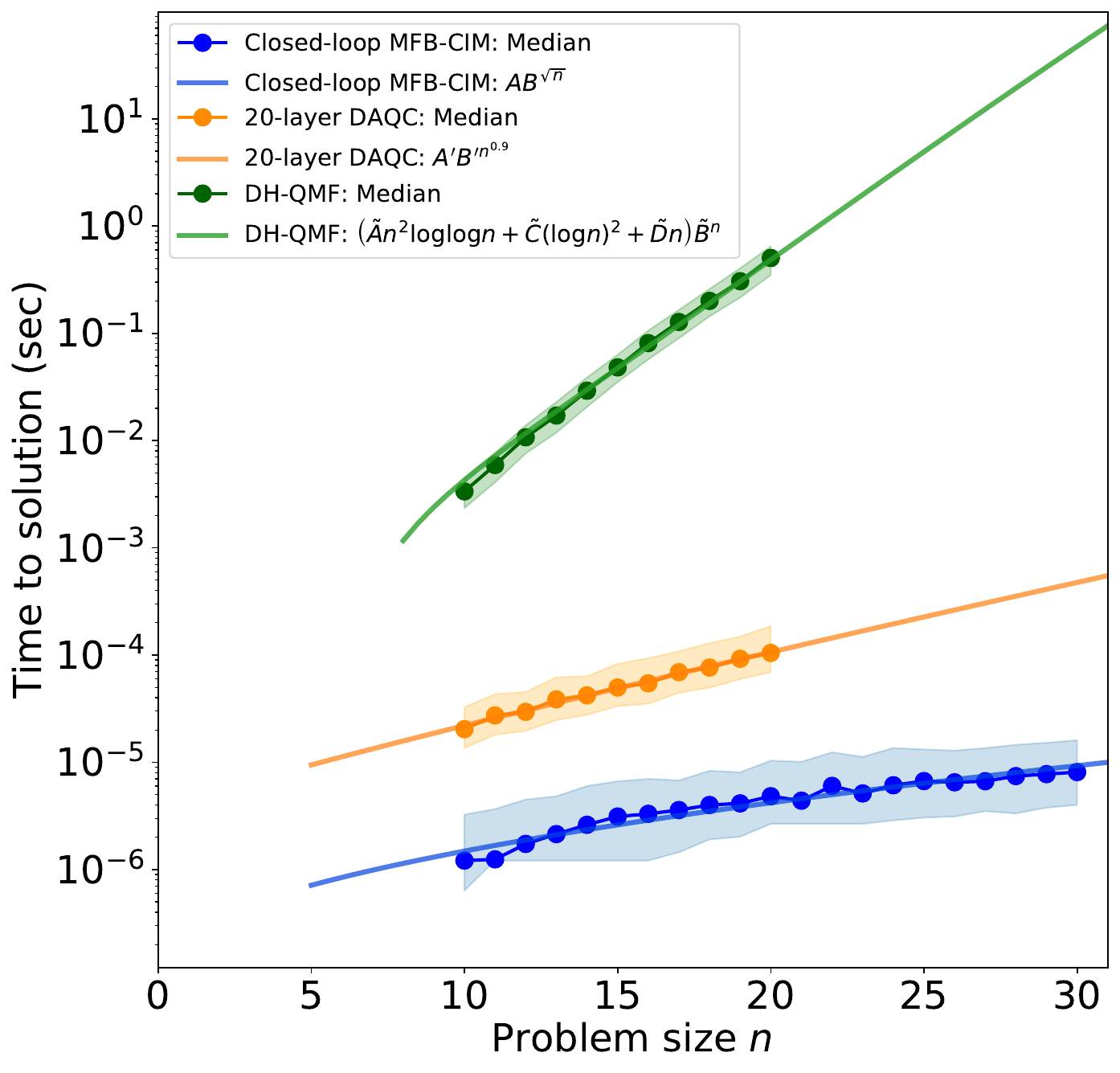}
    & \includegraphics[width=0.47\linewidth]{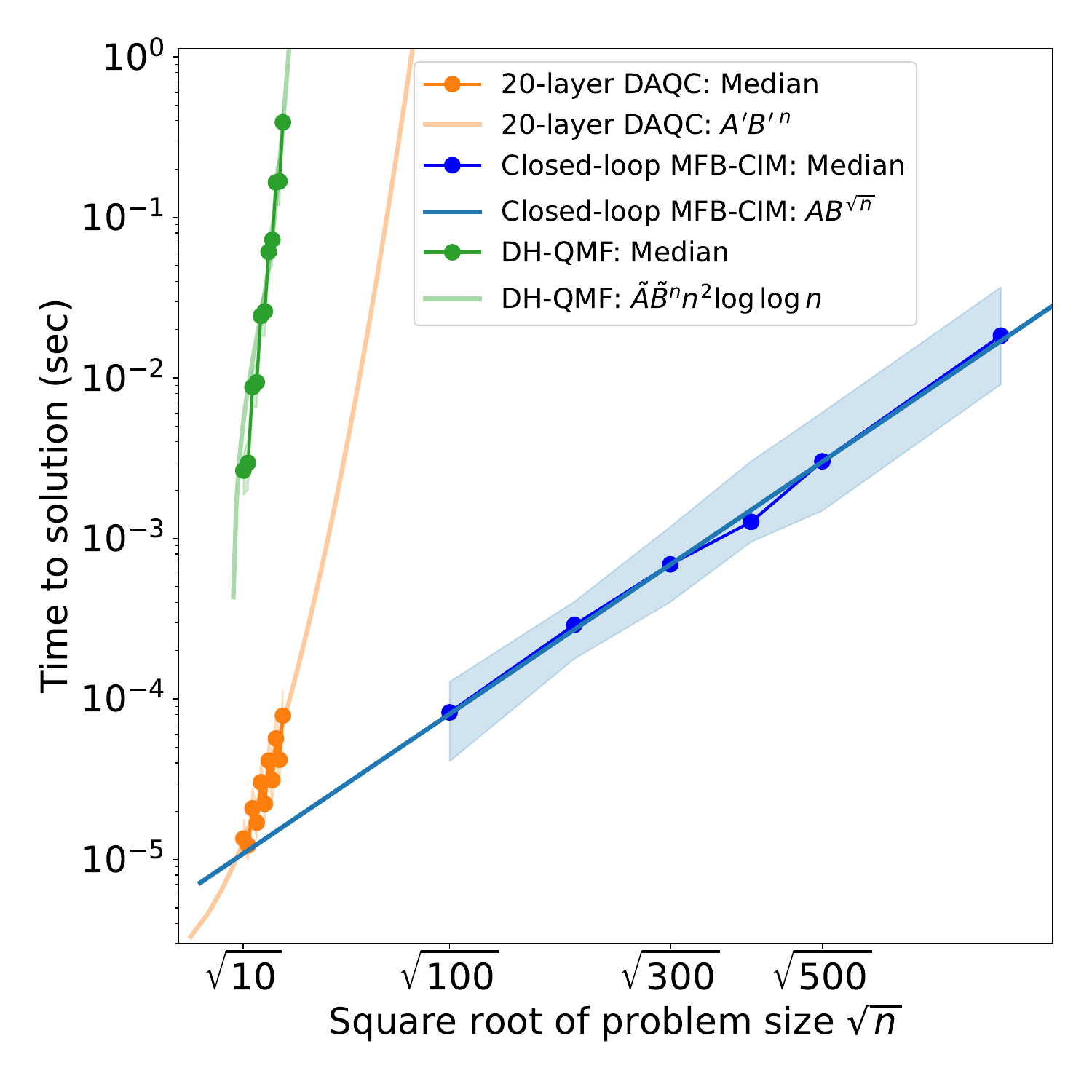}
\end{tabular}
\caption{Comparison of the time-to-solution (TTS) scalings for the \mbox{MFB-CIM}, DAQC, and \mbox{DH-QMF} in solving {\sc
MaxCut} problems. \newline (a) Wall-clock time of a closed-loop CIM with a high-finesse cavity  ($\gamma_{\text{s}} \Delta T_{\text{c}} = 0.1$), DAQC with an optimum number of layers ($p=20$), and \mbox{DH-QMF} with an a priori known number of optimum iterations versus problem size $n$ for fully connected 21-weight graphs. 
(b) TTS of the closed-loop CIM on the fully connected SK model for problem sizes from $n=100$ to $n=800$, in steps of 100. For each problem size, the minimum TTS with respect to the optimization over $t_{\text{max}}$ is plotted.
In comparison, the SK model TTSs are shown for 20-layer DAQC and \mbox{DH-QMF} for problem sizes ranging from $n=10$ to $n=20$.
The straight, lighter-blue line (a linear regression) for the CIM demonstrates a scaling according to $A B^{\sqrt{n}}$.
The lighter-orange and lighter-green best-fit curves for DAQC and \mbox{DH-QMF}  are extrapolated to larger problem instances, illustrating a scaling that is exponential in $n$ rather than in ${\sqrt{n}}$.
In both figures, the shaded regions show the IQRs.}
\label{fig:TTS_comparison}
\end{figure*}

A direct comparison of the three algorithms for solving {\sc
MaxCut} problems
is illustrated in \cref{fig:TTS_comparison}.
In \cref{fig:TTS_comparison} (a),
the median wall-clock TTS of DH-QMF, DAQC, and the closed-loop MFB-CIM are
plotted as a function of problem size $n$ for randomly generated 21-weight {\sc
MaxCut} instances. The solid blue line indicates a best-fitting curve, $f_{\text{CIM}}(n)=A
B^{\sqrt{n}}$, for the closed-loop MFB-CIM, where $A=121$ nanoseconds and $B=2.21$; the solid orange line represents a best-fitting curve, $f_{\text{DAQC}}(n)=A'B'^{\,n^{0.9}}$, for a \mbox{$20$-layer} DAQC, where $A'=3.56$ microseconds and $B'=1.26$; and the solid green  curve represents a best-fitting curve, $f_{\text{QMF}}(n)=\left(\tilde{A} n^2\log\log n + \tilde{C}(\log n)^2 +\tilde{D}n\right)\tilde{B}^n$, for DH-QMF, where
$\tilde{B}=\sqrt{2}$, and
$\tilde{A}$, $\tilde{C}$, and $\tilde{D}$ are equal 
to $3.9$, $5.25\times 10^{2}$, and $-2.97\times 10^{2}$
microseconds, respectively.

In order to see how the performance of a closed-loop MFB-CIM scales with increasing
problem size, we solved \mbox{{\sc MaxCut}} problems with SK instances of
problem sizes $n= 100, 200,\dots, 800$. A total of 100 instances of the SK model for each
problem size were randomly generated. Using a closed-loop MFB-CIM, we solved each instance 100 times to evaluate the success probability $P_{\text{s}}$ of finding a ground state and compute a wall-clock time to achieve a success probability of $\geq 0.99$.
It is assumed that all-to-all spin coupling is implemented
in 10 nanoseconds, which corresponds to a cavity round-trip time.
The signal field lifetime is 100 nanoseconds, that is, $\gamma_\text{s} \Delta T_\text{c} = 0.1$.
We use the continuous-time Gaussian model 
as described in \cref{sec:MFB-CIM_A}.
The results are shown in \cref{fig:TTS_comparison}(b), along with the predicted
performance of DAQC and DH-QMF for the SK model instances.
The minimum wall-clock TTS for the closed-loop MFB-CIM at the optimized runtime
$t_{\text{max}}$ scales as an exponential function of $\sqrt{n}$, while those for
DH-QMF and DAQC  scale as exponential functions of $n$. At a problem size
of $n=800$, the wall-clock TTS for the closed-loop MFB-CIM is $\sim10$ milliseconds, while
those for DH-QMF and DAQC  are $\sim10^{120}$ seconds and $\sim10^{50}$
seconds, respectively. 

For the bimodal SK model, which is known to be \lq\lq easy\rq\rq for many algorithms, and for a limited problem-size range of $100\le n\le 500$, we empirically observe a sub-exponential scaling of $\Theta(2^{\sqrt{n}})$ for the closed-loop MFB-CIM's TTS. Such a sub-exponential scaling in solving the SK model instances using CIM-based algorithms has also been reported in other recent studies~\cite{leleu2021scaling, hamerly2019experimental}. For the 21-weight problem instances, due to the limited problem-size range $5\le n\le 30$ of the data available, we cannot reliably infer the actual asymptotic scaling. While our results for the \text{MFB-CIM} seem to agree well with a sub-exponential scaling (with the same exponent $\sqrt{n}$), extrapolations from numerical findings based on small-sized problem instances can potentially be misleading.

In contrast, the scaling of DAQC appears to be exponential. In the absence of empirical data for large problem sizes (even for the SK instances), we perform a careful regression analysis on our data, which we report in~\cref{app_TTS_scaling}. Our analysis suggests an exponential scaling for solving the SK model problem instances and a slightly sub-exponential scaling with the exponent $n^{0.9}$ for the \mbox{21-weight} problem instances. 
Nevertheless, we remain reluctant to extrapolate any exponential scaling laws from this investigation.

As for the TTS scaling of the DH-QMF algorithm, an exponential law of $\widetilde{\mathcal{O}}\left(\sqrt{2^n}\right)$ for the query complexity is supported by rigorous proofs~\cite{durr1996quantum,boyer1998tight}. 
Our benchmarking study reveals that this exponential scaling is not improved for problem instances based on the SK model. In addition, the query complexity does not account for the cost of a single query to the oracle. Our benchmarking results, 
shown in~\cref{fig:QMF-Regression},
are based on a regression towards the scaling given in~\cref{eq:QMF-circuit-depth}, which includes an additional $\text{poly}(n,\log n)$ factor to account for the scaling of the circuit depth of our oracle implementation.

\section{Conclusion}
\label{sec:conclusion}

In this paper, we have presented the results of our study of the scaling of two types of measurement-feedback
coherent Ising machines (MFB-CIM) and compared this scaling to that of
discrete adiabatic quantum computation~\mbox{(DAQC)} and the  \mbox{D\"urr--H{\o}yer} algorithm for quantum minimum finding \mbox{(DH-QMF)}. We performed this comparative study by testing numerical simulations of
these algorithms on \mbox{21-weight} {\sc MaxCut} problems, that is, weighted
\mbox{{\sc MaxCut}} problems with randomly generated edge weights attaining 21
equidistant values from $-1$ to $1$.
We emphasize that our study was 
a numerical analysis; its results depend on the experimental choices we have empirically made to the best of our abilities.

The MFB-CIM of the first type is an open-loop \mbox{MFB-CIM} with predefined feedback
control parameters and the second  is a closed-loop MFB-CIM with
self-diagnosis and dynamically modulated feedback control parameters. The open-loop MFB-CIM utilizes the anti-squeezed $\hat{X}$ amplitude
near threshold under a positive pump amplitude for finding a ground state but at
larger problem sizes the machine is often trapped in local minima. The
closed-loop MFB-CIM employs the squeezed $\hat{X}$ amplitude under a negative
pump amplitude, in which a finite internal energy is sustained through an
external feedback injection signal rather than through parametric amplification.
This second machine self-diagnoses its current state by performing Ising energy
measurement and comparison with the previously attained
minimum energy. The machine continues to explore local
minima without getting trapped even in a ground state.
We observed that for both the \mbox{21-weight} {\sc MaxCut} problems and
the SK Ising model, the closed-loop MFB-CIM outperforms the
open-loop MFB-CIM. One remarkable result is that a low-finesse cavity machine
realizes a shorter TTS than a high-finesse one.
This fact clearly demonstrates that the dissipative coupling of the
machine to external reservoirs is a crucial computational resource for
MFB-CIMs. The wall-clock TTS of the closed-loop MFB-CIM closely follows
$\mathrm{TTS} \approx 4.32 \times {(1.34)}^{\sqrt{n}}$ microseconds for the SK model instances
of size $n$ ranging from $100$ to $800$, assuming a cavity round-trip time of 10 nanoseconds
and a $1/e$ signal amplitude decay time of 100 nanoseconds   ($\gamma_{\text{s}} \Delta T_{\text{c}} = 0.1$).
The performance of the MFB-CIM shown in ~\cref{fig:TTS_comparison} is already competitive
against various heuristic solvers implemented on advanced digital platforms
such as CPUs, GPUs, and \mbox{FPGAs,} in which massive parallel computation is
performed over many billions of  \mbox{transistors~\cite{aramon2019physics,Yoshimura2017implementation,takemoto20192,
leleu2021scaling,Goto2021high,Tatsumura2021scaling}.} 
Note that the results shown in~\cref{fig:TTS_comparison} are based on the assumption of a \mbox{MFB-CIM} architecture that employs only a single OPO as an active element (i.e., it involves only a single optical resonator, along with a nonlinear optical crystal, pumped by a laser) for processing information encoded in time-multiplexed oscillations of the resonator. It is anticipated that advanced on-chip coherent network computing  technologies (see, e.g.,~\cite{McKenna2022}) will allow the design of highly parallelized \mbox{MFB-CIM} architectures 
involving multiple OPO components operated in parallel, with the potential for massively parallel computation that  would further enhance performance.

We have also studied the scaling of the DAQC algorithm in solving 21-weight and SK model {\sc MaxCut} problem instances.
We considered two schemes for optimizing the quantum gate parameters of DAQC,
denoted in the paper as $(\gamma, \beta)$. In the first scheme, we treat $\gamma$
and $\beta$ as hyperparameters that follow a schedule inspired by the adiabatic
theorem. In this case, DAQC can be viewed as a Trotterization of an adiabatic
evolution from the ground state of a mixing Hamiltonian to the ground state of
a problem Hamiltonian.  The second scheme is a variational hybrid
quantum--classical algorithm (similar to the QAOA approach) wherein a classical optimizer is tasked with optimizing the gate
parameters $\gamma$ and $\beta$. The variational scheme must perform
repeated state preparation and projection measurements to estimate the
ensemble averaged energy, which makes the optimization step not only costly
but vulnerable to the shot noise of these measurements. Another disadvantage of
the variational scheme is that optimizing the ensemble average energy does not
necessarily improve the TTS, which is the more practical measure of performance
for the algorithm (see \cref{app_qaoaopt} for more details).
As shown in~\cref{linearpt_scaling}, the adiabatic schedules
achieve very low $R_{99}$ values, suggesting a challenging bound on the allowed number
of shots for the variational scheme to outperform the adiabatic scheme
for this problem. 
Given these considerations, we used a pre-tuned
adiabatic scheme to assess the performance limits of DAQC.
In contrast, we note that the quantum state in an MFB-CIM
survives through repeated measurements, as the measurements performed on
the OPO pulses are not direct projective measurements but  indirect
approximate measurements. These measurements perturb the internal quantum state
of the OPO network but do not completely destroy it. As a result, the above
drawback of a variational scheme for DAQC does not apply to the closed-loop
MFB-CIM. The wall-clock TTS of DAQC with hypertuned adiabatic schedules is
well-represented by the $\mathrm{TTS} \approx 4.6 \times (1.17)^{n}$ microseconds.
As shown in~\cref{fig:TTS_comparison}, extrapolating this
trend suggests that DAQC will perform poorly compared to MFB-CIM as the
problem sizes increase due to an exponential dependence on the number,
$n$, of vertices in the {\sc MaxCut} problem compared to an exponential
growth with a $\sqrt{n}$ exponent in the case of MFB-CIM.

Finally, we have also studied the scaling of \mbox{DH-QMF} for solving 21-weight and SK model {\sc MaxCut}
problems. As this algorithm is based on Grover's search, it performs
$\tilde{\mathcal{O}}(\sqrt{2^n})$ Grover iterations, implying it makes a number of queries, of the same order, to its
oracle. The algorithm also iterates on multiple values of a classical threshold
index; however, this does not change the dominating factors in the scaling of the
algorithm. We have shown that the wall-clock TTS of DH-QMF is well-approximated by the
$\mathrm{TTS} \approx 17.3 \times 2^{n/2}n^2\log \log n$ microseconds
when extrapolated to larger problem sizes.
As shown in
\cref{fig:TTS_comparison}, \mbox{DH-QMF} requires a computation
time that is many orders of magnitude larger than that for either DAQC or \mbox{MFB-CIM}.
This comparatively poor performance of \mbox{DH-QMF} can be traced back to the linear amplitude
amplification in the Grover iteration in contrast to the exponential amplitude
amplification at the threshold of the OPO network. Our study thus leaves open the question of whether there exist
optimization tasks for which Grover-type speedups are of practical significance.

\section*{Methods}
\vspace{-3.5mm}
\noindent
The methods used to obtain the benchmark results for each of the analyzed algorithms
are provided in the corresponding sections, respectively. Additional details are provided as appendices in the Supplementary Material document.

\section*{Data Availability}
\vspace{-3.5mm}
\noindent
The datasets generated and analyzed as part of our study are available
from the corresponding author on reasonable request.

\section*{Acknowledgements}
\vspace{-3.5mm}
\noindent
The authors thank Marko Bucyk for his careful review and editing the manuscript. A.~S. thanks Shengru Ren for helpful discussions.
All authors acknowledge the support of the
NSF CIM Expedition award (CCF-1918549).
P.~R. furthermore thanks the financial support of Mike and Ophelia Lazaridis, and Innovation, Science and Economic Development Canada.

\section*{Author Contributions}
\vspace{-3mm}
\noindent
K.~S., A.~S., and S.~K. contributed equally to this research.
With P.~R.'s guidance, K.~S., N.~G., and W.~B.~K.\ performed numerical
experiments pertaining to the analysis for discrete adiabatic
quantum computation and the DAQC algorithm, and A.~S.\ developed
and implemented the benchmark analysis for D\"urr--H{\o}yer's
quantum minimum finding algorithm. With Y.~Y.'s guidance,
S.~K., S.~R., Y.~I., E.~N., and T.~O.\ conducted the benchmark study
for the coherent Ising machines. K.~S. and A.~S., with support
from P.~R., derived and formulated the details provided in
Section III and Section IV and their corresponding appendices,
respectively. S.~K., with assistance from Y.~Y., determined and
wrote the content for Section II and its corresponding appendix.
The remaining parts of the paper were written by P.~R. and Y.~Y.,
aided by K.~S., A.~S., and S.~K. All authors contributed to ideation
throughout this work. P.~R. and Y.~Y. led the overall efforts
of this benchmark study.

\noindent
\section*{Competing Interests}
\vspace{-3mm}
\noindent
The authors declare no competing interests.

\vspace{3mm}
\begin{appendices}
\appendix

\section{Review of Related Work}
\label{app_related_work}

The motivation for our work is to benchmark coherent Ising machines (CIM) against circuit model quantum algorithms. In this appendix, we review related literature and discuss what  makes our analysis differ from previous work. 

{\em Coherent Ising Machine} --- The CIM based on a network of  degenerate optical parametric oscillators (DOPO) was originally  proposed in~\cite{wang2013coherent}. Various schemes have been 
developed since then, based on optical delay-line (ODL) coupling~\cite{takata2015quantum,maruo2016truncated} and measurement-feedback (MFB) coupling~\cite{shoji2017quantum,inui2020noise}.
The quantum master equations for \mbox{ODL-CIM} and MFB-CIM can be cast into c-number stochastic differential equations using a standard procedure based on representations over the phase space (e.g., the Wigner function, the Sudarshan--Glauber \lq\lq{}P representation\rq\rq{}, or the Husimi \lq\lq{}Q representation\rq\rq{}), the Fokker--Planck equation, and  It{\^o} calculus. In our work, we use the Gaussian approximation of  MFB-CIM~\cite{inui2020noise,ng2021efficient,clements2017gaussian}. 

The standard CIM, which is based on a linear mutual coupling scheme, suffers from amplitude heterogeneity among the constituent DOPOs~\cite{wang2013coherent}, resulting in a performance degradation due to an incorrect mapping of the Ising Hamiltonian to the DOPO network loss. To overcome this drawback, a self-diagnosis scheme, along with dynamical feedback control, has been devised~\cite{leleu2019destabilization,kako2020coherent,reifenstein2021coherent}. 
Our  work demonstrates the first analysis of such a scheme 
using the Gaussian approximation of  MFB-CIM, and   
 provides for the first time a benchmark against the circuit model quantum algorithms suitable for solving Ising problems. 
 Our benchmarking analysis complements previous  benchmarking studies against other solvers, including quantum annealing, digital annealing, simulated annealing, breakout local search, parallel tempering, and simulated bifurcation machines~\cite{hamerly2019experimental,leleu2019destabilization,leleu2021scaling,reifenstein2021coherent}.

{\em DH-QMF algorithm} --- 
The quantum minimum finding algorithm (QMF)  
was originally proposed by D{\"u}rr and H{\o}yer~\cite{durr1996quantum} shortly after Grover invented his algorithm for unstructured search~\cite{grover1996fast,grover1997quantum}. This randomized algorithm uses the quantum 
exponential searching algorithm by Boyer et al.~\cite[Sec.~4 and Theorem 3]{boyer1998tight},  which can be viewed as a generalization of Grover's search algorithm. 
Quantum exponential searching is a randomized method for finding a solution to a search 
problem that has potentially multiple solutions even when their number is not known ahead of time 
(which is the typical scenario at each
threshold-index update step during the progression of 
the DH-QMF algorithm). Denoting this unknown number of solutions by $t$, the algorithm returns 
one of the solutions with equal probability after an expected number $\mathcal{O}(\sqrt{N/t})$ of Grover iterations if $t\ge 1$, and runs forever if $t=0$. 
Assuming there is no time out in the execution of the DH-QMF algorithm, any state whose 
energy is strictly smaller than the energy corresponding to the threshold index may be chosen with some probability as the new threshold index. D{\"u}rr and H{\o}yer~\cite{durr1996quantum} 
show that this probability is given by the inverse of the {\em rank} of the state 
(with respect to its energy value), but is independent of the dimension of the search space. 
This fact is then used to show that the algorithm finds the state of the minimum energy value 
with a success probability of $\mathcal{O}(1)$ using at most  $\mathcal{O}(\sqrt{N})$ Grover iterations. 

Our work presents for the first time a Monte Carlo simulation of the progression of the \mbox{DH-QMF} 
algorithm aimed at the inference of {\em concrete} lower bounds on the optimistic performance of the algorithm (in terms of the TTS). 
Our Monte Carlo analysis (see \cref{sec:meth_qmf}) is based on sampling according to \mbox{energy--frequency} (equivalently, cut-size--frequency) histograms associated with weighted graphs. In particular, 
the energy (or cut-size) histogram generated for each weighted graph instance 
allows us to infer the number of marked states, which in turn allows us to determine the success and failure probabilities, as well as to predict both the optimal number 
of Grover iterations at each threshold-index update step and the number of
Grover searches needed to boost the success probability beyond a target  threshold.
Our approach thus allows us to emulate the progression 
of the \mbox{DH-QMF} algorithm and hence to infer a concrete optimistic 
lower bound on the TTS for each given problem instance. In our studies, we have not encountered any previous work that uses a similar approach to simulating the progression of the \mbox{DH-QMF} algorithm.

{\em DAQC algorithm} ---
The second circuit-based algorithm  analyzed in our benchmarking study is 
discrete adiabatic quantum computation (DAQC). Its circuit ansatz is derived from  implementing Hamiltonian simulation for a given adiabatic evolution using the well-established  first-order Suzuki–Trotter expansion. Its application to solving various optimization problems has been extensively explored since quantum adiabatic evolution was first introduced to the solving of NP-complete problems by Farhi et al.~\cite{farhi2001quantum}. A related concept is the quantum approximate optimization algorithm 
(QAOA)~\cite{farhi2014quantum,guerreschi2019qaoa}, which can be viewed as a diabatic counterpart to DAQC.  Indeed, the circuit ansatz of QAOA is very much akin to a Trotterized analogue of quantum adiabatic evolution: it uses an iterative unitary evolution of pure states in a quantum circuit according 
to a mixing Hamiltonian and a problem Hamiltonian, which in the framework 
of adiabatic quantum computation correspond to the initial and final 
Hamiltonians of evolution, respectively. 
QAOA is considered a promising candidate for solving combinatorial optimization problems on noisy, intermediate-scale quantum (NISQ) devices. Although its formulation is not restricted to implementations on such devices, QAOA has become associated with NISQ algorithms, due to its use of shallow (i.e., low-depth) quantum circuits, along with a variational method for optimizing the set of parameters specifying the unitary gates in those circuits. 
Its performance as a NISQ algorithm has been extensively studied in recent work~\cite{guerreschi2019qaoa, zhou2020quantum, crooks2018performance,willsch2020benchmarking}. 

We have studied two schemes for optimizing Trotterization parameters.
The first scheme treats gate parameters as hyperparameters that follow a tuned schedule for Trotterized adiabatic evolution.  The second scheme uses a variational hybrid \mbox{quantum--classical} protocol to optimize the 
gate parameters. We found a performance advantage for the first scheme 
over the second. The challenges of the latter scheme are discussed in~\cref{app_qaoaopt}.
For this reason, we have used  pre-tuned DAQC schedules for our benchmarking analysis. 
Moreover, to obtain the theoretical performance limit, we dropped the requirement of having to use only low-depth circuits that are necessary in the case of NISQ devices. That is, our results pertain to the implementation of DAQC using quantum circuits of arbitrary depth. 
Our new insights with respect to the optimization of quantum gate parameters, and 
our results on the {\em concrete} lower bounds on the optimistic performance 
of the \mbox{DAQC} algorithm (in terms of the TTS) in solving \textsc{MaxCut} 
problem instances, complement previous studies on the performance 
of QAOA. In particular, our extensive analysis of hyperparameter tuning for DAQC parameter schedules does not resemble any previous numerical results in the literature, and may serve as a new baseline for future benchmarking studies based on DAQC.

\section{Discrete-Map Gaussian Model of the CIM}
\label{app_discmodel}

In this Appendix, we summarize the discrete-map Gaussian model of the CIM presented in~\cite{ng2021efficient}, and we adapt the feedback step to include the dynamic feedback control used for the closed-loop MFB-CIM. This discrete-map model is used to study the optimization performance of the MFB-CIM in ~\cref{sec:cim-DTmodel}.
In the discrete Gaussian quantum model of MFB-CIM, the density operator of the
$i$-th OPO pulse is fully characterized by the mean amplitude $\mu_{i}$ and
covariance matrix $\Sigma_{i}$ defined by \cref{eq:disc-mu,eq:disc-sigma}. The
total density operator before all pulses start their $\ell$-th round trip is
expressed by $\otimes_{i=1}^{n} \hat{\rho} (\mu_{i}(\ell), \Sigma_{i}(\ell))$.
Propagation of the state of the $i$-th pulse through $\ell$-th roundtrip from
$\hat{\rho} (\mu_{i}(\ell), \Sigma_{i}(\ell))$ to $\hat{\rho} (\mu_{i}(\ell+1),
\Sigma_{i}(\ell+1))$ is described by performing the following discrete maps
consecutively.

1. \emph{Background linear loss:} The lumped background linear loss transforms the
density operator as
\begin{align}
   \hat{\rho} (\mu_{i}, \Sigma_{i})
   \mapsto \mathrm{tr}_{c} \left( \mathcal{B}
   \left[ \hat{\rho} (\mu_{i}, \Sigma_{i})
   \otimes \hat{\rho} (0_{c}, \Sigma_{c}^{0}) \right] \right),
\end{align}
where $\Sigma_{c}^{0} = \mathrm{diag}(1/2,1/2)$ is the covariance of a coherent state. The beamsplitter map $\mathcal{B}$ is defined by
\begin{align}
   \label{eq:map-B}
   \mathcal{B} \left[ \hat{\rho} (\mu, \Sigma) \right]
   = \hat{\rho} (S \mu, S \Sigma S^{\mathrm{T}}),\\
   \label{eq:mat-bs}
   S =
   \begin{pmatrix}
       t & 0 & -r & 0 \\
       0 & t & 0 & -r \\
       r & 0 & t & 0 \\
       0 & r & 0 & t
   \end{pmatrix}.
\end{align}
Here, $t=\sqrt{1-r^2}$ is the amplitude transmission coefficient of a fictitious
beamsplitter which represents background linear loss. Physically, $\hat{\rho}
(0_{c}, \Sigma_{c}^{0})$ is a reservoir vacuum state and it is traced out after
mixing with the signal pulse at the beamsplitter.

2. \emph{Parametric amplification/deamplification during OPO crystal propagation:} The
propagation through a second-order nonlinear crystal with the pump pulse
transforms the density operator as
\begin{equation}
\hat\rho(\mu_i, \Sigma_i) \mapsto \mathrm{tr}_b\left[\chi\Bigl(\hat\rho(\mu_i,\Sigma_i)
\otimes \hat\rho(\mu_b,\Sigma^0_b)\Bigr)\right],
\end{equation}
where $\mu_b$ and $\Sigma^0_b = \mathrm{diag}(1/2,1/2)$ describe the initial
condition of the (Gaussian) pump pulse, and the map $\chi$ abstractly represents
their joint propagation through the crystal, that is,
\begin{equation}
\chi: \hat\rho(\mu_i,\Sigma_i) \otimes \hat\rho(\mu_b,\Sigma^0_b) \mapsto \hat\rho(\mu_{i,b},\Sigma_{i,b}),
\end{equation}
where $\hat\rho(\mu_{i,b},\Sigma_{i,b})$ is a joint two-mode Gaussian state
of the signal and pump at the output. This joint output is determined
by the equations of motion for the mean-field and covariance matrix:
\begin{align}
\frac{d \langle \hat{X_{i}} \rangle }{dt}
&= \epsilon \langle \hat{X_{b}} \rangle \langle \hat{X_{i}} \rangle
+ \epsilon \langle \delta \hat{X_{b}} \delta \hat{X_{i}}
+ \delta \hat{P_{b}} \delta \hat{P_{i}} \rangle\\
\frac{d \langle \hat{X_{b}} \rangle}{dt}
&= - \frac{\epsilon}{2} \langle \hat{X_{i}}^2 \rangle
- \frac{\epsilon}{2} \langle \delta \hat{X_{i}}^2
-\delta \hat{P_{i}}^2 \rangle \\
\frac{d \langle \delta \hat{X_{i}}^2 \rangle}{dt}
&= 2 \epsilon \langle \hat{X_{b}} \rangle \langle \delta \hat{X_{i}}^2 \rangle
+ 2 \epsilon \langle \hat{X_{i}} \rangle \langle
\delta \hat{X_{b}} \delta \hat{X_{i}} \rangle\\
\frac{d \langle \delta \hat{P_{i}}^2 \rangle}{dt}
&= -2 \epsilon \langle \hat{X_{b}} \rangle \langle \delta \hat{P_{i}}^2 \rangle
+ 2 \epsilon \langle \hat{X_{i}} \rangle \langle \delta \hat{P_{b}}
\delta \hat{P_{i}} \rangle \\
\frac{d \langle \delta \hat{X_{b}}^2 \rangle}{dt}
&= - 2 \epsilon \langle \hat{X_{i}} \rangle \langle \delta \hat{X_{b}}
\delta \hat{X_{i}} \rangle\\
\frac{d \langle \delta \hat{P_{b}}^2 \rangle}{dt}
&= -2 \epsilon \langle \hat{X_{i}} \rangle \langle \delta \hat{P_{b}}
\delta \hat{P_{i}} \rangle\\
\frac{d \langle \delta \hat{X_{b}} \delta \hat{X_{i}} \rangle}{dt}
&= \epsilon \langle \hat{X_{i}} \rangle \langle \delta \hat{X_{b}}^2
- \delta \hat{X_{i}}^2 \rangle
+ \epsilon \langle \hat{X_{b}} \rangle \langle \delta \hat{X_{b}}
\delta \hat{X_{i}} \rangle\\
\frac{d \langle \delta \hat{P_{b}} \delta \hat{P_{i}} \rangle}{dt}
&= \epsilon \langle \hat{X_{i}} \rangle \langle \delta \hat{P_{b}}^2
- \delta \hat{P_{i}}^2 \rangle
- \epsilon \langle \hat{X_{b}} \rangle \langle \delta \hat{P_{b}}
\delta \hat{P_{i}} \rangle
\end{align}

Here $\epsilon$ is the parametric coupling rate defined by the Hamiltonian
$\mathcal{H} = i \frac{\hbar \epsilon}{2} (\hat{b} \hat{a}^{\dag^2} -
\hat{b}^{\dag} \hat{a}^2)$, where $\hat{a}$ and $\hat{b}$ are signal and pump
annihilation operators. We assume that the input state into the crystal
satisfies $\langle \hat{P_{i}} \rangle = \langle \hat{P_{b}} \rangle = 0$ (i.e., there is no
coherent excitation along the quadrature-phase) and $\langle \{ \delta \hat{X_{i}},
\delta \hat{P_{i}} \} \rangle = \langle \{ \delta \hat{X_{b}}, \delta
\hat{P_{b}} \} \rangle =0$ (both the signal and pump have no correlation between
in-phase and quadrature-phase fluctuations). Note that $\langle \hat{P_{i}}
\rangle = \langle \hat{P_{b}} \rangle = 0$ is satisfied at all times under the
above conditions. The defined map $\chi $ thus describes all such effects as
linear parametric amplification/deamplification, signal-pump entanglement
formation, and back conversion from the pump to the signal.

3. \emph{Out-coupling and homodyne detection:} The out-coupling of the internal signal
pulse is described by the map
\begin{align}
\label{eq:map-Bout}
   \hat{\rho} \left(\mu_{i}, \Sigma_{i} \right)
   &\mapsto \hat{\rho} \left(\mu_{i,h}, \Sigma_{i,h} \right) \nonumber\\
   &= \mathcal{B}_\mathrm{out} \left[ \hat{\rho} \left(\mu_{i}, \Sigma_{i} \right)
   \otimes \hat{\rho} \left(0_{h}, \Sigma_{h}^{0} \right)\right],
\end{align}
where the beamsplitter map $\mathcal{B}_\mathrm{out}$ is defined by
\cref{eq:map-B,eq:mat-bs} with an out-coupling rate of $r_\mathrm{out}$. In
\cref{eq:map-Bout}, a probe mode $h$ is prepared in a vacuum state and mixed
with the signal pulse. This process creates a joint correlated state (entangled
state) between the internal pulse and external (out-coupling) pulse. Suppose a
homodyne measurement for the out-coupled pulse reports a result $m_{i}(l)$
for the $i$-th signal pulse at the $\ell$-th round trip. Such an indirect
approximate measurement projects the internal state to a new state by the map
\begin{align}
\label{eq:map-H}
   \hat{\rho} \left(\mu_{i,h}, \Sigma_{i,h} \right)
   &\mapsto \mathcal{H} \left[ \hat{\rho}
   \left(\mu_{i,h}, \Sigma_{i,h} \right) \right] \nonumber \\
   &= \hat{\rho} \left(\mu_{i}^{(m_{i})}, \Sigma_{i}^{(m_{i})} \right),
\end{align}
where the homodyne detection map $\mathcal{H}$ is defined by
\begin{align}
\label{eq:H-mu}
\mu_{i}^{(m_{i})} &= \mu_i + \left( \frac{w_{i} - \mu_h}{\Sigma_{XX}} \right) v_X\\
\label{eq:H-Sigma}
\Sigma_{i}^{(m_{i})} &= \Sigma_{i} - \frac{v_X {v_X}^{\mathrm{T}}}{\Sigma_{XX}}.
\end{align}
Here, $v_X$ is the $X$ off-diagonal component (the degree of signal-probe
correlation) of the matrix $\Sigma_{i, h}$ and $\Sigma_{XX}$ is the $X$ diagonal
element of $\Sigma_{h}$. The second terms of the right-hand sides of
\cref{eq:H-mu,eq:H-Sigma} express the mean-field shift and variance reduction
induced by the homodyne measurement.

4. \emph{Feedback injection:} We implement the Ising coupling by applying
the displacement operation for the internal pulse amplitude based on the
measurement results of the $\ell$-th round trip, $m_{j}(\ell)$,
for all pulses except for the $i$-th pulse. The displacement magnitude is given
by
\begin{equation}
   v_{i} (\ell) = J_0 e_{i} (\ell) \sum_{j \neq i} J_{ij} m_{j}(l),
\end{equation}
where $e_{i} (\ell)$ is the feedback-field amplitude of the $\ell$-th round trip
which is determined by the equation of motion, \cref{eq:error-equation}, for the
closed-loop CIM. 
The feedback gain $J_0$ scales with the inverse of $\sqrt{\frac{N_{\text{decay}}}{N} \sum_{i,j} \left| J_{ij} \right|}$, where $N_{\text{\tiny decay}}$ is the number of round trips required for the signal field amplitude to attenuate by a factor of $1/e$.
The feedback injection map is thus determined by
\begin{multline}
\label{eq:map-D}
\hat{\rho} \left(\mu_{i}, \Sigma_{i} \right)
\mapsto \mathcal{D}_{v_{i}}
\left[ \hat{\rho} \left(\mu_{i}, \Sigma_{i} \right) \right]
= \hat{\rho} \left(\mu_{i} + v_{i}, \Sigma_{i} \right).
\end{multline}

\noindent The above four steps are applied to all pulses \mbox{($i= 1, \dots, n$)}, completing one
round trip through the CIM cavity.

\section{Discussion of Optimal Loss Parameters for the MFB-CIM}
\label{app_loss_opt}
\begin{figure}[t]
\centering
\begin{tabular}{c @{\hskip 0.13in} c}
    \text{\scriptsize{(a) Closed-Loop CIM, $n=30$}}
    & \text{\scriptsize(b) Open-Loop CIM, $n=30$} \\ \\
    \includegraphics[width=0.49\linewidth]{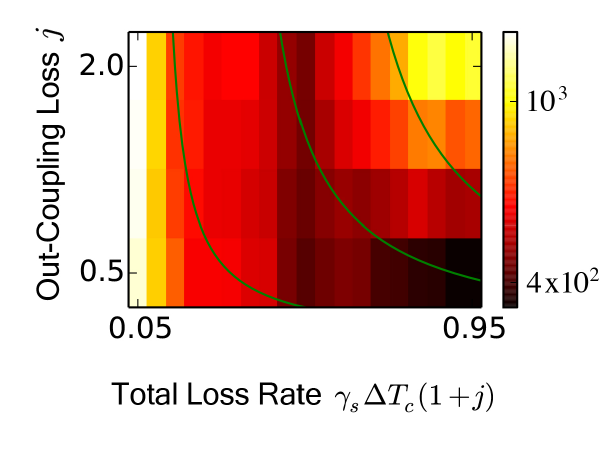}
    & \includegraphics[width=0.49\linewidth]{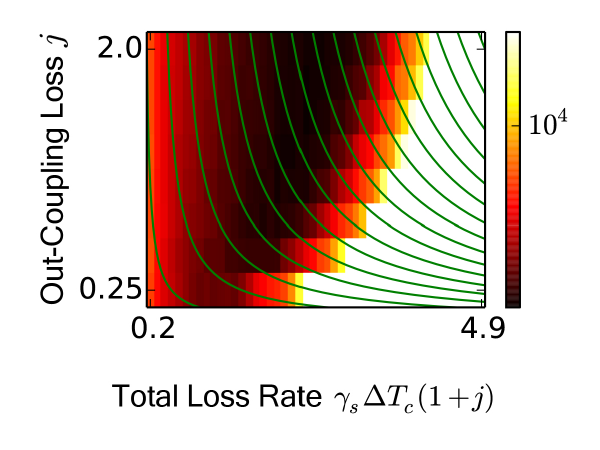} \\ \\
    \text{\scriptsize(c) Closed-Loop CIM, $n=100$}
    & \text{\scriptsize(d) Open-Loop CIM, $n=100$} \\ \\
    \includegraphics[width=0.49\linewidth]{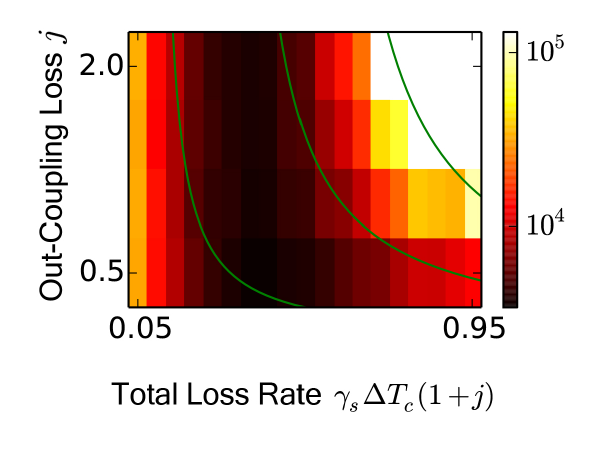}
    & \includegraphics[width=0.49\linewidth]{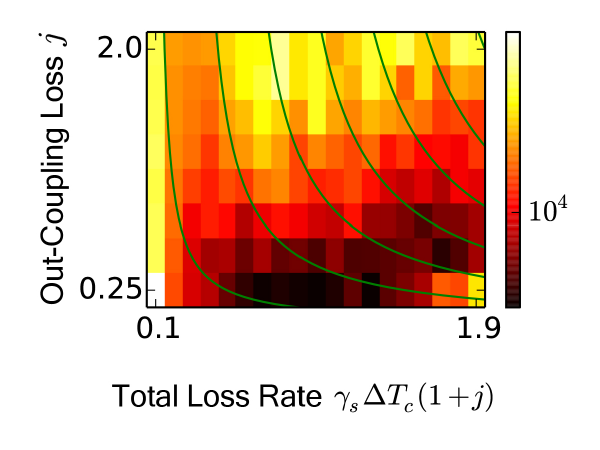}
\end{tabular}
\caption{Heat maps of the TTS for the Sherrington--Kirkpatrick model with the x-axis representing the total loss rate $\gamma_\text{s} \Delta T_\text{c} (1 + j)$ and the y-axis representing the out-coupling loss $j$. (a) and (b) Heat maps for the closed-loop and open-loop CIMs for $n=30$. (c) and (d) Heat maps for the closed-loop and open-loop CIMs for $n=100$. The colours indicate the value of the TTS in terms of the number of round trips, where a darker colour represents a shorter TTS. The green contour lines correspond to fixed values for \mbox{$J = j\gamma_\text{s}$}.}
\label{fig:app_heatmap}
\end{figure}

In \cref{fig:TTS-finesse}, the effect of changing the total loss rate \mbox{$\gamma_\text{s} \Delta T_\text{c} (1 + j)$} on the TTS by using the discrete-time model of the MFB-CIM is shown.
There are various ways the total loss rate can be varied.
For the results displayed in \cref{fig:TTS-finesse}, we kept $j$ constant at the value $1$ (recall that $j$ is a parameter that corresponds to the escape efficiency in \cite{ng2021efficient}, which
is the ratio of the out-coupling loss associated with the optical homodyne measurement to the total cavity loss) and varied $\gamma_\text{s} \Delta T_\text{c}$. There is a sweet spot around $\gamma_\text{s} \Delta T_\text{c} (1 + j) \approx 1$.

In \cref{fig:app_heatmap}(a) and \cref{fig:app_heatmap}(b), heat maps are shown of the TTS for a problem instance of size $n=30$ are shown. Here, the \mbox{x-axis} represents the total loss rate $\gamma_\text{s} \Delta T_\text{c} (1 + j)$ and the \mbox{y-axis} represents the out-coupling loss $j$. In these plots,  $j=1$ on the y-axis corresponds to the TTS curves plotted in~\cref{fig:TTS-finesse}. The green contour lines correspond to fixed values for \mbox{$J = j\gamma_\text{s}$}.
As evident from these plots, at least in the case of the open-loop CIM, an increase in the value of the total loss rate, moving along the horizontal axis, results in the optimal region becoming larger, while moving along a green contour line, the optimal region becomes sharper.
In the case of the closed-loop CIM, there appear to be two optimal regions.
We believe that the more accurate optimal region in this case is the region along the vertical line given by $\gamma_\text{s} \Delta T_\text{c} (1 + j) \approx 0.5$ or ($N_{\text{decay}} = 2$), even though in this region the TTS is longer, because, as the total loss rate becomes sufficiently large,
the nonlinearity increases in strength such that the error correction mechanism can no longer stabilize the amplitude to the desired target amplitude.
The reason there is a short TTS in this region for $n=30$ is that the problems are small enough that they can still be solved despite the unstable behaviour of the solver.
However, in the case of the problem size $n=100$, as shown in \cref{fig:app_heatmap}(c), this second region no longer has a short TTS, and the optimal TTS occurs in the region  around the vertical line defined by $\gamma_\text{s} \Delta T_\text{c} (1 + j) \approx 0.3$.

\section{Hyperparameter Tuning for DAQC Parameter Schedules}
\label{app_hptuning}

As described in \cref{sec:tts-qaoa}, we have a recipe for generating
DAQC parameter schedules for any problem Hamiltonian $H_\text{P}$ and number of layers
$p$. We consider two hyperparameters for these schedules:

\begin{itemize}

\item
The number $L=T/p$ is the evolution time in each Trotterized layer of the
associated annealing schedule. A larger value of $L$ corresponds to a slower and
therefore better associated annealing schedule, but also
brings along a greater Trotterization error;

\item
The number $a$ is the coefficient of the cubic term in the adiabatic
schedule. When $a=0$ the schedule is linear, and when $a=4$ the schedule is
cubic, with $f'(T/2)=0$. We therefore only consider $a \in [0, 4]$, because for
$a>4$ the schedule would be decreasing at $t=T/2$.

\end{itemize}

\begin{figure}[b]
\centering
\includegraphics[trim=0 0 0 30, width=0.47\textwidth]{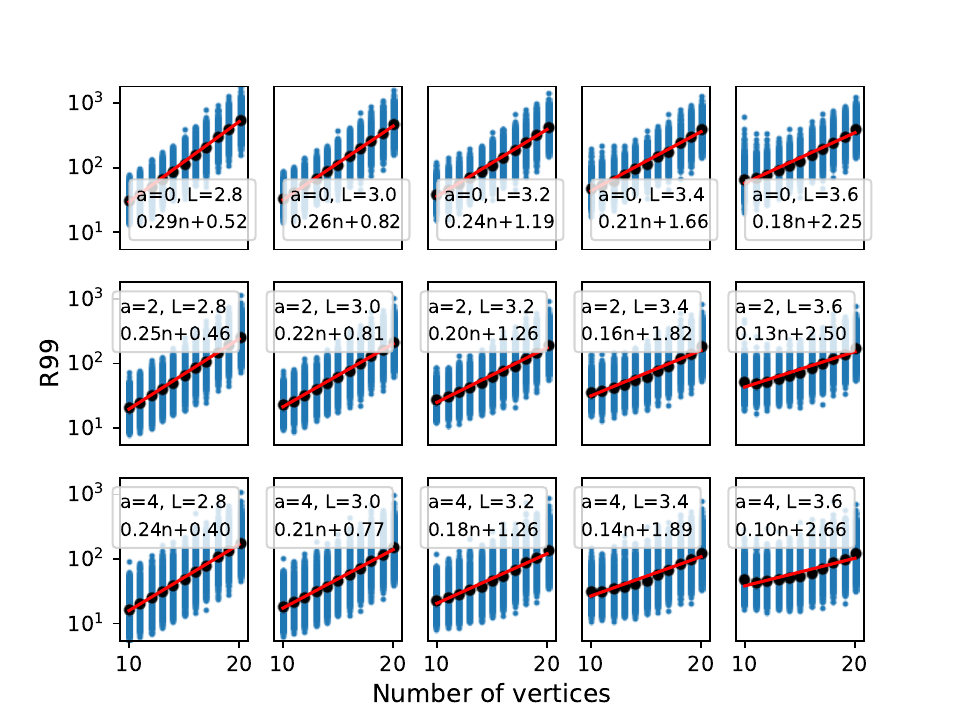}
\caption{
$R_{99}$ of the good initial DAQC parameters at $p=4$ layers for various values of
$a$ and $L$, on all 1000 graph instances of each size ranging from 10 to 20.}
\label{hptuning}
\end{figure}

\begin{figure}[t]
\centering
\includegraphics[width=0.47\textwidth]{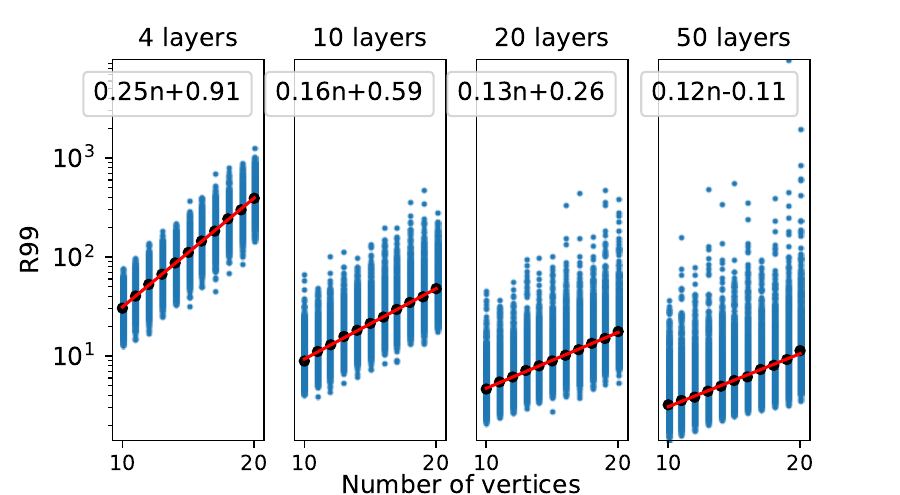}
\includegraphics[width=0.47\textwidth]{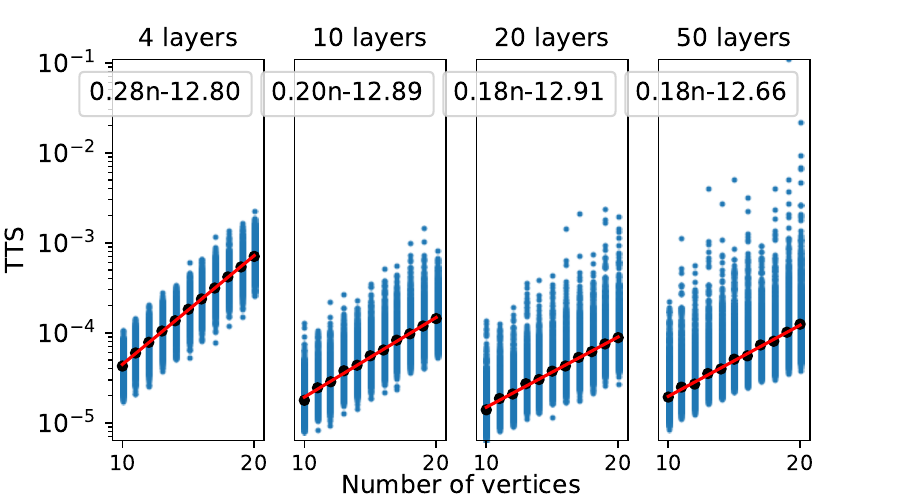}
\caption{
$R_{99}$ and TTS of a linear schedule for $10 \le n \le 20$,
$p \in \{4, 10, 20, 50\}$,
with hyperparameters $a=0.0$ and $L = 1.6 + 0.1n$.}
\label{linearpt_scaling}
\end{figure}

\begin{figure}[t]
\centering
\includegraphics[width=0.5\textwidth]{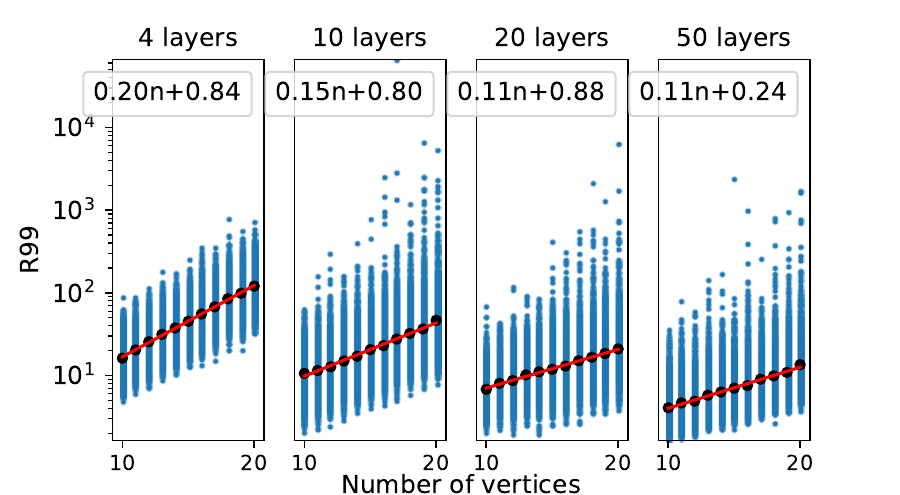}
\includegraphics[width=0.5\textwidth]{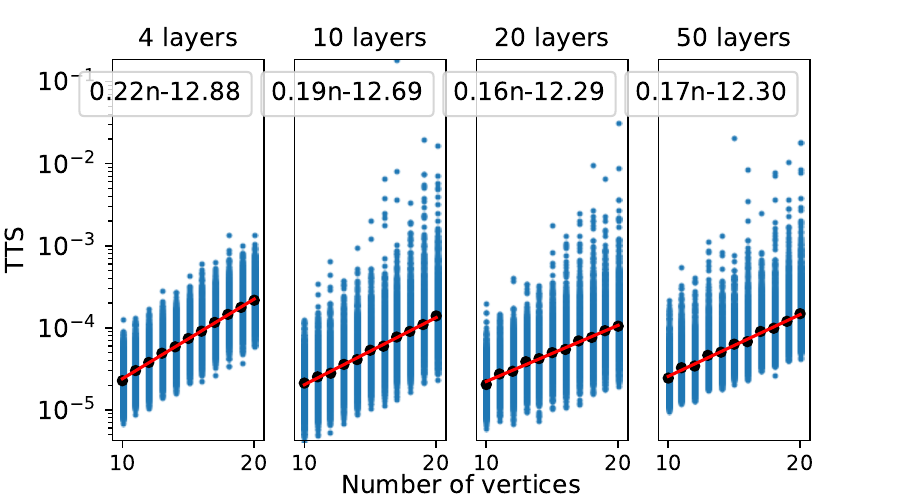}
\caption{
$R_{99}$ and TTS of a cubic schedule for
$10 \le n \le 20$, $p \in \{4, 10, 20, 50\}$,
with hyperparameters $a=4.0$ and $L = 1.6 + 0.1n$. The performance is better than that of the
linear schedule for shallow circuits, but stops improving as the number of layers becomes larger.}
\label{cubicpt_scaling}
\end{figure}

Here, we compile our results on the performance of DAQC with cubic schedules
for various values of the hyperparameters $a, L,$ and $p$. In
\cref{hptuning,linearpt_scaling,cubicpt_scaling},
the horizontal axis displays the number of vertices for the problem
instance, and the vertical axis displays the $R_{99}$ or TTS (in logarithmic scale).
Each blue dot represents a single problem instance. All plots depict a
total of 11,000 problem instances varying from 10 to 20 nodes in size.
Each black point represents the geometric mean of all values of $R_{99}$ or TTS for
problem instances of a given size. Finally,
the red line indicates the best linear fit to the black points. The
equation corresponding to the best-fit line is written in each subplot, where $n$ is the
number of vertices.

We empirically found that a value of $L$ between 2.6 and 3.6 worked best. In
\cref{hptuning}, we plot the  $R_{99}$ values of the good parameter schedule with
hyperparameters \mbox{$a \in \{0, 2, 4\}$} and $L \in \{2.8, 3.0, 3.2, 3.4, 3.6\}$.
Note that $a=4$ (a cubic schedule with a derivative of $0$ at the inflection point)
outperforms $a=0$ (a linear schedule). We observed that, as the number of vertices
$n$ increases, the optimal value of the scaling constant $L$ increases.
Therefore, our tuned hyperparameter value used in
\cref{linearpt_scaling,cubicpt_scaling} is \mbox{$L = 1.6  +  0.1n$.}

In \cref{linearpt_scaling,cubicpt_scaling}, we present the
scaling of a linear schedule opposite to that of a cubic schedule. As the
number of layers increases, performance as measured by $R_{99}$ improves, as
expected. However, with more layers, more time is required to perform a single
circuit shot, and therefore the scaling of TTS is actually \emph{worse} at 50
layers than it is at 20 layers. For large numbers of layers, the linear schedule
and cubic schedule perform similarly, which is expected because both are
Trotterizations of a very slow adiabatic schedule.


\section{Challenges Encountered with the Variational Optimization Protocol}
\label{app_qaoaopt}

When DAQC parameter schedules are tuned variationally, the energy measurements
from the quantum device are used to decide the next parameters to try via a
hybrid quantum--classical process. A single ``shot'' with the
parameters $(\gamma, \beta)$ consists of running the DAQC circuit
once with parameters $(\gamma, \beta)$, and measuring the energy of
the prepared state $|\psi(\gamma, \beta)\rangle$, which destroys the
prepared state and returns a single measurement outcome. We perform a large
number of shots using $(\gamma, \beta)$, and the results are
averaged to estimate the expected energy
\begin{equation}
    EE(\gamma, \beta):= \langle\psi(\gamma, \beta)|H_\text{P}|\psi(\gamma, \beta)\rangle.
\end{equation}
This expected energy is treated as a loss function which is minimized by a
classical optimizer. This approach suffers from two major challenges.

\begin{figure}[ht]
\centering
\includegraphics[trim=0 0 0 10, width=0.477\textwidth]{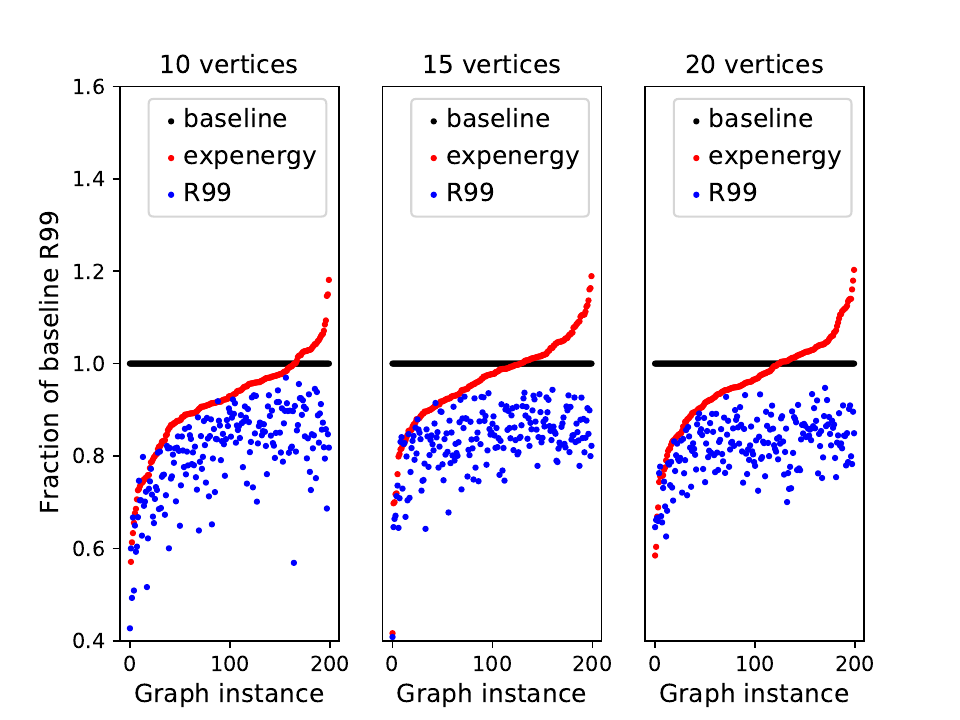}
\caption{
Plot depicting the fraction of the baseline $R_{99}$ achieved when optimizing for
expected energy with no shot noise (red) versus optimizing for $R_{99}$ (blue).
Baseline $R_{99}$ (black) is given by the cubic parameter schedule, as described in
\cref{sec:tts-qaoa}. Even when shot noise is absent, optimizing for
expected energy can \emph{increase} the $R_{99}$ about a third of the time, as is evidenced
by the fact that a third of the red points are above the black line. We
performed this optimization using 100 function evaluations using the Nelder--Mead method, and
due to imperfect optimization, a few blue points landed above the red curve. The
$x$-axis is the graph instance number from 0 to 199, where graphs have been
sorted according to the $y$-value of the red point.}
\label{R99vsexpenergy}
\end{figure}

Firstly, we want parameters $(\gamma, \beta)$ which minimize the $R_{99}$,
rather than the expected energy. Although these two loss functions are related,
they are not perfectly correlated, and this difference becomes more apparent as
we move closer to the parameters which minimize $R_{99}$. Unfortunately, it is
impossible to optimize the ansatz with respect to $R_{99}$, as this would
require knowledge of the ground state.

Secondly, because projective measurements are stochastic, our estimate of the
expected energy is approximate, and this makes parameter optimization difficult.
To overcome this issue, we would need to use a large number of shots per point
$(\gamma, \beta)$, which makes the variational algorithm costly.

In \cref{R99vsexpenergy}, we illustrate the implications of the first challenge. We
consider a four-layer DAQC circuit on graphs of size 10, 15, and 20. For each
graph $G$, the following analysis is performed. First, the cubic schedule $\theta_G$ (see
\cref{sec:tts-qaoa}) is found and its $R_{99}$ is calculated. The Nelder--Mead method
is then used to optimize the expected energy, with its parameter schedule initialized as
$\theta_G$ and given access to 100 perfect evaluations of expected energy (which
ordinarily can only be approximated). The $R_{99}$ of the result is divided by
the $R_{99}$ of the cubic schedule, and these ratios have been plotted in red. Finally, the Nelder--Mead method is used to optimize $R_{99}$, with a schedule
initialized with $\theta_G$ and access to 100 perfect evaluations of $R_{99}$
(which is ordinarily impossible to calculate). The $R_{99}$ of the result is
divided by the $R_{99}$ of the cubic schedule, and these ratios have been plotted in blue.
For better visibility, the graph instances along the $x$-axis have been sorted
by the $y$-values of the red points. We observe that even with \emph{perfect}
estimation of the expected energy, optimization results in a \emph{worse} final
$R_{99}$ in 15 to 40 percent of graph instances. This is the case despite the fact that the cost (in shots) of performing this optimization has been discarded. The effect of including the cost would have been substantial.

\section{Regression Analysis on the Scaling of DAQC}
\label{app_TTS_scaling}

\begin{figure}[t!]
\centering
\begin{tabular}{c}
    \text{(a) Regression Model Confidence for 21-Weight Graphs}\\
    \includegraphics[width=0.8\linewidth]{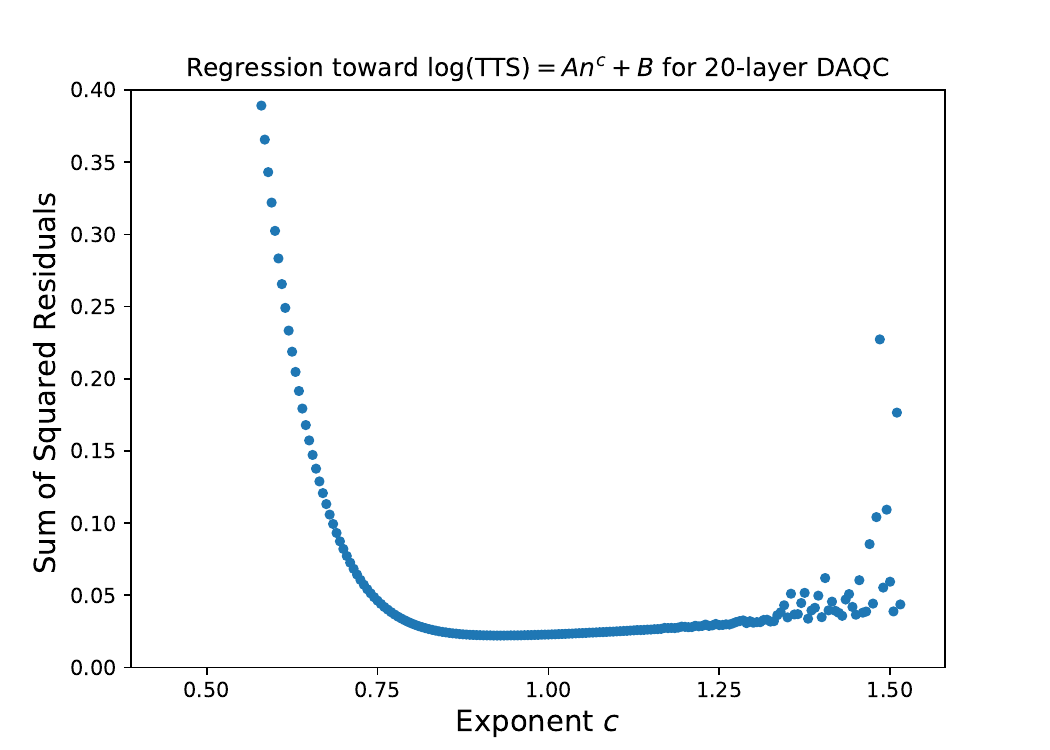}\\
    \\
     \text{(b) Regression Model Confidence for SK Model}\\
    \includegraphics[width=0.8\linewidth]{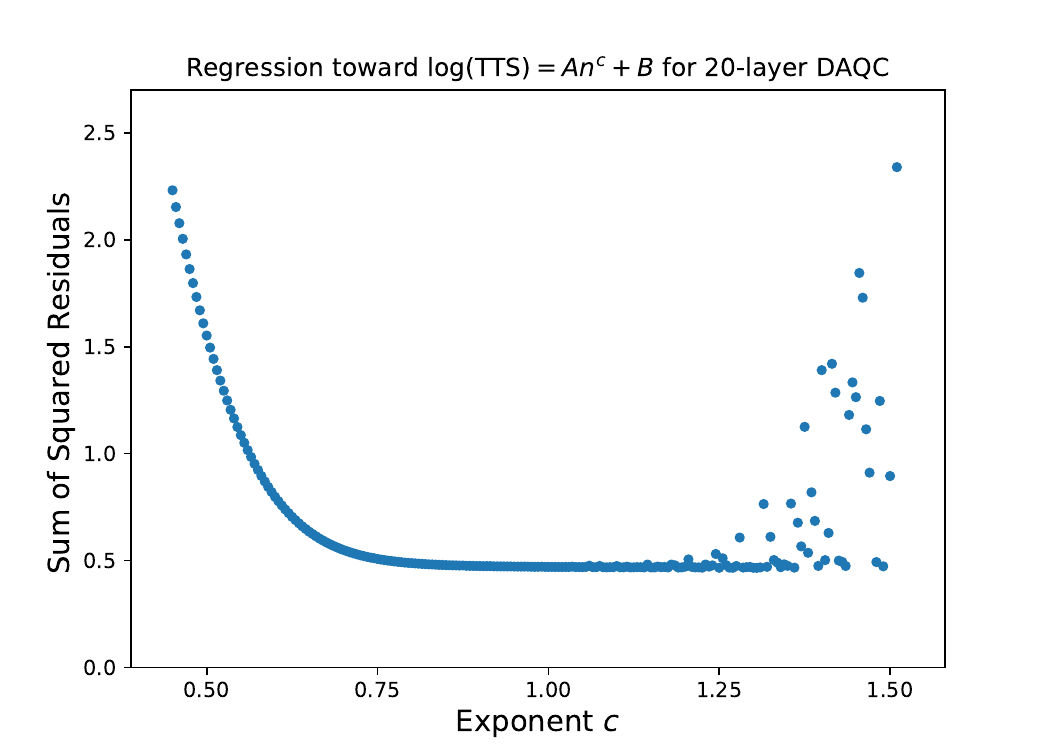}
\end{tabular}
\caption{Quality of the regression model $\log(\text{TTS})=An^c+B$ for different values of the exponent $c$, for DAQC with an optimum number of layers ($p=20$). (a) Sum of squared residuals (SSR) as a function of $c$ for 21-weight graphs; (b) SSR as a function of $c$ for the fully connected bimodal SK model. The smaller the SSR value is, the better the model fits the data. For each value of $c$, the regression was performed using the stochastic gradient descent optimization algorithm with injected Gaussian noise. For 21-weight problem instances, the best regression fit is observed for $c \approx 0.9$, which suggests a slightly sub-exponential scaling.}
\label{fig:app_scaling_daqc}
\end{figure}

To investigate the validity of assuming an exponential scaling of the TTS for DAQC, we conducted a regression analysis for the more general scaling law 
\begin{equation}
\log(\text{TTS}) = An^c+B.
\label{eq:general-scaling}
\end{equation}
In \cref{fig:app_scaling_daqc}, we display the minimum sum of squared residuals,  which is a direct measure of how well the regression model fits the data, for the range of values $0.5\le c\le 1.5$ of the exponent $c$. For each value of $c$ within this range, 
the method of stochastic gradient descent 
is used to find the optimum parameter values of $A^*$ and $B^*$ that minimize  the sum of squared residuals. Our regression analysis includes additional Gaussian noise injection at each iteration of the update. For both the 21-weight problem instances and the bimodal SK model instances, we observe the highest confidence  with respect to the quality of the regression fit for exponent values that are close to $c = 1$.  This suggests that DAQC indeed scales exponentially in solving these problem instances. For the 21-weight problem instances, the scaling is actually slightly sub-exponential at the value $c \approx 0.9$.

\section{Grover's Search as a Subroutine \\ of DH-QMF}
\label{app:app_qmf}

Grover's search algorithm~\cite{grover1996fast,grover1997quantum} has been
extensively studied and applied since its invention more than twenty years ago.
This appendix provides more details with a focus on its implementation as a
subroutine of DH-QMF. In~\cref{app:Grover}, we start with a brief review of how
Grover's search algorithm works. In~\cref{app:QMF-oracle}, we expand on the
quantum circuits used to implement the QMF oracle, which is   required when Grover's
search is employed as a subroutine of DH-QMF, and explain the contributions to its
resource requirements.

\subsection{A Brief Review of Grover's Search Algorithm}
\label{app:Grover}

\begin{figure}[htb]
  \centering
  \includegraphics[width=1.0\linewidth]{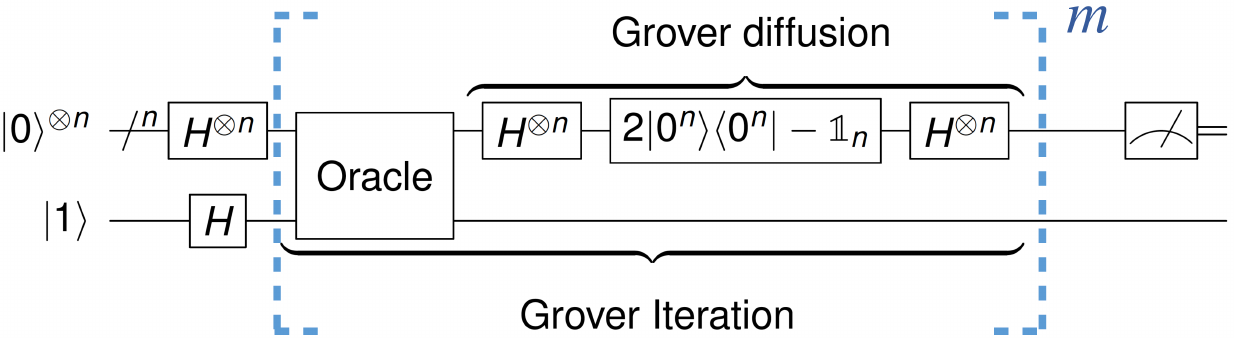}
  \caption{Quantum circuit for Grover's search. The key components are
  an oracle, which marks the solution states, and the Grover diffusion,
  which implements a reflection about the mean amplitude.
  The composition of the oracle followed by Grover diffusion forms the
  so-called Grover iteration, which is repeated
  $m\in\mathcal{O}\left(\sqrt{2^n}\right)$ times.
  Here, $H$ denotes the Hadamard gate.}
  \label{fig:Grover-circuit}
\end{figure}

The circuit of Grover's search algorithm is illustrated in ~\cref{fig:Grover-circuit}.
The quantum circuit takes as inputs an $n$-qubit register  $\texttt{vertex}$ and
a single-qubit register $\texttt{flag}$, where $n=\lceil \log_2 N \rceil $. The
$\texttt{vertex}$ register is used to encode the possible spin configurations
(and any superpositions of them);  it is initialized in the state
$\ket{0}^{\otimes n}$ and transformed into a uniform superposition
\mbox{$\frac{1}{\sqrt{2^n}}\sum_{x=0}^{2^n-1}\ket{x}_{\texttt{\tiny vertex}}\in
\left(\mathbb{C}^2\right)^{\otimes n}$} by applying a Hadamard gate
(denoted by $H$) to each qubit. The  $\texttt{flag}$ qubit is prepared in the
state $\ket{-} \equiv\frac{1}{\sqrt{2}}\left(\ket{0}-\ket{1}\right)= HX\ket{0}$.
Grover's search is implemented by  repeatedly applying the
``Grover iterations'' a number of times specified by $m$.
After $m$ Grover iterations, the register $\texttt{vertex}$ is measured
in the computational basis. The measurement result ($n$ classical bits) is
intended to yield a solution to the problem.

The effect of the Grover iteration is the combined effect of an oracle query
followed by the Grover diffusion. To explain the key role of the quantum
oracle, it us useful to formulate the search problem  as follows. Let
$\{x_1,\dots , x_{N} \} $ denote the set of the $N$ unordered items. We define
a classical function \mbox{$f:\{x_1,\dots , x_{N} \} \rightarrow \{0, 1\}$} such that
$f(x)=1$ if and only if $x$  has the property we are looking for, and $f(x)=0$
otherwise. The problem thus consists in finding an item  \mbox{$x\in \{x_1,\dots ,
x_{N} \}$} such that $f (x) = 1$. The quantum oracle $O_f$
corresponding to the classical function $f$ is  a unitary implementation of $f$.
It is commonly defined as
\begin{equation}
O_f: \, \ket{x}_{\texttt{\tiny vertex}}\ket{z}_{\texttt{\tiny flag}}
\mapsto \ket{x}_{\texttt{\tiny vertex}}\ket{z\oplus f(x)}_{\texttt{\tiny flag}},
\label{Eq:Oracle_f}
\end{equation}
where $z\in\{0,1\}$ and  $\oplus$ represents a bitwise XOR. If we choose $z=0$,
the $\texttt{flag}$ qubit  outputs the value $1$ if and only if  $x$ is a
solution to the search problem. We say the oracle {\em marks} the solution
states. The crucial property is that the oracle can be queried on a
superposition of $N$ input states, and to compute the corresponding function
values it needs to be queried only once:
\begin{equation}
\frac{1}{\sqrt{2^n}}\sum_{x=0}^{2^n-1}\ket{x}_{\texttt{\tiny vertex}}
\ket{0}_{\texttt{\tiny flag}}\overset{O_f}\longmapsto
\frac{1}{\sqrt{2^n}}\sum_{x=0}^{2^n-1}\ket{x}_{\texttt{\tiny vertex}}
\ket{f(x)}_{\texttt{\tiny flag}}.
\label{Eq:Oracle_on_superposition}
\end{equation}
In Grover's algorithm, we prepare the $\texttt{flag}$ qubit in the $\ket{-}$
state. The resulting effect is a ``phase kick-back'', which  gives rise to a
minus sign as a phase whenever the input is a solution state:
\begin{equation}
\ket{x}_{\texttt{\tiny vertex}}\ket{-}_{\texttt{\tiny flag}}
\overset{O_f}\longmapsto (-1)^{f(x)}
\ket{x}_{\texttt{\tiny vertex}}\ket{-}_{\texttt{\tiny flag}}.
\label{Eq:phase-kickback-effect}
\end{equation}
Observe that the state of the $\texttt{flag}$ qubit remains unaffected and we
effectively implement the transformation $\ket{x}_{\texttt{\tiny vertex}}
\mapsto (-1)^{f(x)}\ket{x}_{\texttt{\tiny vertex}}$, which is the definition
of a ``phase oracle''. However, the $\texttt{flag}$ qubit plays a
crucial role in inducing this transformation. While the factor $(-1)^{f(x)}$
seems like a global phase for a single term, it becomes a relative phase for a
superposition of inputs:
\begin{equation}
\frac{1}{\sqrt{2^n}}\!\!\sum_{x=0}^{2^n-1}\!\ket{x}_{\texttt{\tiny vertex}}
\ket{-}_{\texttt{\tiny flag}}\overset{O_f}\longmapsto \frac{1}{\sqrt{2^n}}
\!\!\sum_{x=0}^{2^n-1}\!(-1)^{f(x)}\ket{x}_{\texttt{\tiny vertex}}
\ket{-}_{\texttt{\tiny flag}}.
\label{Eq:phase-kickback-effect_superposition}
\end{equation}

The following Grover diffusion implements a reflection about the mean amplitude.
If $\alpha_x$ denotes the amplitude of the $\ket{x}$ component prior to
applying the Grover diffusion, the effect of the latter is  $\alpha_x \mapsto
2\bar{\alpha}- \alpha_x$, where $\bar{\alpha}:=\frac{1}{N}\sum \alpha_x$.
Observe that the amplitudes of the marked components (those that pick up a
negative phase after the oracle query) are amplified while the amplitudes of all
other components decrease. The combined effect of an oracle query followed by
the Grover diffusion thus results in {\em amplitude amplification} of the
solution states, while shrinking the amplitudes of all other states in the
superposition. When repeated numerous times, the amplitudes of the solution
states eventually become significantly larger than those of the non-solution
states. The quadratic speedup with respect to classical search can be understood
as coming about from adding amplitudes $\Omega\left(\frac{1}{\sqrt{N}}\right)$
to the marked items with each query, which results in an
$\mathcal{O}\left(\sqrt{N}\right)$ convergence. This convergence rate was shown
by Grover to be also optimal. Hence, the query complexity is actually
$\Theta\left(\sqrt{N}\right)$.

\subsection{The QMF Oracle}
\label{app:QMF-oracle}

\begin{figure}[ht]
\centering
    \begin{tabular}{l}
        \text{(a) Coarse profile of the QMF oracle:} \\
        \\
        \includegraphics[width=0.97\linewidth]{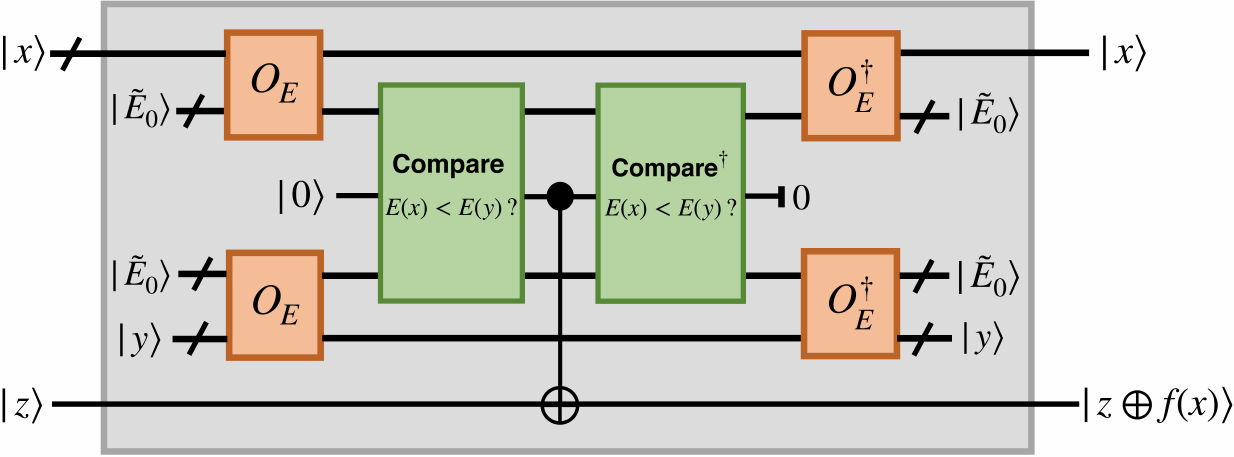}\\
        \\
        \\
        \text{(b) Energy oracle $O_E$:} \\
        \\
        \includegraphics[width=0.97\linewidth]{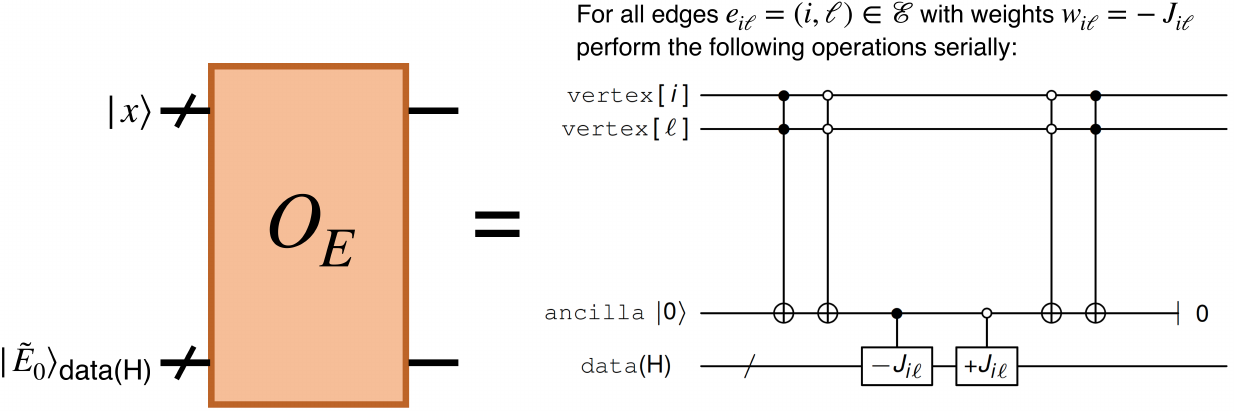}\\
    \end{tabular}
\caption{Quantum oracle as a key component of the Grover step as part of \mbox{DH-QMF}.
The oracle  marks every state whose energy is strictly smaller than the
threshold value $E(y)$, which is computed given the latest threshold index  $y$.
The result is recorded in a single-qubit $\texttt{flag}$: given its input state
$|z\rangle$ (where $z\in\{0,1\}$), the oracle outputs $|z\oplus f(x)\rangle$,
where $f(x)=1$ if, and only if, $E(x)<E(y)$, and $f(x)=0$ otherwise. (a)~The circuit
consists of several queries to the energy oracle $O_E$, which reversibly
computes the energy corresponding  to a given input state, and applications
of a unitary module called ``Compare'', which compares the values held by two registers
and records the result (0 or 1) in a single-qubit ancilla.
To infer if $E(x)<E(y)$ for a given input
$|x\rangle$, we prepare the quantum state $|y\rangle$ corresponding to the
known threshold index $y$, then independently compute $E(x)$ and $E(y)$
by separately employing $O_E$, respectively, and compare their values
using ``Compare''. The computational registers for holding the energy values
are initialized in $|\tilde{E}_0\rangle$, where  $\tilde{E}_0$ is a constant
energy shift chosen so as to avoid negative energies.
If $E(x)<E(y)$ is \texttt{TRUE}, a $1$ is recorded in an ancilla qubit that was
initialized in $|0\rangle$; the ancilla remains unaltered otherwise.
Using a \textsc{Cnot} gate, we copy out the result of the comparison to the
single-qubit $\texttt{flag}$ and reverse the whole circuit producing
this result. (b) $O_E$ is implemented by serially executing the shown circuit
template for every vertex pair $(i,\ell)$. Depending on whether
$\texttt{vertex}[i]$ and $\texttt{vertex}[\ell]$ carry the same or different
values, we respectively subtract or add the value $J_{i\ell}$ in the
$\texttt{data}(H)$ register.
}
\label{fig:QMF-Oracle}
\end{figure}

The search for a ground state of an Ising Hamiltonian
$H=-\sum_{i<\ell}J_{i\ell}Z_i Z_\ell$ (corresponding to an undirected weighted
graph with weights $w_{i\ell}=-J_{i\ell}$) requires an oracle which marks all
states whose energies are strictly smaller than the energy corresponding to the
latest updated threshold index value, respectively, which we refer to as the
``QMF oracle'' in this paper. Its quantum-circuit implementation is
shown in~\cref{fig:QMF-Oracle}. Note that here, instead of using the weights
$w_{k\ell}=\pm 0.1 j\in\left[-1,1\right]$ for $j\in\{0,1,\dots, 10\}$, we take
the weights to be the integers $-10 \le w_{k\ell}\le 10$; this facilitates the
quantum circuit implementation of arithmetic operations without altering the
underlying $\textsc{MaxCut}$ problem.

In addition to the $n$-qubit register $\texttt{vertex}$ for encoding the
possible spin configurations and any superpositions of them and a single-qubit
register $\texttt{flag}$ for holding the result of the oracle, several other
computational registers as well as ancillae  are required to reversibly compute
the energies $E(x)$ and $E(y)$ and compare their values. More concretely, we
need another $n$-qubit register to encode the value $y$ of the threshold index
as a quantum state $\ket{y}$. Furthermore, we need two computational registers
to store the computed values $E(x)$ and $E(y)$; we call these registers
``$\texttt{data}(H)$''  to indicate that they hold the computed data
related to the Hamiltonian. Both are initialized such that they initially hold
an integer $\tilde{E}_0$ that is an upper bound on
the maximum possible absolute value of an energy
eigenvalue, $\tilde{E}_0\ge\max_x \left|E(x)\right|$. This energy shift by a
constant value allows us to have a nonnegative energy spectrum, which
facilitates the implementation of the energy comparison. The maximum possible
absolute energy eigenvalue, $\max_x \left|E(x)\right|$, is bounded by the
product of the total number of edges in the graph times the maximum absolute
edge weight in the weighted graph. The registers $\texttt{data}(H)$ must thus be
able to store a value twice as large as this bound. Since generic weighted
graphs have full connectivity, the total number of edges in such graphs is
$\binom{n}{2}=n(n-1)/2$, where $n$ is the number of vertices, while the maximum
absolute edge weight  in our analysis is $\max_{(i,\ell)}
\left|w_{i\ell}\right|=10$. Hence, we may use $\tilde{E}_0:= 10\binom{n}{2}=5n(n-1)$ and choose the registers $\texttt{data}(H)$
to be of size $\lceil \log_2 \left(10n(n-1)\right)\rceil\in\mathcal{O}(\log n)$.

The energy values $E(x)$ and $E(y)$ are computed using two separate energy
oracles, whose quantum circuit implementation is provided
in~\cref{fig:QMF-Oracle}(b). For a given input
$\ket{x}=\ket{\xi_{0}}\otimes\dots\otimes\ket{\xi_{n-1}}$ held by the
$\texttt{vertex}$ register, we serially execute the shown circuit template for
every vertex pair $(i,\ell)$ in the graph whose edge $e_{i\ell}$ is nonzero.
Each such circuit subtracts or adds the value $J_{i\ell}$ in the
$\texttt{data}(H)$ register, depending on whether $\xi_i=\xi_\ell$ or
$\xi_i\not=\xi_\ell$, respectively, effectively contributing the term
$(-1)^{\xi_i}(-1)^{\xi_\ell}\left(-J_{i\ell}\right)$ to the overall energy.
The series for all pairs of vertices accumulates the sum
$\sum_{ij}(-1)^{\xi_i}(-1)^{\xi_\ell}\left(-J_{i\ell}\right)$, which together
with the initial value $\tilde{E}_0$ results in the value \mbox{$E(x)=\tilde{E}_0 -
\sum_{i\ell}(-1)^{\xi_i}(-1)^{\xi_\ell}J_{i\ell}$} held by the $\texttt{data}(H)$
register as output of the energy oracle $O_E$. Similarly, we obtain the value
$E(y)=\tilde{E}_0 - \sum_{i\ell}(-1)^{\eta_i}(-1)^{\eta_\ell}J_{i\ell}$ for the
quantum state $\ket{y}=\ket{\eta_{0}}\otimes\dots\otimes\ket{\eta_{n-1}}$
corresponding to the  threshold index $y$. For generic weighted graphs with full
connectivity, this serial implementation contributes a factor $\mathcal{O}(n^2)$
to the overall circuit depth scaling. Moreover, there is an additional
contribution from the arithmetic operations needed to implement addition and
subtraction of the constant integer $J_{i\ell}$ within the $\texttt{data}(H)$
register. Our circuit implementations and resource estimates have been obtained
using projectQ~\cite{projectQ}. The implementation of addition or subtraction
of a constant $c$, that is,  $\ket{E}\mapsto \ket{E\pm c}$, in projectQ
\cite{steiger2018projectq} is based on Draper's addition in
Fourier space \cite{draper2000addition}, which allows for optimization when
executing several additions in sequence, which applies to our circuits.
Due to cancellations of  the quantum Fourier transform (QFT) and its inverse,
$\text{QFT}\:\text{QFT}^{-1}=\mathds{1}$, for consecutive additions or subtractions
within the sequence given by the serial execution of circuits shown
in~\cref{fig:QMF-Oracle}(b), the overall sequence contributes a multiplicative
factor scaling only as $\mathcal{O}(\log\log n)$ to depth, and a multiplicative
factor in $\mathcal{O}(\log n\,\log\log n)$ to the gate complexity. To
understand these contributions, recall that the registers $\texttt{data}(H)$
are of size $\mathcal{O}(\log n)$. The remaining initial QFT and the final
inverse QFT, which transform into and out of the Fourier space in that scheme (cp.~\cite{steiger2018projectq}), contribute an additional additive term
$\mathcal{O}\left((\log n)^2\right)$ to both the depth and the gate complexity
of the overall sequence. Hence, the implementation of the energy oracle $O_E$
contributes the factors $\mathcal{O}(n^2\log\log n+ (\log n)^2)$ to the overall
circuit depth and $\mathcal{O}(n^2\log n\log\log n+(\log n)^2)$ to the overall
gate complexity.

The energy computation is followed by a unitary operation called
``Compare'', which compares the energies $E(x)$ and $E(y)$.
Using methods developed in \cite{berry2018improved},
we can implement this comparison by a circuit with a depth only logarithmic
in the number of qubits, that is, with a depth in $\mathcal{O}(\log \log n)$,
while its gate complexity is $\mathcal{O}(\log n)$.
An additional single-qubit ancilla  is used to store
the result of the comparison. Concretely, initialized in state
$\ket{0}$, the ancilla is output in the state $\ket{f(x,y)}$, where
\begin{equation}
f(x,y) = \begin{cases}
      0, & \text{if}\ E(x)\ge  E(y) \\
      1, & \text{if}\ E(x)<  E(y)\;.
    \end{cases}
\end{equation}
Using a $\textsc{Cnot}$ gate, we copy out this result to the single-qubit
$\texttt{flag}$ (bottom wire) and reverse the whole circuit used to compute
the result so as to uncompute the entanglement with the garbage generated
along the way.

In summary, the QMF oracle is a quantum circuit of depth
$\mathcal{O}\left(n^2\log \log n+ (\log n)^2\right)$
and gate complexity $\mathcal{O}\left(n^2\log n\log \log n+(\log n)^2\right)$.
The Grover diffusion requires an $n$-controlled $\textsc{Not}$ gate to implement
the reflection, which is a circuit of depth and gate complexity both scaling as
$\mathcal{O}(n)$ in terms of elementary gates. Putting all contributions
together, a single Grover iteration in our implementation has a circuit of 
depth in $\mathcal{O}\left(n^2\log\log n+ (\log n)^2 +n\right)$, while its gate
complexity is $\mathcal{O}(n^2\log n\log\log n+ (\log n)^2 +n)$.
While we have not explicitly shown it, we note that the growth rates of circuit
depth and gate counts are lower-bounded by the same scalings, meaning that
in the above expressions we may replace the $\mathcal{O}(\cdot)$ notation
by $\Theta(\cdot)$.

As an additional final remark, we note that it is possible to achieve
a slightly better circuit depth scaling for the Grover iteration,
namely as $\mathcal{O}\left(n + (\log n)^3 +\log\log n\right)$, by a parallel
(instead of serial) execution of the circuit components shown
in~\cref{fig:QMF-Oracle}(b) pertaining to each vertex pair $(i,\ell)$ in the
graph. However, this parallelization would come at an unreasonably high
additional space cost, as it would necessitate the use of $n(n-1)$
computational registers of size $\mathcal{O}\left(\log n\right)$
instead of only two. The number of qubits required would scale as
$\mathcal{O}\left(n+n^2\log n\right)$. In contrast, our serial implementation
above requires only $\mathcal{O}\left(n+\log n\right)$ qubits.

\end{appendices}

\bibliography{main}

\end{document}